\documentclass[twocolumn]{aastex62}
\pdfoutput=1 %for arXiv submission
\usepackage{amsmath,amssymb,amstext}
\usepackage{gensymb}
\usepackage{url}
\usepackage{tabularx}
\usepackage{array}

\submitjournal{ApJ}
\accepted{April 9, 2021}

\shorttitle{How Cosmic Rays Navigate the Multiphase ISM}
\shortauthors{Bustard et al.}

\begin{document}

\title{Cosmic Ray Transport, Energy Loss, and Influence in the Multiphase Interstellar Medium}

\correspondingauthor{Chad Bustard}
\email{bustard@ucsb.edu}
\author{Chad Bustard}
\affil{Department of Physics, University of Wisconsin-Madison, 1150 University Avenue, Madison, WI 53706, USA}
\affil{Kavli Institute for Theoretical Physics, University of California - Santa Barbara, Kohn Hall, Santa Barbara, CA 93107, USA}

\author{Ellen G. Zweibel}
\affil{Department of Physics, University of Wisconsin-Madison, 1150 University Avenue, Madison, WI 53706, USA}
\affiliation{Department of Astronomy, University of Wisconsin - Madison, 475 North Charter Street, Madison, WI 53706, USA}

\begin{abstract}

The bulk propagation speed of GeV-energy cosmic rays is limited by frequent scattering off hydromagnetic waves. Most galaxy evolution simulations that account for this confinement assume the gas is fully ionized and cosmic rays are well-coupled to Alfv{\'e}n waves; however, multiphase density inhomogeneities, frequently under-resolved in galaxy evolution simulations, induce cosmic ray collisions and ionization-dependent transport driven by cosmic ray decoupling and elevated streaming speeds in partially neutral gas. How do cosmic rays navigate and influence such a medium, and can we constrain this transport with observations? In this paper, we simulate cosmic ray fronts impinging upon idealized, partially neutral clouds and lognormally-distributed clumps, with and without ionization-dependent transport. With these high-resolution simulations, we identify cloud interfaces as crucial regions where cosmic ray fronts can develop a stair-step pressure gradient sufficient to collisionlessly generate waves, overcome ion-neutral damping, and exert a force on the cloud. We find that the acceleration of cold clouds is hindered by only a factor of a few when ionization-dependent transport is included, with additional dependencies on magnetic field strength and cloud dimensionality. We also probe how cosmic rays sample the background gas and quantify collisional losses. Hadronic gamma-ray emission maps are qualitatively different when ionization-dependent transport is included, but the overall luminosity varies by only a small factor, as the short cosmic ray residence times in cold clouds are offset by the higher densities that cosmic rays sample.

\end{abstract}


\section{Introduction}

%Developing a predictive theory of galaxy formation and evolution is a long-standing challenge in astrophysics, and a main barrier to further progress lies in the extremely multi-scale nature of feedback -- the process by which galaxies regulate their star formation through accretion and expulsion of gas. Through a combination of analytic, semi-analytic, and computational work, the community has undertaken a Hurculean effort to understand feedback by connecting physical processes across scales ranging mainly from Mpc down to pc scales, roughly the current resolution limit of galaxy-scale simulations employing various refinement and/or optimization techniques (e.g. computation on GPUs). 

%THIS INTRO NEEDS TO SHED SOME WEIGHT PROBABLY...
%\textcolor{blue}{I think there is a lot of good material here, but it could focus on making the case for why cosmic ray transport is important for understanding the role of cosmic rays in feedback and more broadly in ISM dynamics. This would mean mentioning that collisionless heating occurs in the self confinement model but not the balanced extrinsic turbulence model, that transport seems necessary for cosmic rays to drive a wind, that coupling in dense cool gas affects the mass flux and the star formation rate, a la Farber et al., and the role of bottlenecks. Then go on to discuss whether mock observations of simulations done with various transport models can be used to discriminate between them. What do you think of this approach? }

Highly energetic cosmic rays, despite representing only a microscopic fraction of particles by number density ($n_{CR}/n_{gas} \approx 10^{-9}$), are roughly in energy equipartition with thermal and magnetic energy in the Milky Way interstellar medium (ISM) \citep{Boulares1990}. Because hadronic cosmic rays do not suffer from radiative losses (though hadronic collisions, Coulomb collisions, and ``collisionless" scattering off magnetic perturbations can be large energy sinks), and they don't lose as much energy to adiabatic expansion compared to thermal gas ($P \propto \rho^{4/3}$ for a relativistic gas instead of $P \propto \rho^{5/3}$ for a non-relativistic gas), they fundamentally alter ISM dynamics and are believed to be an important component of feedback, which regulates star formation in galaxies. 

%While thermal pressure and radiation pressure can alone drive outflows in appropriate regimes \citep{ChevalierClegg1985, Nath2009, Murray2011}, both suffer from major setbacks that can be alleviated by non-thermal pressure support from cosmic rays. 

%Thermal pressure must compete with radiative cooling near the base of the wind to sustain a pressure gradient, while radiation pressure, except for in an optically thick medium, can squeeze through under-dense channels in the ISM without entraining ISM gas in a wind.

For GeV cosmic rays, which make up the peak of the cosmic ray energy spectrum and therefore contain most of the energy and momentum of the cosmic ray population, the dominant transport mode is self-confinement via the streaming instability \citep{Wentzel1968, Kulsrud1969, Wentzel1969}. In this picture, cosmic rays with a bulk drift speed greater than the Alfv{\'e}n speed can excite Alfv{\'e}n waves through gyroresonance and pitch angle scatter off the waves until the cosmic rays attain isotropy in the wave frame. Scattering transfers energy from the cosmic ray population to the waves, and subsequent wave damping heats the background gas. When the scattering mean free path is short\footnote{In fully ionized Milky Way ISM conditions, this is a good approximation, since the mean free path is $\approx 0.1$ pc}, cosmic rays are well-coupled to waves, and the bulk population advects at the Alfv{\'e}n speed down the cosmic ray pressure gradient directed along the local magnetic field. Extrinsic turbulence can alternatively confine cosmic rays, in which case cosmic rays scatter off the turbulent cascade. The resulting transport is field-aligned diffusion, but unlike the self-confinement model, there is no net transfer of energy to the thermal gas \citep{ZweibelReview2017}. 

To date, most galaxy evolution simulations that include a relativistic cosmic ray fluid assume some combination of diffusive and streaming transport in addition to cosmic ray advection with the non-relativistic thermal gas. Given the canonical picture that $10\%$ of supernova energy is converted to cosmic ray energy through first-order Fermi acceleration, simulations with diffusive or streaming transport included generally find colder, smoother, more extended outflows than thermally driven winds, and in many cases, the non-thermal support is necessary to launch an outflow (e.g. \citealt{Ipavich1975, breitschwerdtwinds1991, everettwinds2008, uhligwinds2012, 2013ApJ...777L..38H, SalemCROutflows2014, ruszkowskiwinds2017, MaoOstriker2018, Buck2020, 2020A&A...638A.123D, Bustard2020, HopkinsLStar2020}). Theories and observations alike also suggest the existence of a long-lived cosmic ray reservoir in the circumgalactic medium (CGM), where they may further influence the gas cycle in and out of galaxies \citep{SalemCROutflows2014, SalemCGM2016,girichidis2018, butskyWinds2018, Kempski2019, 2019PhRvL.122e1101B, Heintz2020, JiCRHalos2020}, but the extent to which cosmic rays navigate through the disk-halo interface, carry mass-loaded winds, and influence the CGM are transport-dependent. Given the myriad possibilities for cosmic rays to narrow gaps between observations and simulations of galaxy evolution, more work is needed to extend and apply first-principles fluid models of cosmic rays to galaxy evolution simulations and, likewise, to constrain these models with mock observations.

An important but mostly overlooked complication, worthy of more detailed study, is that the waves that scatter cosmic rays are not always guaranteed to be present. In particular, density inhomogeneities, frequently under-resolved in simulations of galaxy evolution, can induce alternating regions where cosmic rays are coupled or decoupled from waves. Let's consider a front of self-confined cosmic rays impinging upon a cloud. As the cosmic rays stream down their pressure gradient and encounter a drop in Alfv{\'e}n speed in the cloud, their pressure build-up at the cloud interface exerts a force on the cloud \citep{Wiener2017, Wiener2019}. Upstream of this ``bottleneck", however, the steady-state solution is a flat cosmic ray pressure profile unable to excite hydromagnetic waves and unable to exert a force along the direction of the magnetic field \citep{Skilling1971}. Cosmic rays similarly become decoupled when the confining waves are heavily damped, forcing steady-state cosmic ray streaming speeds to be much higher than the gas Alfv{\'e}n speed in order for the wave excitation rate ($\propto v_{st}$) to be comparable to the damping rate. In partially neutral clouds, where damping due to ion-neutral friction can be strong, this effect is believed to be especially significant \citep{Kulsrud1969}. Additionally, even when cosmic rays are coupled in partially neutral gas, the wave velocity that they move at is the \emph{ion} Alfv{\'e}n speed, $v_{A}^{ion} = v_{A}/\sqrt{f_{ion}}$, which may be much larger than the gas Alfv{\'e}n speed in media with low ion fraction, $f_{ion}$. For the remainder of this paper, we will use the term ``ionization-dependent" transport to refer to these collective modifications to the standard streaming transport picture.

%, even if another wave excitation mechanism besides drift anisotropy exists (such as extrinsic turbulence or pressure anisotropy).  

%\textcolor{blue}{This is a good place to cite Skillling's 1971 paper in which he predicted the existence of bottlenecks, and Josh's bottleneck papers, and mention that the the latter two studies found that bottlenecks can lead to cloud acceleration. I don't know if it's worth pointing out that even in the presence of bottlenecks, cosmic rays could be confined by extrinsic turbulence or by pressure anisotropy instabilities, but when ion-neutral damping - or any other damping mechanism - is strong the waves will be wiped out whatever their source.In that sense, the two cases are not equivalent.}

As it's been shown that cosmic ray bottlenecks can theoretically accelerate warm, fully ionized clouds in galaxy halos up to 100s of km/s, it's of great interest to understand the conditions in the \emph{ISM} rather than CGM where cosmic rays are similarly well-coupled and capable of doing work through their pressure gradients. This problem is not trivial and may have important consequences for feedback.

%Given the added complication that cloud interiors may be partially neutral, in which decoupling or fast, ``ionization-dependent transport" serves to flatten cosmic ray pressure gradients, this problem is not trivial and may have important consequences for feedback. {\bf For the remainder of this paper, we will refer to this (typically) fast transport in partially neutral gas as ``ionization-dependent" transport.}

\cite{Farber2018} ran magnetohydrodynamic (MHD) ISM ``patch" simulations of cosmic ray driven winds with a locally adaptive diffusion coefficient boosted in cold, predominantly neutral gas to mock up the effects of ion-neutral damping. This led to different gas properties, a broader spatial distribution of cosmic rays, and higher wind speeds as cosmic rays preferentially pushed on the \emph{hot} gas and escaped more easily into the halo. These pioneering ``two-$\kappa$" models, in which the diffusion coefficient shifted between a low and high value depending on gas temperature, underscored the possible implications of decoupling, but the transition from fully ionized to partially neutral transport is highly nonlinear with further dependencies on magnetic field strength and cosmic ray pressure gradient \citep{Everett2011}. An idealized study focused on teasing out some of the complicated interplay between bottlenecks, variable transport, and collisional losses in the multiphase ISM is still needed and will be presented in this paper.  

%\textcolor{olive}{Can mock observations of simulations be used to discriminate transport models?}

%Reconciling this first-principles but highly nonlinear transport, which depends on local gas conditions (e.g. density, magnetic field strength, and ionization fraction), with observational constraints such as cosmic ray grammage and gamma-ray emission will be an important challenge in the coming years.

This study is further motivated by recent full-galaxy simulations from the FIRE collaboration \citep{Chan2019, Hopkins2020} that attempt to constrain cosmic ray transport with gamma-ray emission and grammage estimates. Intriguingly, the best-fit simulations necessitate very fast cosmic ray transport speeds, especially in \emph{diffuse} gas, to lower gamma-ray emission to observed values, while fast transport just in cold, partially neutral clouds is insufficient. Such a high gamma-ray overproduction (a factor of 10 or greater) in simulated dwarf galaxies, also found in LMC-specific simulations of \cite{Bustard2020}, demands a follow-up on smaller scales. In this paper, we do not include gravity or feedback, but this step back in complexity is necessary in order to focus on the typically unresolved and already complex cosmic ray - cloud interactions without confounding results with changes in star formation rates, etc.

%, where we take a more focused, idealized approach to tease out some of the complicated interplay between bottlenecks, super-Alfv{\'e}nic streaming, and collisional losses in the multiphase ISM.

%\textcolor{olive}{Need focused, higher resolution simulations to address the true influence of cosmic rays...}
%Despite these promising advances and the rapidly growing literature on cosmic ray driven winds, a number of puzzles and opportunities for improvement remain. Specifically, gaining a clearer picture of how cosmic rays navigate and affect the multiphase ISM, frequently under-resolved in simulations of galaxy evolution, is a pressing issue.

Utilizing the \cite{JiangCRModule} cosmic ray framework implemented in the Athena++ MHD code, we run high-resolution (pc to sub-pc) 1D, 2D, and 3D simulations of cosmic ray fronts running through various multiphase ``obstacle courses" -- either single clouds or a lognormal distribution of clumps -- to probe cosmic ray transport speeds, energy loss, and momentum transfer in idealized multiphase ISM environments, focusing on transport with and without neutral gas effects included.

%, we seek to:   

%\begin{itemize}
%\item Quantify cosmic ray energy loss (or ``calorimetry") from collisional and collisionless processes when a cosmic ray front impinges upon different cloud targets or clump distributions

%\item Determine how super-Alfv{\'e}nic streaming within cold, partially neutral clouds affects gamma-ray emission and how gamma-ray emission depends on ISM conditions

%\item Compare how cosmic rays sample the ISM (avoiding clouds vs penetrating them) 

%\item Explore momentum exchange between the cosmic ray front and cold clouds, with implications for mass-loading of galactic winds

%\item Explore momentum exchange between the cosmic ray front and cold clouds. Namely, we want to see whether cosmic rays can carry mass with them, or whether they shoot through the gaps in the ISM as radiation pressure does in an optically thin medium. We expect (and show) that this is sensitive to the magnetic field strength. 
%\end{itemize}

%In addition, within the constraints of our idealized setup, we address various other questions related to cosmic ray content in clouds and simulation implementation. E.g. is the cosmic ray energy within clouds the same as outside clouds? Are the results sensitive to the assumed ionization function, grid resolution, and maximum allowed advective and diffusive fluxes? Is the cosmic ray mean free path short enough that we can safely model cosmic ray - cloud interactions with a fluid model? 

The paper is outlined as follows. In \S \ref{background}, we provide more background on the nonlinear cosmic ray transport and energy loss mechanisms we study in this work. We present our assumptions and analytic expectations for cosmic ray transport in \S \ref{analytic} and our implementation of these effects in the Athena++ MHD code \citep{AthenaRef} in \S \ref{implementation}. In \S \ref{sec:singleCloud}, we present and compare 1D and 2D simulations of cosmic ray fronts impinging upon representative ISM clouds. In \S \ref{mainSims}, we present 2D and 3D simulations of cosmic ray transport through a layer of lognormally distributed multiphase clumps in a mock ISM. For each ``obstacle course", we quantify cosmic ray energy loss (both collisional and collisionless), the momentum imparted to the gas, and probability distribution functions of cosmic ray energy and collisional energy loss  (with and without taking into account neutral gas
). In \S \ref{discussion}, we discuss the results and limitations of our work, and we conclude in \S \ref{conclusions}. 

Given the range of applications covered in this paper, the reader who is primarily interested in cosmic ray energy loss and implications for gamma-ray observations will want to read \S \ref{mainSims}. Readers who are interested in cosmic ray influence on galaxy evolution should also read \S \ref{sec:singleCloud}. 

Simulation visualizations can be found at \url{https://bustardchad.wixsite.com/mysite/visualizations}

\section{Background}
\label{background}
\subsection{Cosmic Ray Coupling and Decoupling}
\label{LitReview}

To first order, cosmic rays comprise a second, relativistic fluid that advects with the nonrelativistic thermal gas. We additionally know from the near isotropy of observed cosmic rays and their long confinement times in the galaxy, estimated from spallation measurements, that cosmic rays are constantly being scattered (see recent reviews by e.g. \citealt{ZweibelReview2017, Amato2018, Tjus2020}). Their resulting transport is well-described by a random walk along magnetic field lines with an energy-dependent diffusion coefficient. For cosmic rays at energies above a few hundred GeV, scattering off waves generated by an external turbulent cascade likely dominates the transport. Cosmic rays are coupled to the waves through resonant interactions, and transport is akin to magnetic field aligned diffusion with a coefficient depending on the amplitude of scattering waves. Unless the left and right propagating Alfv{\'e}n waves in the turbulent cascade have unequal intensities, though, there is no net transfer of energy between cosmic rays and the thermal gas. 

For GeV-energy cosmic rays, however, the confining waves can be generated by the cosmic rays themselves. Cosmic rays with even a tiny amount of drift anisotropy can resonate with magnetic perturbations and become self-confined: they exchange energy with Alfv{\'e}n waves through gyroresonance, pitch angle scatter off those waves, and become locked to the wave frame, therefore advecting at the local ion Alfv{\'e}n speed $v_{A}^{ion} = B/\sqrt{4\pi \rho_{ion}}$ (in addition to advecting with the background gas flow). In a steady-state, the conversion of cosmic ray energy to wave energy is balanced by wave damping, which heats the background gas at a rate $\propto v_{A}^{ion} \cdot \nabla P_{CR}$. We refer to this energy transfer as collisionless because it is mediated by magnetic fields, not direct collisions between particles.

In a multiphase ISM punctuated by density irregularities, however, cosmic ray coupling can quickly break down. For instance, consider a decrease in Alfv{\'e}n speed along a magnetic field line. One can show that, in steady-state, streaming cosmic rays must conform to a constant cosmic ray pressure profile for an extended region upstream \citep{Skilling1971}. Under such conditions, no waves are excited \footnote{Cosmic ray pressure anisotropy may provide an alternative confinement mechanism by exciting magnetic waves in the upstream region, where the cosmic ray streaming instability outlined here does not act. This could lock cosmic rays back to the thermal gas (instead of allowing them to decouple), but appears not to  transfer much momentum or energy to the gas \citep{Zweibel2020Anisotropy}. For this work, we will only consider cosmic ray confinement through drift anisotropy and save an analysis of pressure anisotropy for future work.}, and cosmic rays no longer transfer energy or momentum to the thermal gas. We will refer to these regions as cosmic ray ``bottlenecks" \citep{Wiener2017, ZweibelReview2017, Wiener2019}, a subset of a larger class of decoupled regions that \cite{Skilling1971} referred to as ``free-zones."

This bottleneck effect has been studied with 1D and 2D numerical simulations under simplified conditions, namely a cosmic ray front impinging on either a 1D slab or 2D cylindrical cloud. While these simulations promisingly show that cosmic ray bottlenecks may actually accelerate these cold clouds \citep{Wiener2017, Wiener2019, Bruggen2020}, thereby helping entrain them into hot outflows, they most importantly expose the complex relationship between cosmic rays and the multiphase ISM. Although cosmic ray transport is determined by local conditions on scales of a cosmic ray gyroradius, the \emph{global} environment determines these local conditions. In the bottleneck scenario, the presence of a cold cloud creates a traffic jam, decoupling cosmic rays from thermal gas, extending for macroscopic distances upstream of the cloud. One can then imagine a checkerboard of these clouds, each triggering cosmic ray bottlenecks. How a steady source of cosmic rays navigates this series of traffic jams depends on the global cloud and magnetic field structure.

Another example of ``free-zones" can occur in colder, partially neutral gas, where ion-neutral collisions decouple ions and neutrals at scales above the cosmic ray gyroradius and thereby damp the Alfv{\'e}n waves that play the crucial scattering role \citep{Kulsrud1969}. In a steady-state, the resonant streaming instability growth rate, which is proportional to the cosmic ray drift velocity, must balance this increased damping rate, effectively increasing the streaming velocity (sometimes by orders of magnitude). If ion-neutral damping is strong enough, the cosmic ray mean free path may in fact be larger than the partially neutral gas region, leading cosmic rays to free-stream. Even if the cosmic rays can drive waves at a rate fast enough to couple them to the gas, cosmic rays only scatter off Alfv{\'e}n waves that propagate in the ions, meaning the streaming velocity is rigorously the \emph{ion} Alfv{\'e}n speed, which can be orders of magnitude larger than the gas Alfv{\'e}n speed when $\rho_{ion} << \rho$. \footnote{For the remainder of this paper, we will distinguish between these two important effects, reserving the frequently-used term ``fast transport" specifically for elevated diffusive transport arising from cosmic ray decoupling. We will refer to the combination of decoupling and elevated ion Alfv{\'e}n speeds when ion fraction $f_{ion} < 1$ as ``ionization-dependent transport".}

%\cite{Kulsrud1969} pointed out the importance of ion-neutral friction for damping cosmic-ray-generated waves, leading the cosmic rays to free-stream through partially neutral clouds. \cite{Skilling1971} called these ``free-zones", where cosmic rays would rapidly spread out along field lines to form uniform conditions. This is opposed to ``wave zones", where cosmic rays are in resonance and confined by hydromagnetic waves: as \cite{Wentzel1969} described it, cosmic rays are then ``engaged by the hydromagnetic clutch." As noted in \cite{Skilling1971}, the multiphase ISM represents a series of such wave zones and free zones, with additional, extended free zones induced by density irregularities, where cosmic rays experience a decrease in Alfv{\'e}n speed and, in a steady-state, 

\subsection{Collisional Losses}

In addition to instigating complex, nonlinear transport, cold clouds are also targets for cosmic ray hadronic and Coulomb collisions, the former leading to a catastrophic decay of pions into gamma-rays. These collisional losses, depending on environment, can represent large energy sinks that suppress cosmic ray influence in galaxies \citep{HopkinsWinds2020, Crocker2020a}. This is most true for starburst galaxies, which have high volume-filling factors of dense, neutral gas and by all inferences from their gamma-ray emission appear to be good proton calorimeters \citep{Lacki2011, Tang2014, YoastHull2015, YoastHull2016, Wang2018, Krumholz2020}, meaning that all cosmic rays produced from supernovae are consumed by collisions. Observations of $L_{\star}$ and dwarf galaxies, however, paint a different picture where the vast majority of protons escape hadronic collisions presumably due to a combination of diffusion and advection in supernova-driven winds \citep{Lacki2011, FermiLMC2015, Fu2017, Lopez2018SMC}. While we only have limited gamma-ray observations of $\approx 10$ external star-forming galaxies (e.g. \citealt{Ajello2020}), and while it is difficult to exactly separate the diffuse hadronic gamma-ray emission from point-source emission, the resulting scaling relations between gamma-ray luminosity and star formation rate (SFR) are already in severe disagreement with the current iteration of MHD + cosmic ray dwarf and $L_{\star}$ galaxy simulations -- that is, simulations that include cosmic ray diffusive or streaming transport but assume that transport is not affected by ionization. 

\cite{Chan2019, Hopkins2020} vary cosmic ray transport models in their state-of-the-art full-galaxy simulations and find that gamma-rays will be overproduced compared to Local Group dwarf galaxy observations unless the cosmic ray diffusion coefficient is large ($ > 10^{29}$ cm$^{2}$/s) compared to the canonical value of a few $\times 10^{28}$ cm$^{2}$/s inferred from cosmic ray propagation models. Testing a variety of proposed first-principles transport models, \cite{Hopkins2020} constrain the issue further: propagation appears to be rate-limited by the warm ionized medium (WIM) and inner CGM, where the current paradigm of self-confinement models predicts relatively slow transport speeds. Fast transport in partially neutral gas is not sufficient to decrease gamma-ray emission, which is instead dominated by small patches of the WIM where the authors believe a runaway increase in self-confinement could be occurring. That is, large cosmic ray pressure gradients could generate confining waves that further trap cosmic rays, increase the cosmic ray pressure gradient, and so on. %Instead of cosmic ray collisions primarily occurring in dense, neutral gas, maybe this runaway causes the collisions to occur in diffuse gas or in cloud interfaces, where cosmic rays become trapped. 

Clearly, the interplay between cosmic ray transport, wave damping, and collisional losses in the multiphase ISM is very rich and has implications for gamma-ray constraints as well as general ISM and galactic wind dynamics. The goal of this paper is to shine a light on this interplay with idealized but high resolution simulations.

\section{Analytic Expectations}
\label{analytic}

Within the self-confinement picture of cosmic ray transport, the streaming velocity is

\begin{equation}
\label{vstream}
\mathbf{v_{st}} = \frac{\mathbf{B} \cdot (\mathbf{\nabla \cdot P_{CR}})}{|\mathbf{B} \cdot (\mathbf{\nabla \cdot P_{CR}})|} v_{A}^{ion}
\end{equation}
where $v_{A}^{ion} = B/\sqrt{4 \pi \rho_{ion}} = B/\sqrt{4\pi\rho f_{ion}}$, where $f_{ion}$ is the ion fraction by number, and $\rho = m n$ is the total gas density. Here, and in the following shortened derivation (see also \citealt{JiangCRModule, Hopkins2020}), we've been careful to write the Alfv{\'e}n speed as the ion Alfv{\'e}n speed, $v_{A}^{ion}$ to reflect that cosmic rays resonate with Alfv{\'e}n waves that propagate in the ions. Note that although Equation (\ref{vstream}) is written in terms of the cosmic ray pressure tensor $\mathbf{P}_{CR}$, we will assume isotropic pressure and replace the divergence by a gradient. As seen from Equation \ref{vstream}, self-confinement only occurs in the presence of a cosmic ray pressure gradient directed along the magnetic field. Equation \ref{vstream} holds, as written, when there is no wave damping, but more generally, the steady-state velocity is obtained by equating the growth rate of cosmic ray - excited Alfv{\'e}n waves $\Gamma_{CR}$ with the wave damping rate $\Gamma$:

\begin{equation}
    \Gamma_{CR}(\gamma) \approx \frac{\pi}{4} \frac{\alpha - 3}{\alpha - 2} \Omega_{0}\frac{n_{CR}( > \gamma)}{n_{i}} \left(\frac{v_{D}}{v_{A}^{ion}} - 1 \right) = \Gamma 
\end{equation}

$n_{i}$ is the ion number density, $n_{CR}( > \gamma)$ is the number density of cosmic rays with Lorentz factor greater than $\gamma$, and we'll assume $\alpha = 4$ is the power-law exponent of the cosmic ray distribution function in momentum space.

Because the growth rate of the streaming instability scales linearly with the bulk drift speed of the cosmic rays, a higher drift speed is needed when wave damping is present. The net drift with respect to the Alfv{\'e}n wave frame is $v_{D} - v_{A}^{ion}$, and we write this as a diffusive flux \citep{JiangCRModule}:

\begin{equation}
    F_{\rm diffuse} = \kappa_{||} \nabla f \approx (v_{D} - v_{A}^{ion}) f
\end{equation}

Then the (purely parallel) diffusivity is 

\begin{equation}
    \kappa_{||} = \frac{f}{\nabla f} \frac{4}{\pi} \frac{\alpha - 3}{\alpha - 2}
    \frac{\Gamma n_{i} v_{A}^{ion}}{\Omega_{0} n_{CR}( > \gamma)}
    \label{diffusivityEqn}
\end{equation}

Following \cite{Hopkins2020}, we write $f/ \nabla f = l_{CR}$, a cosmic ray scale length, and we can formulate this diffusion coefficient as an effective boost to the streaming velocity, $v_{st}^{eff} = v_{A}^{ion} + \kappa_{||}/(\gamma_{CR} l_{CR})$ where $\gamma_{CR} = 4/3$. Note that in our simulations, we implement the diffusive flux $\kappa_{||}$ and model streaming with speed $v_{st} = v_{A}^{ion}$, \emph{not} $v_{st} = v_{st}^{eff}$. We then make some substitutions ($e_{B} =  1/2 (v_{A}^{ion})^2 n_{i} m_{p}$, $e_{CR} = m_{p} n_{CR} c^{2}$) to re-write Equation \ref{diffusivityEqn} in terms of more typical code quantities to obtain:

\begin{equation}
    \kappa_{||} = \frac{4cr_{L}\Gamma e_{B}l_{CR}}{\pi v_{A}^{ion} e_{CR}}
    \label{diffusivityEqn2}
\end{equation}

\begin{equation}
    v_{st}^{eff} = v_{A}^{ion} + \frac{3cr_{L}\Gamma e_{B}}{\pi v_{A}^{ion} e_{CR}}
    \label{streamVelocity}
\end{equation}
where $r_{L} = c/\Omega$ is the cosmic ray Larmor radius, and we assume cosmic rays occupy a single energy bin at 1 GeV.

%\footnote{When $f_{ion} < 1.0$, which leads to the additional diffusive flux (Equation \ref{diffusivityEqn2}) and correction to the Alfv{\'e}n speed ($v_{A}^{ion}$ instead of $v_{A}$), we will refer to the combined effects as ``fast transport".} As in \cite{Hopkins2020}, we calculate the cosmic ray scale length on the fly assuming $l_{CR} = P_{CR}/ \nabla P_{CR}$, which is a weighted average of the scale length $f/ \nabla f$. 

%\textcolor{olive}{Instead of a factor of 4$\gamma_{CR}$, I seem to get a factor of 8 (assuming $\alpha = 4$). Is that similar to what you're getting? I suppose a factor of 8 vs a factor of 5.33 doesn't make a big difference in any of our results, but it's still kind of unfortunate.}\textcolor{blue}{There are some compromises in the foregoing treatment which need a full explanation. One is that $e_{CR}$ seems to stand for both cosmic ray energy density and the energy of a typical cosmic ray particle. Also, note that if you say $\gamma_{CR}$ is some kind of average or typical Lorentz factor then you can absorb it into $r_L$ (I'm assuming $\gamma_L=\gamma_{CR}$, but maybe that's not what you mean) Anyway, if I set $\alpha = 4$ I then get $\kappa_{||}=(32/9\pi)(c/v_{A}^{ion})(e_B/e_{CR})l_{CR}\Gamma r_L$.}

The damping rate $\Gamma$ can have contributions from multiple processes, including nonlinear Landau damping, which we find to have no bearing on our results (see \S \ref{discussion}) and turbulent damping \citep{Yan2004, Lazarian2016, Holguin2019}. In this work, we will focus on ion-neutral damping. The damping rate for a hydrogen-helium plasma is \citep{Kulsrud1969, Drury1996, 2001ApJ...558..859D, Hopkins2020} 
\begin{equation}
    \Gamma_{in} = \frac{\nu_{in}}{2} \approx 10^{-9} f_{neutral} T_{1000}^{1/2} \rho_{-24} \quad s^{-1}
\end{equation}
where $\nu_{in}$ is the collision frequency between ions and neutrals and $f_{neutral} = 1 - f_{ion}$ is the neutral fraction.

%This frequency can be written as $\nu_{H,i}\rho_{H}/\rho_{n} + \nu_{He,i} \rho_{He}/\rho_{n}$ where $\rho_{H}$ is the mass density of Hydrogen, $\rho_{He}$ is the mass density of Helium, and the collision frequencies are between neutral Hydrogen, Helium, and all other ions \citep{2001ApJ...558..859D}. The collision frequencies are proportional to the ion thermal velocities $\propto T^{1/2}$. 

%Following \cite{Hopkins2020}, we write the damping rate for a hydrogen-helium plasma as approximately

%\begin{equation}
%    \Gamma_{in} \approx 10^{-9} f_{neutral} T_{1000}^{1/2} \rho_{-24} \quad s^{-1}
%\end{equation}
%where $f_{neutral} = 1 - f_{ion}$ is the neutral fraction. 

Obtaining the ion fraction $f_{ion}$ is quite complicated as ionization depends on the local radiation field and the population of low energy (primarily 2 - 10 MeV) cosmic rays that may not follow the same transport as the GeV cosmic rays we model here. Including a more complete treatment of ionization will be the subject of future work, but for now, we remain agnostic about the exact ion fraction for clouds of varying densities and temperatures. We prescribe an ion fraction that, near $10^{4}$ K, matches the tabulated \cite{Sutherland1993} ionization fraction fairly well. Note that, while the Sutherland-Dopita table assumes collisional ionization equilibrium and only extends down to $10^{4}$ K, we extend the neutral fraction to lower temperatures following a hyperbolic tangent function. We then impose a floor on the ion fraction, $f_{ion}^{min}$, ranging from $10^{-4}$ to $10^{-1}$ to parameterize our ignorance and explore the sensitivity of our results to ionization level, although not reaching the $f_{ion}$ characteristic of Galactic molecular clouds, which can be as low as $f_{ion} \sim 10^{-8}$. For the relatively diffuse clouds we focus on, with column densities $\sim 10^{20} - 10^{21}$ cm$^{-2}$, it's likely that there are sufficient UV photons to keep carbon ionized, and ion fractions lower than $10^{-4}$ are unlikely \citep{1999RvMP...71..173H, 2017ApJ...845..163N, Silsbee2019}. The equation we use is

\begin{equation}
\label{fioneqn}
    f_{ion} = 1 - \left(\frac{1}{2} (1-f_{ion}^{min}) \left(1 + tanh\left(\frac{a - T}{c}\right)\right)\right)
\end{equation}
where T is the temperature in Kelvin, $a = 1.6 \times 10^{4}$ K, and $c = 2 \times 10^{3}$ K.

%Note that I did these steps myself and got a factor of 8 instead of 4 out front...Also note that, if following Hopkins et al. 2020, since $\gamma_{L}$ is in GeV and other quantities are (presumably) in CGS units, you can't forget a conversion from GeV to CGS, which gives an extra factor of 1/624.15 out front if you plug in $\gamma_{L} = 1$, as they do in \citep{Hopkins2020}. 

%\subsection{Ionization Fraction}
%Previous MHD simulations of cosmic ray bottlenecks assumed that the gas (both inside and outside the cloud) was fully ionized and that no wave damping mechanism was present. Here we lighten those assumptions and consider a gas with varying ion fraction $f_{ion}$. This changes the Alfv{\'e}n speed ($v_{A}^{ion} = v_{A}/\sqrt{f_{ion}}$) and adds an additional diffusive flux that is needed to balance ion-neutral damping: in a steady-state, the streaming instability growth rate must match the ion-neutral damping rate, which requires a faster bulk cosmic ray drift speed. 

\begin{figure}
    \centering
    \includegraphics[width = 0.48\textwidth]{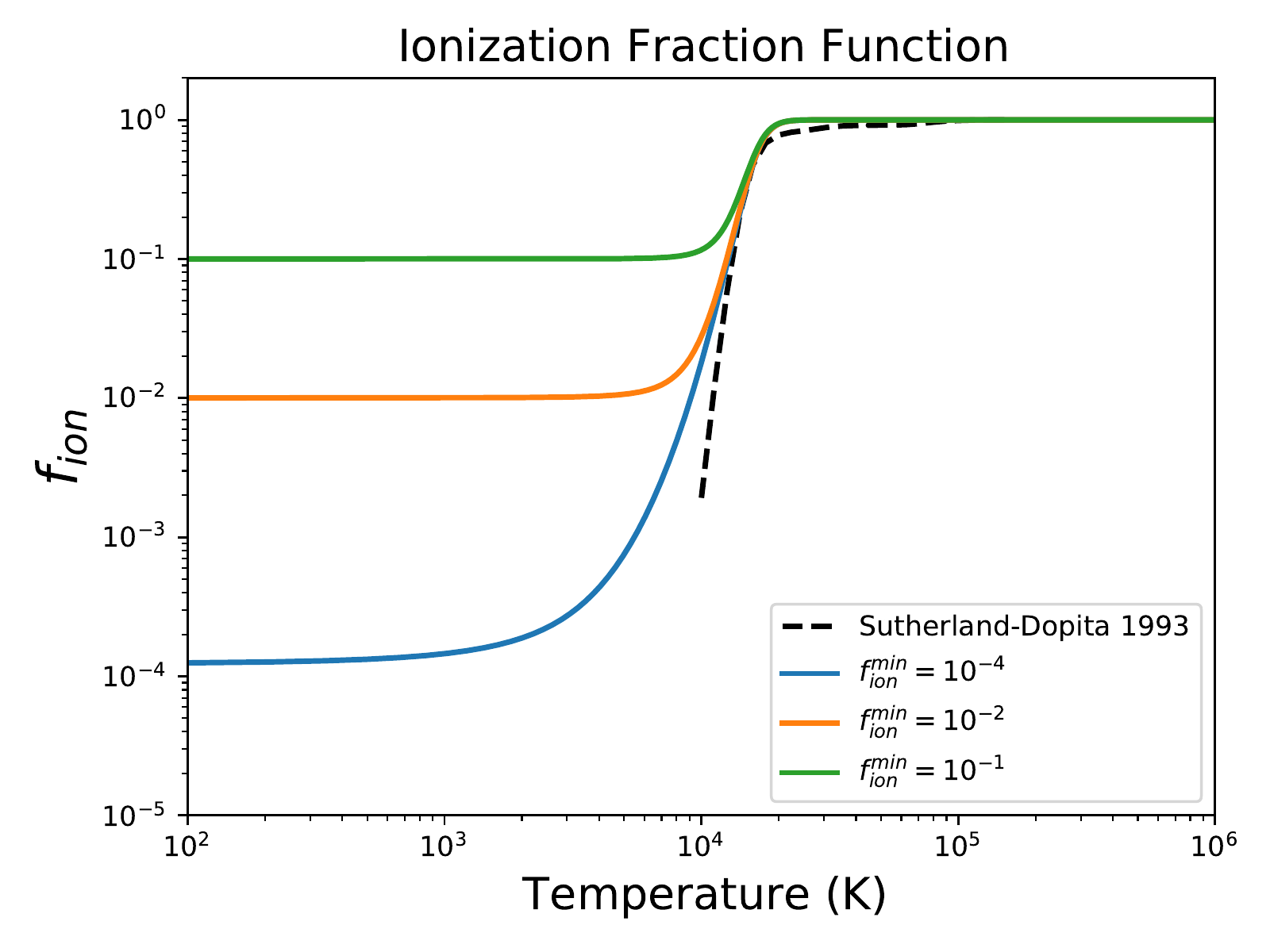}
    \includegraphics[width = 0.48\textwidth]{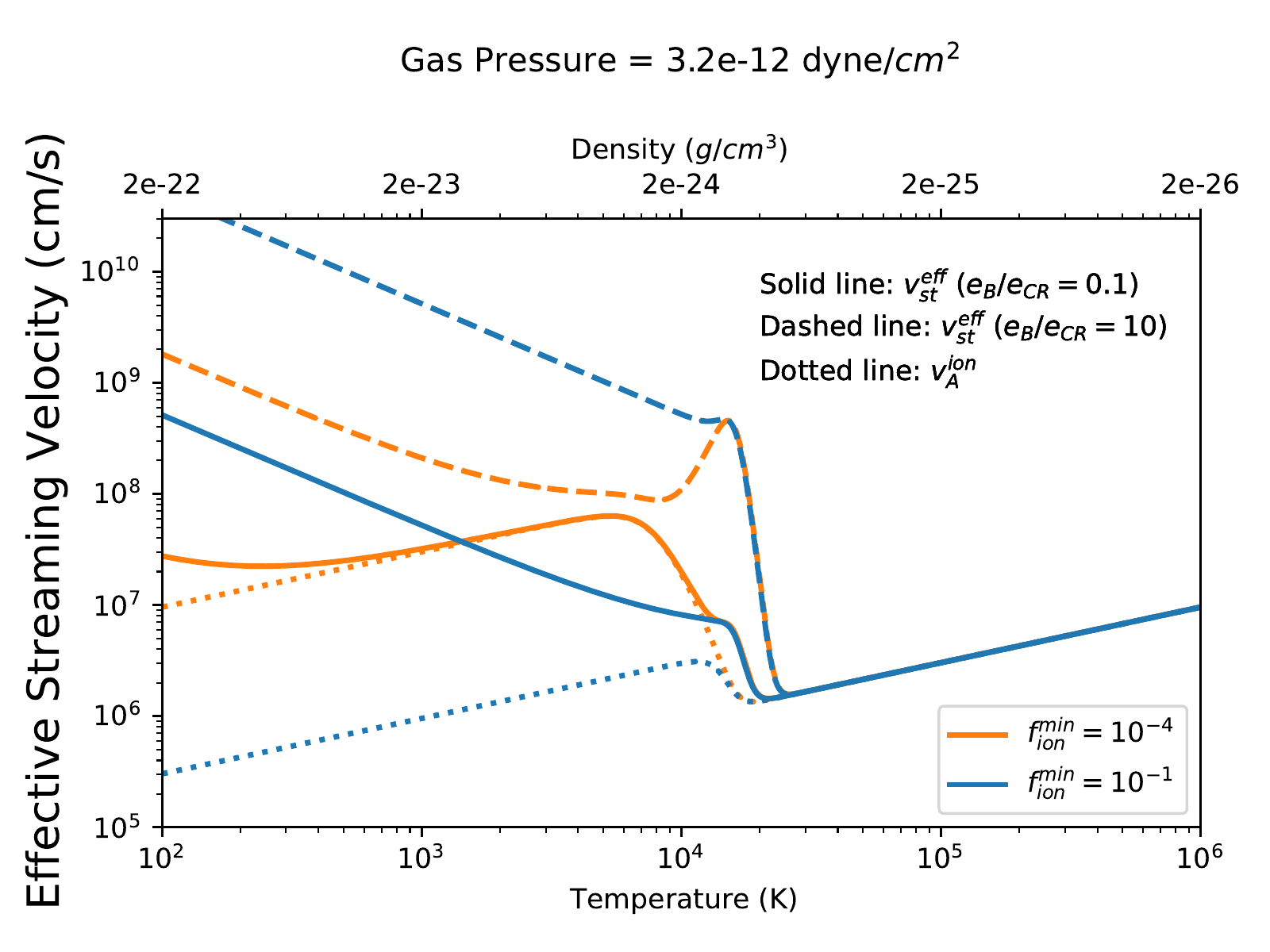}
    \caption{Top panel: Prescribed ion fraction as a function of temperature for three minimum ion values, $f_{ion}^{min} = 10^{-4}, 10^{-2}$ and $10^{-1}$. Bottom panel: Ion Alfv{\'e}n velocities ($v_{A}^{ion}$) and effective streaming velocities ($v_{st}^{eff} = v_{A}^{ion} + \kappa_{||}/(\gamma_{CR} l_{CR})$ from Equation \ref{streamVelocity}) for two different $f_{ion}^{min}$ values and varying ratios of magnetic to cosmic ray energy density. For high cosmic ray energy density, $v_{st}^{eff} \approx v_{A}^{ion}$ when the ion fraction is low, but these velocities can diverge significantly when the ion fraction is higher or cosmic ray energy density is lower. This is especially true near $10^{4}$ K, where ion-neutral damping induces a bump in $v_{st}^{eff}$.}
    \label{fig:analyticFig}
\end{figure}

Figure \ref{fig:analyticFig} shows the results of Equation \ref{streamVelocity} for ionization fraction functions with various $f_{ion}^{min}$. For a large cosmic ray energy $e_{CR}$ relative to magnetic energy $e_{B}$, the streaming velocity tracks pretty closely to the ion Alfv{\'e}n velocity as the last term on the RHS of Equation \ref{streamVelocity} is small. We also see, however, that the ion Alfv{\'e}n velocity and streaming velocity can diverge considerably when the minimum ion fraction is higher or when the cosmic ray energy density is lower. Clearly the streaming velocity is sensitive to ion fraction function, and this is especially true at cloud interface temperatures near $10^{4}$ K. Exploring this sensitivity with more realistic treatments of ionization will be the subject of future work, but as we'll see in our simulations of cosmic ray fronts, overall cosmic ray penetration into clouds is largely insensitive to whether we include or exclude ion-neutral damping (the RHS of equation \ref{streamVelocity}). This is because the dynamics are set by the cosmic ray energy density \emph{outside} the cloud, where $e_{CR}$ is by-design comparable to the thermal and magnetic energy densities in our simulations; this is the case in the average Milky Way ISM and near cosmic ray sources, and, rather than situations where $e_{CR}$ is small and therefore won't significantly affect the ISM, this is our regime of interest. 

Of course, when there is no cosmic ray pressure gradient, cosmic rays are completely decoupled from waves and free-stream, but when a pressure gradient does exist, cosmic rays generate self-confining waves. We show in our simulations that, in our regime of interest when $e_{CR}$ is not much less than $e_{B}$, a steep pressure gradient \emph{can} develop when the streaming velocity decreases in a small region at the cloud interface (compare $v_{st} (T = 3 \times 10^{4}$ K) to $v_{st} (T = 10^{5}$ K), and subsequent transport follows $v_{A}^{ion}$ fairly closely. The necessary pressure gradient to overcome ion-neutral damping can even be created from an initially uniform cosmic ray pressure when collisional losses are strong (see Appendix and \citealt{Skilling1976}). %, but these conditions are not as easily met when the cosmic ray energy density is lower. 

This is all consistent with the work of \cite{Everett2011}, who study the steady-state interaction between a population of GeV energy cosmic ray protons, with non-zero pressure gradient, and cold clouds of varying densities and magnetizations. They find that steeper pressure gradients and higher magnetic field strengths allow cosmic rays to remain coupled to waves deeper into the cloud. For a shallow pressure gradient, though, the cosmic rays eventually decouple from waves and free-stream through the cloud, leading \cite{Everett2011} to hypothesize that the cosmic ray energy inside and outside clouds will be roughly the same. Whether the diffusive flux can compensate for the drop in advective flux, leading to this scenario, depends on the dynamical and thermodynamical effects of the cosmic ray pressure gradient, which is accounted for in our simulations but not in \cite{Everett2011}, in which the cloud properties are held fixed.

\section{Computational Methods}
\label{implementation}
Until recently, a large suite of simulations probing the interaction of cosmic rays with cold, partially neutral ISM clouds would have been infeasible. The heart of the issue is that self-confined cosmic rays can only stream down their pressure gradient. To avoid grid-scale instabilities caused by this abrupt direction change at cosmic ray extrema, the predominantly employed ``regularization" method \citep{2009arXiv0909.5426S} adds artificial diffusion, which necessitates a quadratic timestep restriction ($\Delta t \propto (\Delta x)^{2}$). For fast cosmic ray transport (hence, fast signal speed) in partially neutral gas, this condition makes high-resolution simulations imprudent. 

%For example, due to severe timestep limitations, Farber et al. capped the diffusion coefficient at $3 \times 10^{29} cm^{2} s^{-1}$, while a more realistic, higher effective diffusivity could have enhanced the differences in transport models. 

%In this work, we take a step back in terms of feedback implementation (we use a constant cosmic ray flux boundary condition instead of local sources of cosmic ray and thermal energy), but we take a big step forward in terms of resolution and description of the nonlinear transport. 

We can overcome this limitation using a new cosmic ray treatment based on a two-moment method previously used for radiative transfer \citep{JiangCRModule} which has been implemented in Athena++ \citep{AthenaRef}. Unlike the regularization method, this method has no dependence on smoothing parameter near cosmic ray maxima and boasts a linear stable timestep scaling ($\Delta t \propto \Delta x$) that makes it computationally tractable to simulate very fast transport speeds and resolve the small-scale structures that induce them.  

We refer the reader to \cite{JiangCRModule} for more details (see also a similar method by \citealt{Thomas2019}) and use this space just to outline our additions of ionization-dependent transport. Namely, we implement a flag, which when turned on, calculates the streaming velocity as the \emph{ion} Alfv{\'e}n velocity everywhere (including for the $v_{A}^{ion} \cdot \nabla P_{CR}$ heating term), given the temperature-dependent ion fraction function of Equation \ref{fioneqn}. This flag also includes the additional boost due to ion-neutral friction as a diffusive flux term with diffusion coefficient given by Equation \ref{diffusivityEqn2}.

To properly model streaming with the two-moment method, there is a maximum speed of light parameter, $V_{m}$, that needs to be much greater than the maximum propagation speed. For simulations with $v_{st} = v_{A}$ (assuming a fully ionized gas), we set $V_{m} = 10^{9}$ cm/s, which is more than sufficient since the gas Alfv{\'e}n speed is safely lower than $10^{8}$ cm/s in all simulations. For simulations including ionization-dependent transport in partially neutral gas, we increase $V_{m}$ to $10^{10}$ cm/s. In this case, $v_{A}^{ion}$ approaches $10^{8}$ km/s in our fiducial cloud setup and locally exceeds that in the multi-cloud simulations of \S \ref{mainSims}. To ensure that $V_{m}$ remains safely higher than the fastest propagation speed, we also cap the streaming speed at $10^{9}$ cm/s and $\kappa_{||}$ at $3 \times 10^{30}$ cm$^{2}$ s$^{-1}$. We tested the validity of these caps in 1D simulations by increasing them each by a factor of 10 (and therefore increasing $V_{m}$ by a factor of 10, as well), and we found negligible differences.

A separate issue is whether cosmic rays with high effective streaming speeds can be treated as a fluid. We will revisit this in the Appendix, but for now, we take a detailed look at cosmic ray - cloud interactions within the fluid assumption.

We also include sink terms $-\Lambda_{C}$ and $-\Lambda_{H}$, corresponding to Coulomb and hadronic collisions, respectively, in the cosmic ray energy equation. Corresponding $+\Lambda_{C}$ and $+\Lambda_{H}/6$ terms in the thermal energy equation account for 1) the fact that all cosmic ray energy lost to Coulomb interactions heats the gas, and 2) on average, 1/6th of the energy lost to hadronic collisions ends up in secondary $e^{\pm}$ pairs, the majority of which heat the gas through Coulomb collisions \citep{Pfrommer2017}. The rest of the hadronic energy loss escapes as gamma-rays and neutrinos. We use the equations of \cite{Enblin2007, Pfrommer2017} for the Coulomb and hadronic loss terms:

\begin{equation}
    \Lambda_{\rm coul} = -2.78 \times 10^{-16} \left(\frac{n_e}{\rm cm^{-3}}\right) \left(\frac{e_{c}}{\rm erg cm^{-3}}\right) \rm erg cm^{-3} s^{-1}
\end{equation}
\begin{equation}
    \Lambda_{\rm hadr} = -7.44 \times 10^{-16} \left(\frac{n}{\rm cm^{-3}}\right) \left(\frac{e_{c}}{\rm erg cm^{-3}}\right) \rm erg cm^{-3} s^{-1}
\end{equation}
where n is the number density of nucleons, and $n_e$ is the number density of free electrons. \footnote{During the write-up of this project, we found an error in our code where we calculated the Coulomb loss rate using the total gas density instead of the density modified by ionization fraction. Coulomb losses, which should become negligible in cold, neutral gas, are then overestimated in our simulations. Since hadronic collisions dominate everywhere, this is a small correction to the total collisional loss rate and one that is, in reality, partially offset by our neglect of ionization losses that act primarily in dense gas.}

Note that, while the gas evolution equations contain collisional and collisionless heating terms, we do \emph{not} include radiative cooling. This is a practical choice that allows us to prescribe initially static gas distributions and isolate the already complex interplay with cosmic rays. We discuss the implications of this further in \S \ref{discussion} and plan to follow up with a future project including cooling. 

%The additional terms in our evolution equations ($\mathbf{v_{st}}$, $\mathbf{\sigma_{c}}$, and $V_{m}$) result from cosmic ray transport and its implementation in the two-moment method.  $\mathbf{v_{st}}$ is the streaming velocity and $V_{m}$ is the ``effective speed of light", or maximum propagation speed, which should be much greater than the streaming velocity. $\mathbf{\sigma_{c}}^{-1}$ is the interaction tensor: 

%\textcolor{blue}{It's not your fault, but I find the JO notation confusing and odd. $\sigma^{-1}$ is said to be a tensor, but elsewhere $\kappa$ refers to a single scalar diffusion coefficient, and the second term on the RHS is a vector.}

\section{MHD Simulations: Single Cloud}
\label{sec:singleCloud}

We ran a large number of 1D, 2D, and 3D simulations of different cloud sizes and densities embedded in various environments. For our single-cloud simulations, the parameter choices we present in this paper are given in Table \ref{table1}, including which subsections they apply to. We will focus here on the interaction between a cosmic ray front and one representative cloud with maximum total number density $n_{\rm cold}$ = 10 cm$^{-3}$ in a background density of $n_{\rm hot}$ = 0.1 cm$^{-3}$. The initial density profile takes the same form as in \cite{Wiener2017, Wiener2019}:

\begin{equation}
    \rho(r) = \rho_{\rm hot} + \frac{1}{2}(\rho_{\rm cold} - \rho_{\rm hot})\left (1 - \rm tanh\left(\frac{r - r_{c}}{t_{c}} \right) \right)
\end{equation}

The fiducial cloud radius, $r_{c}$, is 10 pc and tapers off with an interface that is fiducially $t_{c}$ = 5 pc thick. As we'll see, the interface thickness and how many cells resolve the interface are key parameters. The initial setup is a 2 kpc long box threaded by a constant magnetic field in the x-direction (plasma $\beta \approx 1.6$ for a 5 $\mu$G field) and at constant thermal pressure $3.23 \times 10^{-12}$ dyne cm$^{-2}$, giving a cloud temperature of $10^{3}$ K and a background temperature of $10^{5}$ K. We do not include radiative cooling, so such an unstable temperature poses no issue. At a temperature of $10^{5}$ K, the background medium is safely fully ionized, which isolates the onset of super-Alfv{\'e}nic streaming to the cloud region. The cosmic ray energy density is set to an initially negligible value. 

\begin{table*}
  \centering
  \caption{Simulation parameters for the single cloud simulations of \S \ref{1DSection}, \S \ref{2DSection}. Below are the cloud radius, $r_{c}$, and interface thickness, $t_{c}$, cloud distance, $x_{c}$ from the left boundary, cloud number density, $n_{cold}$, and temperature, $T_{cold}$, background number density, $n_{hot}$, and temperature, $T_{hot}$, magnetic field strength, B, which is 5 $\mu$G in \S \ref{1DSection} but varied between 1 and 10 $\mu$G in \S \ref{2DSection}, cosmic ray flux at the left boundary, $F_{CR}^{0}$, duration of the cosmic ray pulse, 3D grid dimensions in kpc (note that only the first one and two dimensions are relevant for 1D and 2D simulations, respectively), maximum resolution (see Appendix \ref{append1} for a convergence study), maximum speed of light parameter, $V_{m}$, which varies depending on minimum ion fraction, $f_{ion}^{min}$, capped streaming speed, and capped diffusivity. Where appropriate, differences between 1D, 2D, and 3D simulations are listed. }
  \begin{tabularx}{\textwidth}{cccccc}
  \toprule

%\begin{tabular}{c|c|c|c|c|c|c}
     \textbf{r$_{c}$}, \textbf{t$_{c}$} & \textbf{x$_{c}$} & \textbf{n$_{cold}$}, \textbf{T$_{cold}$} & \textbf{n$_{hot}$}, \textbf{T$_{hot}$} & $\mathbf{B} = B \hat{x}$ & \textbf{F$_{CR}^{0}$} \\
    
     \hline
    
    10 pc, 5 pc & 1 kpc (1D, \S \ref{1DSection}), & 10 cm$^{-3}$, 10$^{3}$ K  & 0.1 cm$^{-3}$, 10$^{5}$ K  & 5 $\mu$G (1D, \S \ref{1DSection}), & $1.54 \times 10^{-5}$ \\
    
     & 100 pc (1D+2D, \S \ref{2DSection}) &  &  & 1-5,10 $\mu$G (1D+2D, \S \ref{2DSection}) & erg cm$^{-2}$ s$^{-1}$ \\
     
     & 100 pc (3D, \S \ref{2DSection}) & & & 1,5 $\mu$G (3D, \S \ref{2DSection}) & \\
     & & & & & \\
     \hline
     %\toprule
    \textbf{Pulse Duration} & \textbf{Grid Size} & \textbf{Resolution} & \textbf{V$_{m}$} & \textbf{max v$_{A}^{ion}$} & \textbf{max $\kappa_{||}$} \\
    \hline
    
    30 Myrs & (2, 0.25, 0.25) kpc & 0.5 pc (1D),  & $10^{9}$ cm/s & $10^{8}$ cm/s & $3 \times 10^{29}$ $cm^{2}$/s \\
    
    & & 1 pc (2D+3D) & ($f_{ion}^{min} = 1.0$) & ($f_{ion}^{min} = 1.0$) & ($f_{ion}^{min} = 1.0$) \\
    
    & & (see Appendix \ref{append1}) & $10^{10}$ cm/s  & $10^{9}$ cm/s & $3 \times 10^{30}$ $cm^{2}$/s \\ 
    
    & & & ($f_{ion}^{min} = 10^{-4}$) & ($f_{ion}^{min} = 10^{-4}$) & ($f_{ion}^{min} = 10^{-4}$) \\
    \hline
\end{tabularx}
\label{table1}
\end{table*}

For the top and bottom grid boundaries ($\hat{y}$ and $\hat{z}$), we use standard outflow boundaries (values are copied over between the ghost cells and edge cells), while we use user-defined boundary conditions along the direction of the magnetic field ($\hat{x}$). Cosmic rays are injected at the left boundary by setting the cosmic ray energy flux, which follows a time profile that gradually increases and then gradually shuts off after 30 Myrs:

\begin{equation}
\begin{aligned}
F_{CR} = F_{CR}^{0} (1-e^{-t/3}) (1-e^{(t-30)/3})
\end{aligned}
\label{fluxprof}
\end{equation}
where our fiducial value of $F_{CR}^{0}$ is $1.54 \times 10^{-5}$ erg cm$^{-2}$ s$^{-1}$ and $t$ is in Myrs. By considering a finite injection period, we envision a local energy burst from a star cluster, with a characteristic lifetime of order 30 Myrs, acting on a cloud in the disk or inner galactic halo. In the outer galactic halo, where the collective effects of supernovae in the disk may be described as an approximately constant, vertical energy flux, a continuous energy injection over a few hundred Myrs may also be appropriate. For pure streaming transport, this yields a cosmic ray front with pressure of order the thermal and magnetic pressures, which in 2D and 3D is significant enough to warp magnetic field lines in the lateral direction when cosmic ray bottlenecks form at cold cloud interfaces. This is exaggerated further in simulations with lower magnetic field strength or with larger cosmic ray influxes, as we show in \S \ref{2DSection}. All other variables abide by outflow boundary conditions at this left boundary. Note that this method of injecting cosmic rays differs from that of \cite{Bruggen2020}, which held the cosmic ray energy density fixed at the left boundary, but is very similar to \cite{Wiener2019}, which implemented a cosmic ray energy source term at the left boundary. 

At the right boundary, all quantities follow outflow boundaries except for the cosmic ray energy density: we set this value to be slightly smaller in the ghost cells than in the neighboring domain cells to ensure that cosmic rays leave the grid due to the negative pressure gradient. In practice, our right boundary is always far enough away that the Alfv{\'e}n travel time to the boundary is long compared to the evolution time we consider, and the exact boundary condition does not matter. We confirmed this by running some test simulations with varying fractions of the ghost cell to edge cell cosmic ray energy densities and some with the boundary placed twice as far away, and we found our results to be insensitive to these choices. As a test, we re-ran the simulations of \cite{Wiener2019} and exactly matched their results. 

%Note that this is not the case if the boundary is placed too close to the cloud and if the cosmic ray energy density follows a true outflow condition, in which case the cloud morphology did not match that of \cite{Wiener2019} because cosmic rays don’t leave the domain.

\begin{figure*}
\centering
\includegraphics[width = 0.98\textwidth]{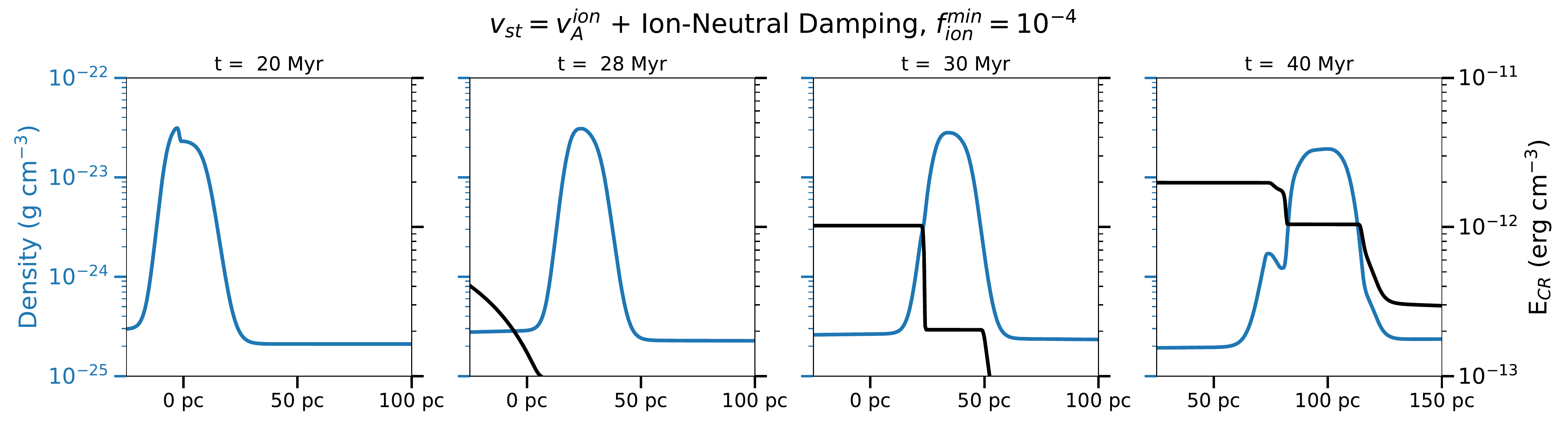}
\includegraphics[width = 0.98\textwidth]{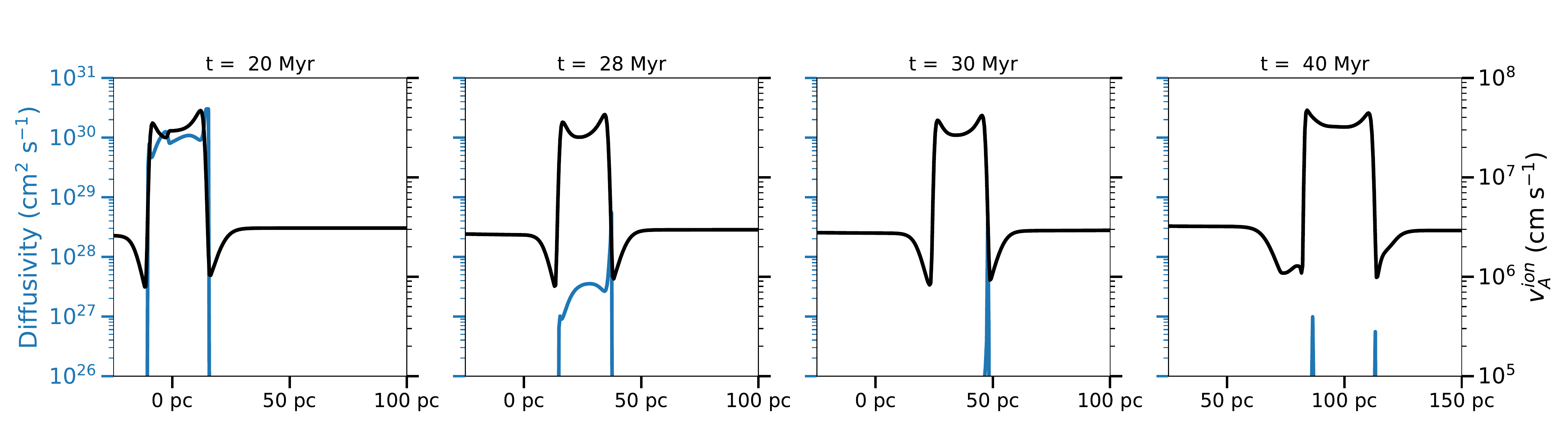}
\caption{Time series of a cosmic ray front impacting an initially stationary, partially neutral cloud of radius 10 pc, interface width of 5 pc, and threaded by a constant 5 $\mu$G magnetic field. The blue lines show the density, while the black lines show the cosmic ray energy density. The minimum ion fraction we consider is $f_{ion}^{min} = 10^{-4}$, resulting in an increase to the ion Alfv{\'e}n speed and time-varying diffusivity within the cloud (both shown in the bottom panel). The cloud initially is at 1.0 kpc away from the left boundary, which corresponds to 0 pc in these plot coordinates, but the combination of cosmic ray pressure and preceding acoustic wave pushes the cloud to the right over time. Note that the right-most panel is shifted to follow the cloud as it moves $\approx 100$ pc over 40 Myrs. Even with ionization-dependent transport included, cosmic rays bottleneck at the leading edge, where they encounter a dip in transport speed. The ensuing pressure gradient drives waves and suppresses the diffusivity due to ion-neutral damping. The resulting cosmic ray pressure gradient is steep at the front and back cloud edges. } 
\label{fig:1D_timeseries}
\end{figure*}

%\textcolor{blue}{Comment: John and I could have gotten a bottleneck because our $\rho_i$ increased by a factor of 10 going into the cloud. I don't know if we didn't because it was precluded by our boundary conditions or because of the diffusivity, which breaks the conservation law that demands bottleneck formation. } 

As the cosmic ray front builds up, cosmic rays stream down their pressure gradient towards the cloud at the Alfv{\'e}n speed ($\approx 29$ km/s for a 5 $\mu$G field). The sound speed outside the cloud is $\approx 37$ km/s, and as found in \cite{Wiener2017, Wiener2019}, the sound wave that outpaces the cosmic ray front distorts and begins to push the cloud before the cosmic rays arrive. We also ran some tests with a background temperature of $3 \times 10^{4}$ K, thereby decreasing the sound speed below the Alfv{\'e}n speed, and we found our main conclusions about cosmic ray - cloud interactions to be robust. Of course, decreasing the temperature while keeping the same cloud density means the cloud is now less pressurized, so there are some differences in the resulting cloud morphology.

\subsection{1D Simulations}
\label{1DSection}
\begin{figure*}
\includegraphics[width = 0.98\textwidth]{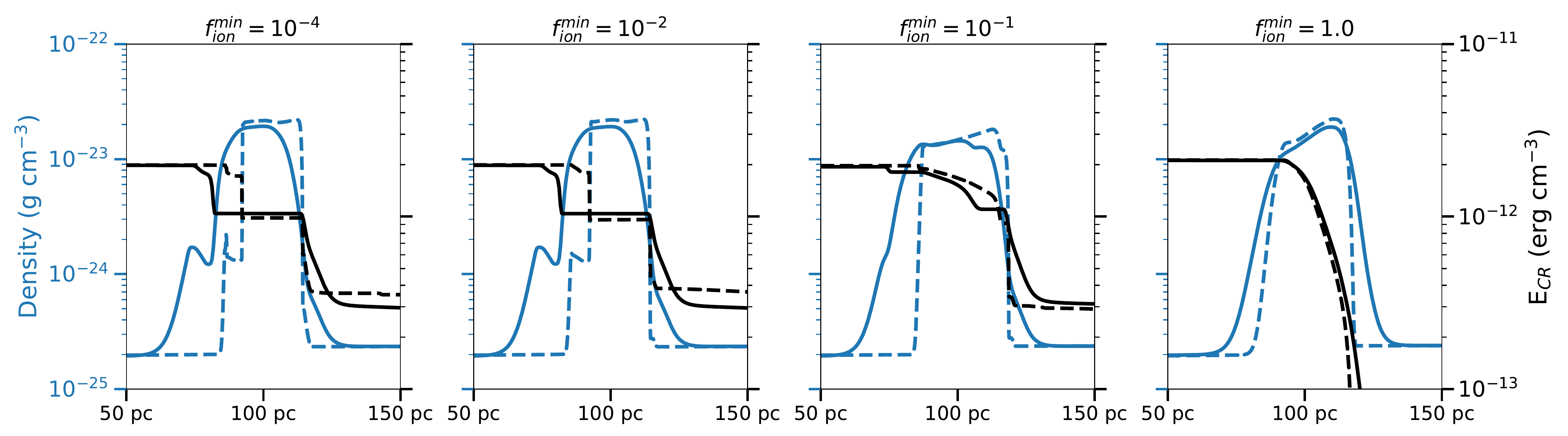}
\caption{Density (blue lines) and cosmic ray energy density (black lines) for the fiducial cosmic ray front impinging on a cloud of radius 10 pc with magnetic field strength 5 $\mu$G. We vary the interface width from 5 pc (solid lines) to 0.5 pc (dashed lines), keeping the interface resolved by 10 cells in each case (0.5 pc and 0.05 pc resolution, respectively). The panels correspond to different $f_{ion}^{min} = 10^{-4}, 10^{-2}, 10^{-1}$, and $1.0$ ($v_{st} = v_{A}$) going from left to right.}
\label{fig:1D_change_fion}
\end{figure*}

We begin with 1D simulations, which therefore force the cosmic rays to eventually enter the cloud. We place the fiducial $n_{cloud} = 10$ cm$^{-3}$ cloud at the center of the box a distance x = 1 kpc from the left boundary. We also ran simulations with the cloud placed closer to the source and find qualitatively similar results. In Figure \ref{fig:1D_timeseries}, we focus on a cloud with interface width 5 pc, and in Figure \ref{fig:1D_change_fion}, we show the effect of decreasing the interface width to 0.5 pc. In each set of simulations, the resolution is set to 1/10 of the interface width, i.e. 0.5 pc and 0.05 pc, respectively. All other parameters are kept fixed (see Table \ref{table1}).

Figure \ref{fig:1D_timeseries} shows a time-series of the cosmic ray front approaching and entering the cloud for a simulation with $f_{ion}^{min}$ set to $10^{-4}$. The top panel shows the gas density and cosmic ray energy density, while the bottom panel shows $v_{A}^{ion}$ and $\kappa_{||}$. Although the cosmic ray pressure within the cloud is initially constant, spatially dependent collisional losses sustain a small pressure gradient directed towards the cloud center, i.e. $l_{CR} > 0$. Since the cosmic ray energy density within the cloud is orders of magnitude less than the thermal and magnetic energy densities, the cosmic ray pressure gradient is still very small, thereby giving a huge effective diffusivity (see Equation \ref{diffusivityEqn2}). 

Cosmic ray propagation quickly changes when the cosmic ray front approaches the cloud. The bottom panel of Figure \ref{fig:1D_timeseries} shows that, in the cloud interface, before the temperature decreases below $\approx 10^{4}$ K, $v_{A}^{ion}$ decreases considerably. Here, the gas is still mainly ionized, and the increase in density outpaces the decrease in ion fraction. Once the temperature decreases further, the neutral fraction and, hence, $v_{A}^{ion}$ increase significantly, but not before this decrease in propagation speed has started a cosmic ray traffic-jam. This bottleneck amplifies the pre-existing cosmic ray pressure gradient (leading to small $l_{CR}$) at the cloud interface, which suppresses the diffusive flux due to ion-neutral damping. The resulting cosmic ray propagation is then dominated by advection at the ion Alfv{\'e}n speed, as cosmic rays are locked to the wave frame. This is seen in the bottom panel of Figure \ref{fig:1D_timeseries}, where the diffusivity drops from the capped value of $3 \times 10^{30}$ cm$^{2}$ s$^{-1}$ down to $\approx 10^{25} - 10^{26}$ cm$^{2}$ s$^{-1}$.

\begin{figure}
\centering
\includegraphics[width = 0.47\textwidth]{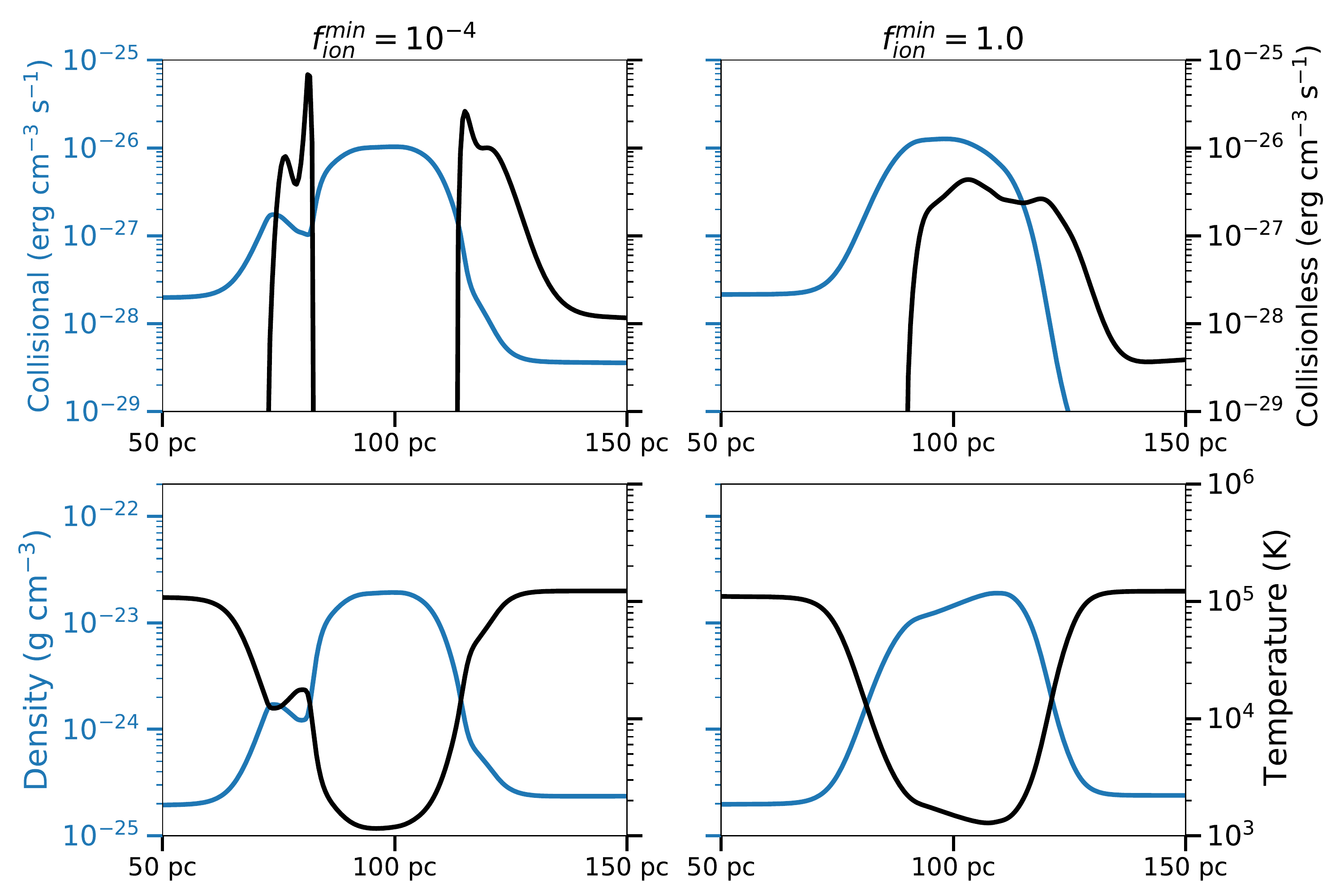}
\caption{Fiducial cosmic ray front impinging on a cloud of radius 10 pc and interface width 5 pc with $f_{ion}^{min} = 10^{-4}$ (left) and $v_{st} = v_{A}$ (right). The top row shows energy losses (both collisional and collisionless). The bottom row shows density and temperature.}
\label{fig:1D_energyloss}
\end{figure}

We check the effects of ion-neutral damping further by varying $f_{ion}^{min}$ between $10^{-4}$ and 1.0 in Figure \ref{fig:1D_change_fion}. Higher $f_{ion}^{min}$ values correspond to lower advective fluxes and higher diffusive fluxes, as expected from \S \ref{analytic} and Figure \ref{fig:analyticFig}, so we expect the effects of ion-neutral damping to be amplified at higher $f_{ion}^{min}$. This appears true in localized regions, but the advective flux still dominantly determines the transport. Compared to $f_{ion}^{min} = 10^{-4}$ and $10^{-2}$, which have almost identical density and cosmic ray energy profiles, for $f_{ion}^{min} = 10^{-1}$, the diffusivity is boosted, especially at the cloud interface region near $T = 10^{4}$ K. This partially suppresses bottleneck formation at the leading cloud edge (the cloud morphology no longer shows a pronounced bump) and drives a flat cosmic ray pressure profile in isolated regions near the cloud edges. Cosmic ray pressure gradients still drive waves throughout the cloud interior, though, locking cosmic rays to waves and suppressing the diffusivity. Because $v_{A}^{ion}$ is now lower, there is an overall slower cosmic ray passage through the cloud and a cosmic ray profile intermediate between $f_{ion}^{min} = 10^{-2}$ and $f_{ion}^{min} = 0$ (fully ionized). Clearly the nonlinear transport is especially complex in this case, but the cosmic ray energy downstream of the cloud is surprisingly very similar to that with lower $f_{ion}^{min}$ values. 

We also vary the cloud interface width from 5 pc to 0.5 pc, shown by the dashed lines in Figure \ref{fig:1D_change_fion}, keeping the interface resolved by 10 cells (0.05 pc resolution). We find that even this thin interface induces a severe bottleneck, and the amount of cosmic ray energy leaving the back side of the cloud is comparable to the thicker interface (0.5 pc) case. In light of the limited resolution of most galaxy-scale simulations, it's instructive to point out that degrading the resolution does lead to a larger cosmic ray flux out the back side of the cloud; in these cases, what should be steep cosmic ray pressure gradients at both the front and back interfaces are smoothed out, leading to an enhancement of the diffusive flux. We expand on this further in the Appendix.

We ran the same simulation assuming the medium is fully ionized and compare results in Figure \ref{fig:1D_energyloss}. The top row shows both collisional and collisionless energy loss. As expected, and as seen in Figure \ref{fig:1D_timeseries}, the steep pressure gradient that forms at the front of the cloud collisionlessly transfers cosmic ray energy to waves at a rate $\propto v_{A}^{ion} \cdot \nabla P_{CR}$. Wave damping then heats the cloud edge, as seen in the bottom panel of Figure \ref{fig:1D_energyloss}, and broadens the interface, resulting in a thumb-like density compression that moves upstream. 

In the fully ionized case, this heating tracks the cosmic ray front which, at this timestamp, has slowly moved to the middle of the cloud. In the partially neutral case, fast transport quickly flattens the cosmic ray pressure in the cloud, so heating is negligible in the cloud center. When cosmic rays hit the back of the cloud, though, they encounter a drop in streaming speed as the medium transitions from neutral to ionized (high $v_{A}^{ion}$ to low $v_{A}^{ion}$). This second bottleneck persists for a long time while cosmic ray energy slowly increases downstream of the cloud, and the steep pressure gradient there induces collisionless energy loss, just as it does on the front cloud edge. This heating may have interesting implications for ion abundances and kinematics \citep{Wiener2017}; however, we note that the energy loss rate is only of order $10^{-26}$ erg cm$^{-3}$ s$^{-1}$, which for a density of $1 - 10$ cm$^{-3}$, is far less than the expected cooling rate of order $10^{-23}$ $10^{-26}$ erg cm$^{-3}$ s$^{-1}$ or higher at temperatures of a few x $10^{4}$ K. In future work, we will include radiative cooling to address this directly and determine realistic interface widths, which are set by a balance between cooling, conduction, and cosmic ray heating. Whether low-energy cosmic rays also modify this environment by increasing the ionization fraction is an open question similarly beyond the scope of this paper.

%Plots to include:
%\begin{itemize}
%    %\item Cosmic ray sea/bath for Everett cloud, lower dens cloud, and for two interface widths -- each would show a few times
%    \item Cosmic ray front with stair-step structure -- showing a few different times. Both CR and density, and gamma-ray emission
    
%    \item subsection on ``sensitivity to ion fraction function" -- showing the same cosmic ray energy and density plots for 4 different ion fraction floors. 
    
%    \item ``Importance of the cloud interface" -- few different interface widths with plots showing the ion Alfv{\'e}n speed and diffusivity (which drops quickly once cosmic rays leak in) -- this leads perfectly into the next figure on mfp. 
    
%    \item Figure showing the cosmic ray scale length and the mean free path -- ``is this instead kinetic?"
%\end{itemize}

%Things to mention:
%\begin{itemize}
%    \item How we capture a lot of the same trends that Everett and Zweibel found with B field strength, cosmic ray pressure, etc.
%    \item Transport tracks the ion Alfv{\'e}n speed pretty closely, but this isn't true if we have a higher ion fraction (more effect from ion-neutral damping). 
%    \item The importance of the interface region
%    \item How much does CR energy density change within vs outside the cloud? Reference to resolution study in the appendix. 
%\end{itemize}

\subsection{2D and 3D Simulations}
\label{2DSection}

To explore the effects of higher dimensions and build intuition for our main suite of simulations in \S \ref{mainSims}, we study the same cosmic ray front interacting with a 2D cloud and, in two of our simulations with B = 1 $\mu$G and 5 $\mu$G, with a 3D cloud. In the cases we will show, the cloud density (10 cm$^{-3}$), radius (10 pc), and interface width (5 pc) are the same as in the 1D simulations (\S \ref{1DSection}), with an initially constant magnetic field in the x-direction, but we move the cloud to a closer distance of 100 pc from the boundary. This is more akin to a cosmic ray front impacting dense clouds within an ISM scale height of a few hundred pc. To make a fair comparison between 1D, 2D, and 3D, we re-ran our 1D simulations with the cloud at a distance of 100 pc instead of 1 kpc. The 2D and 3D simulation resolution is $\approx 1$ pc out to a distance of 1 kpc, where the grid coarsens to a resolution of 16 pc. As in the 1D simulations, the left boundary is placed 2 kpc away, and here we focus our analysis on the closest 1 kpc, where the cloud is most highly resolved.

\begin{figure}
%\centering
\hspace{-2.5em}\includegraphics[width = 0.53\textwidth]{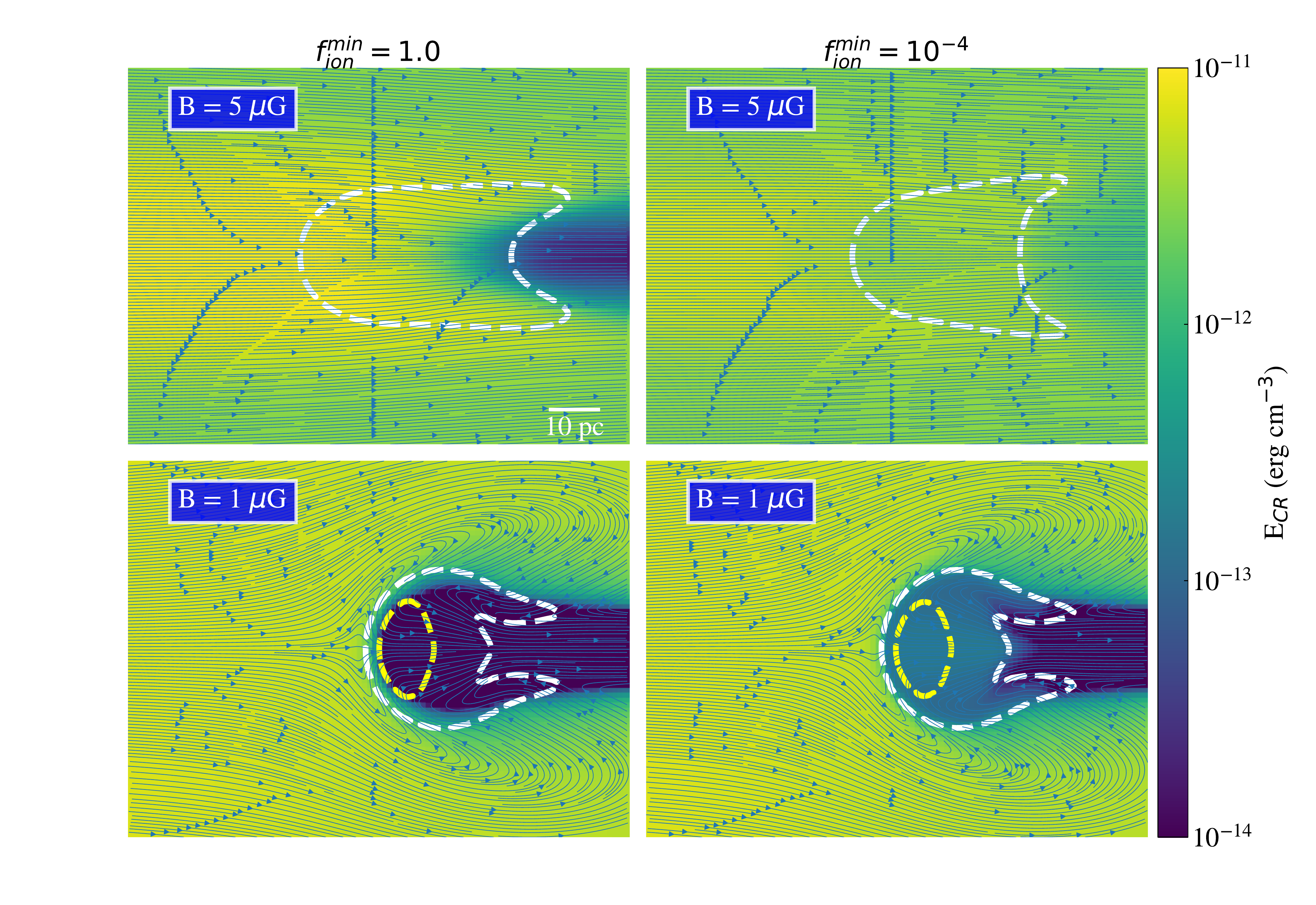}
\caption{Slices of cosmic ray energy density for various 2D, single-cloud simulations. The cloud density initially peaks at 10 cm$^{-3}$, and white, yellow contours represent densities of 1, 10 cm$^{-3}$, respectively. Varying the magnetic field strength from 5 $\mu$G (top row) to 1 $\mu$G (bottom row), we see distinct differences in magnetic field topology created by the pressure wave that precedes the cosmic ray front. In the fully ionized case (left), cosmic ray propagation traffic-jams at the front cloud edge, casting a shadow in $E_{CR}$ downstream, while allowing for ionization-dependent transport through the cloud (right) leads to a slightly less disrupted cloud, greater $E_{CR}$ downstream, and a second bottleneck at the back edge seen most clearly in the lower right panel. In the upper right panel, there are still bottlenecks at the front and back cloud edges, but the cosmic ray pressure drops are less noticeable, reflecting the faster transport that allows cosmic rays to more easily fill into the cloud along straight magnetic field lines.}
\label{fig:2D_ecr}
\end{figure}

\begin{figure*}
\centering
\includegraphics[width = 0.98\textwidth]{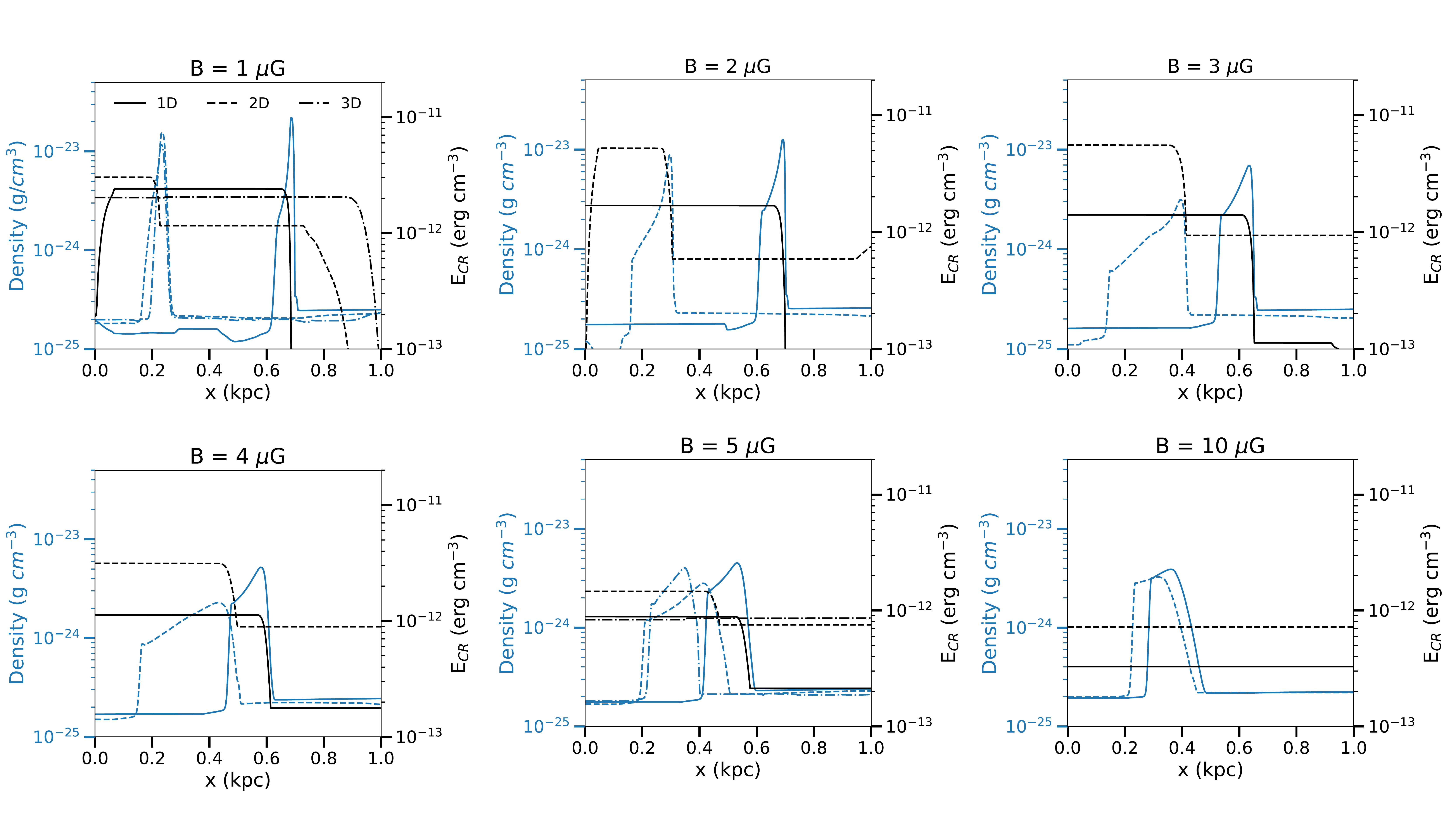}
\caption{Profiles at t = 60 Myrs down the cloud axis showing the evolution of density and cosmic ray energy density for the fiducial cloud in 1D (solid lines), 2D (dashed lines), and for two simulations with B = 1 $\mu$G and 5 $\mu$G, 3D (dot-dashed lines). Each panel is for a different magnetic field strength, and each simulation is without ionization-dependent transport ($f_{ion}^{min} = 1.0$). The cloud is initially placed 100 pc from the cosmic ray source but accelerates to varying degrees depending on magnetic field strength and dimensionality. In 1D, the cloud acceleration monotonically decreases with increasing magnetic field strength, but in 2D, there is a sweet spot around B = $3-5 \mu$G where the cosmic ray front is most effective at pushing the cloud.}
\label{1D_vs_2D}
\end{figure*}

\begin{figure*}
\centering
\includegraphics[width = 0.98\textwidth]{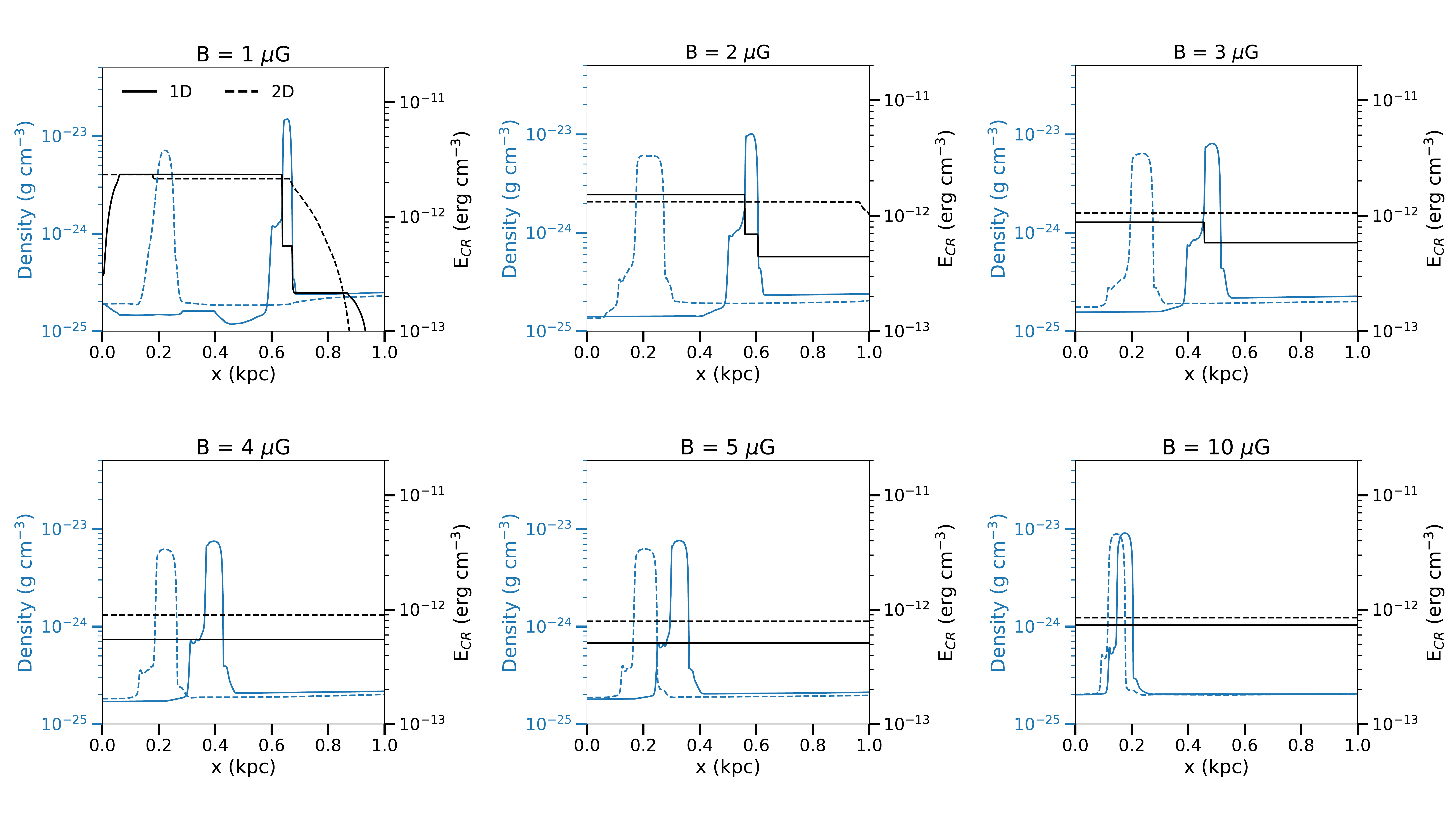}
\caption{Same as Figure \ref{1D_vs_2D} but with $f_{ion}^{min} = 10^{-4}$.}
\label{1D_vs_2D_vaion}
\end{figure*}

\begin{figure*}
\centering
\includegraphics[width = 0.32\textwidth]{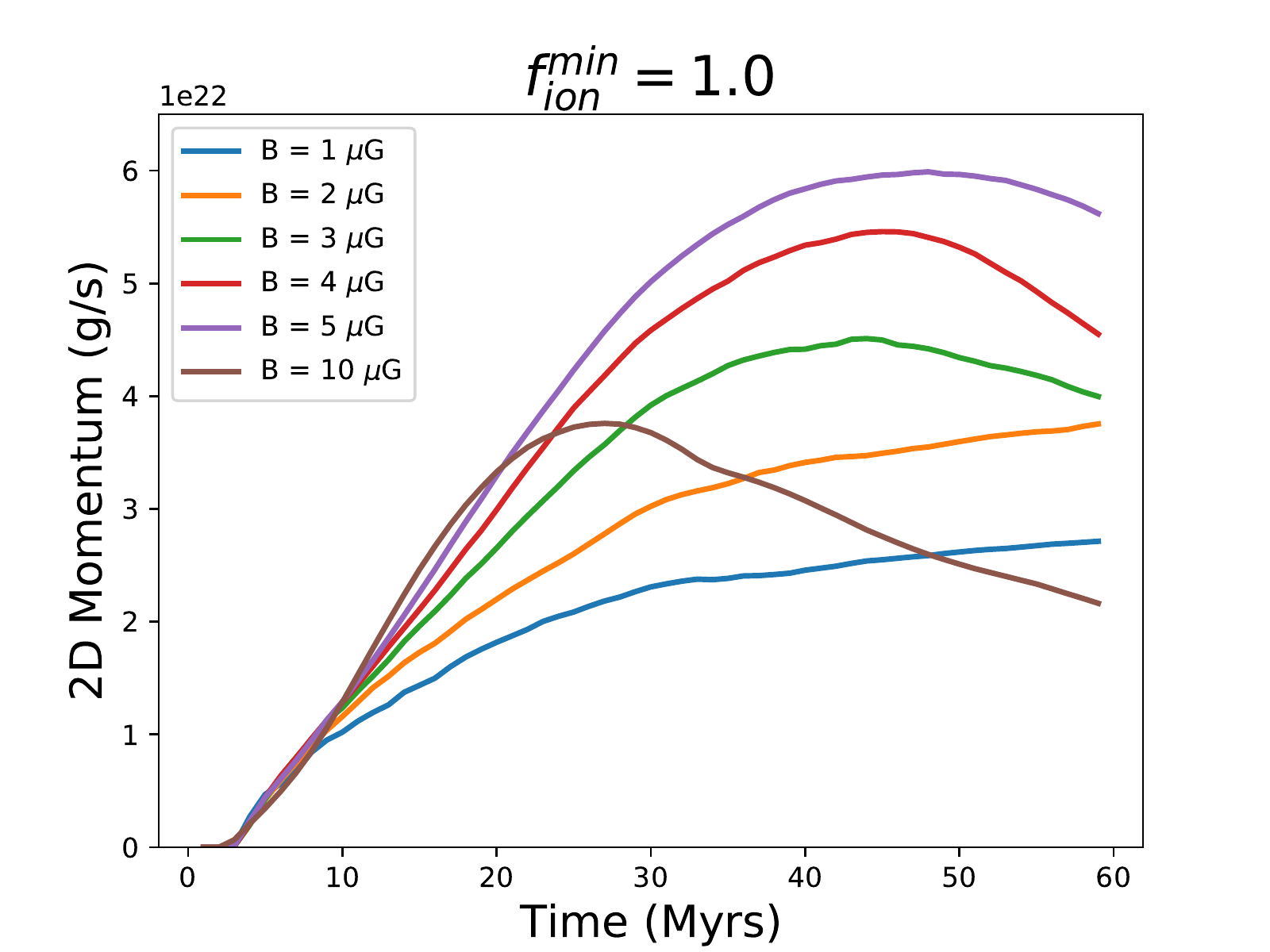}
\includegraphics[width = 0.32\textwidth]{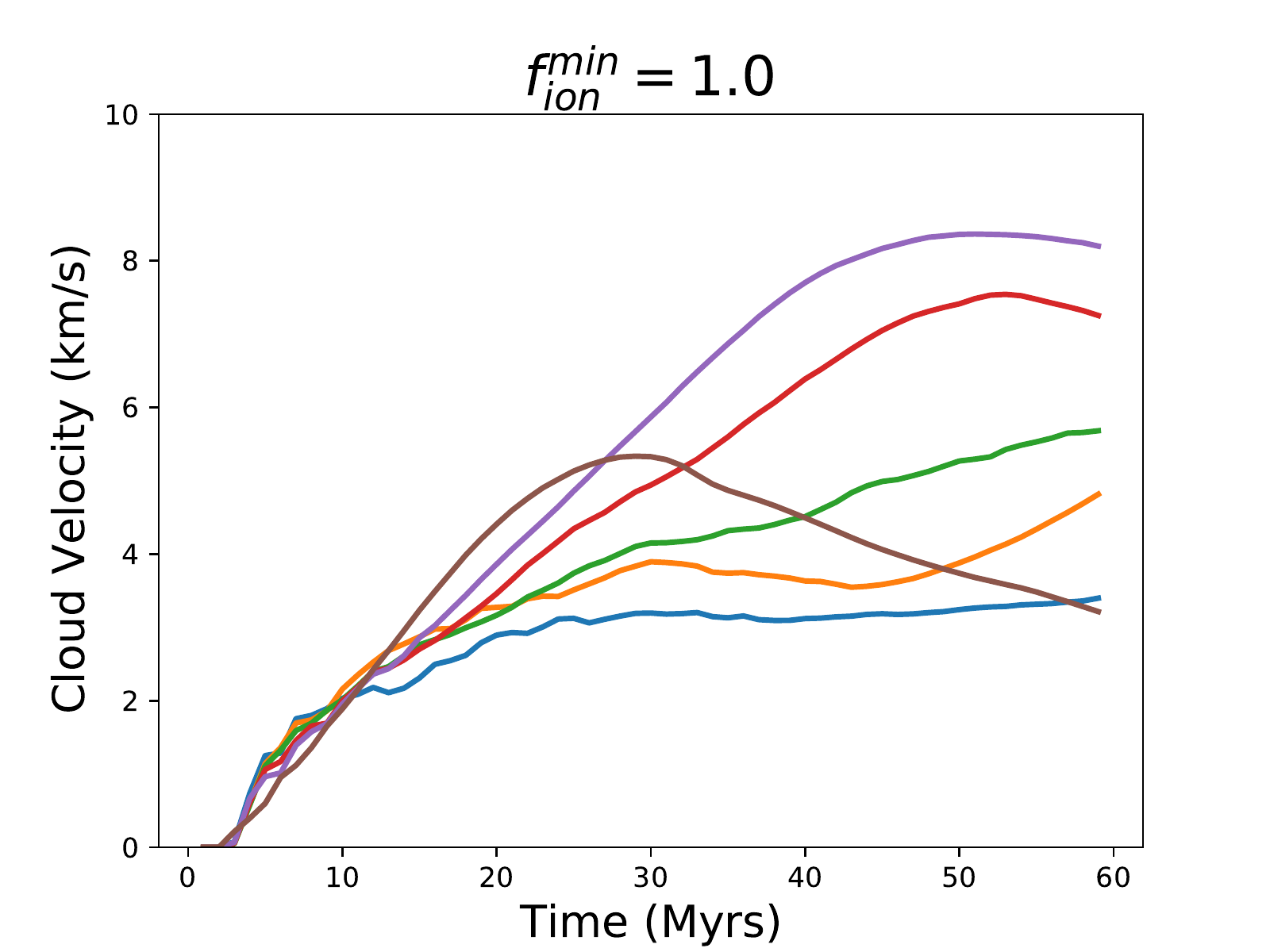}
\includegraphics[width = 0.32\textwidth]{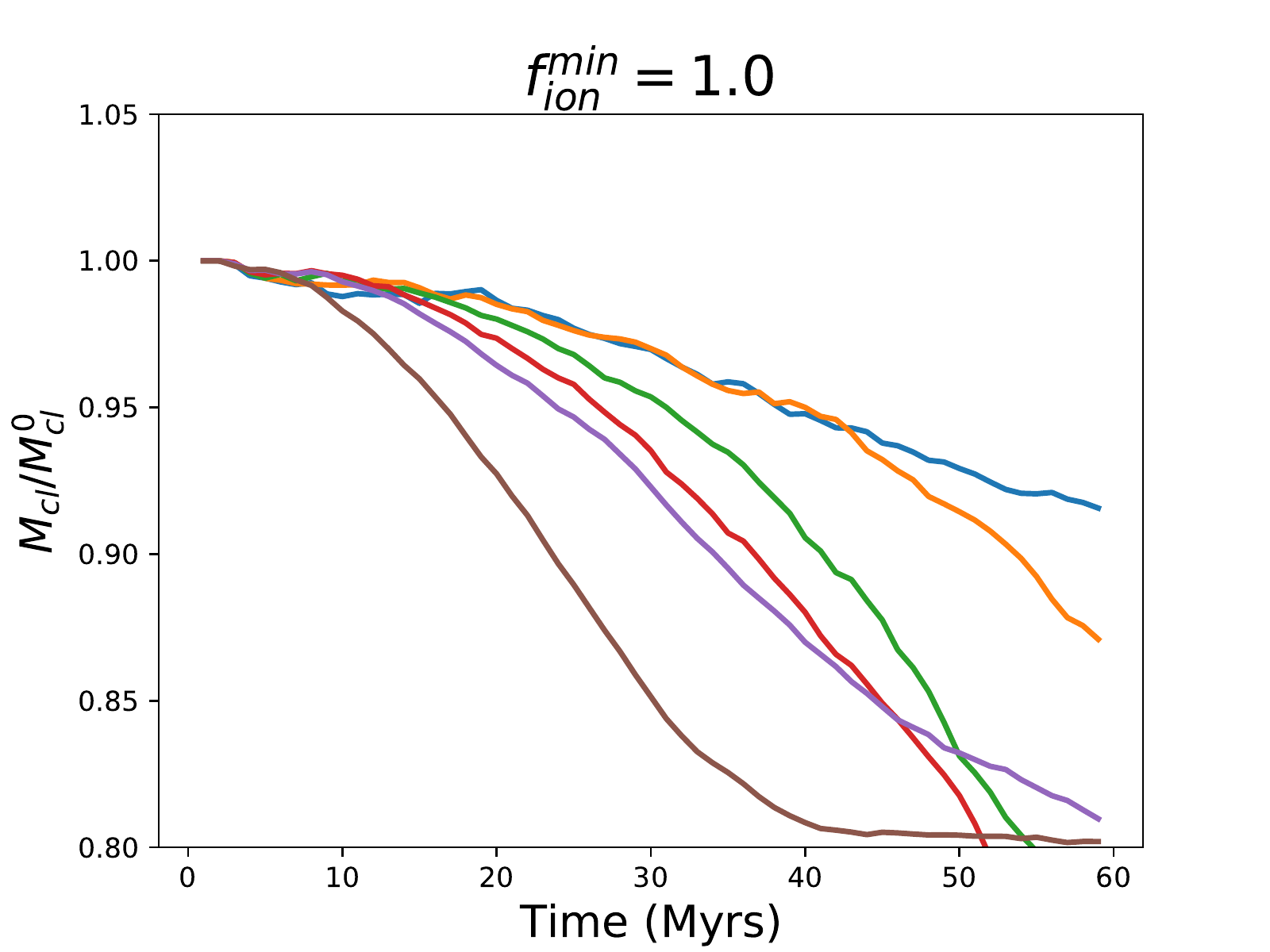}
\caption{For the fiducial 2D cloud, placed 100 pc from the left boundary, these panels show the total 2D momentum, volume-averaged velocity, and change in cloud mass $M_{cl}/M_{cl}^{0}$ over time, defining the cloud as all gas with a temperature $T < 2 \times 10^{4}$ K. For each simulation, we vary the magnetic field strength between 1 and 10 $\mu$G, and $f_{ion}^{min} = 1.0$ in all cases. The greatest cloud acceleration occurs when the magnetic field is intermediate strength $\approx 5 \mu$G, in which case the field line warping is small, but the Alfv{\'e}n crossing time between the source and the cloud, which partially determines the duration of the cosmic ray pressure gradient, is longer than the 10 $\mu$G simulation. Lower magnetic field strengths yield more field line warping around the cloud, which inhibits acceleration. In all cases, the cloud mass decreases over time, which is to be expected since we don't include radiative cooling.}
\label{2DCloud_Acceleration}
\end{figure*}

\begin{figure*}
\centering
\includegraphics[width = 0.32\textwidth]{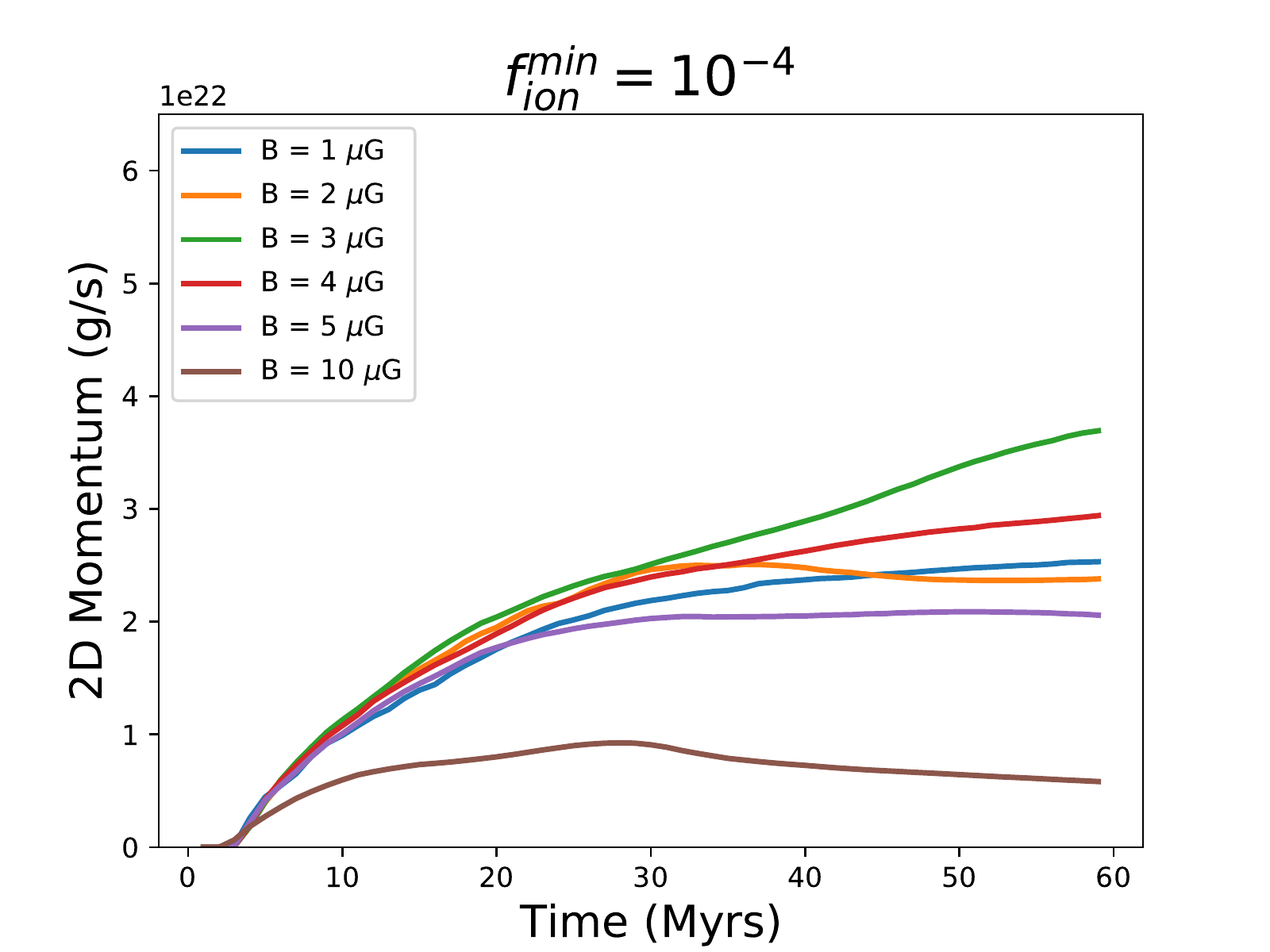}
\includegraphics[width = 0.32\textwidth]{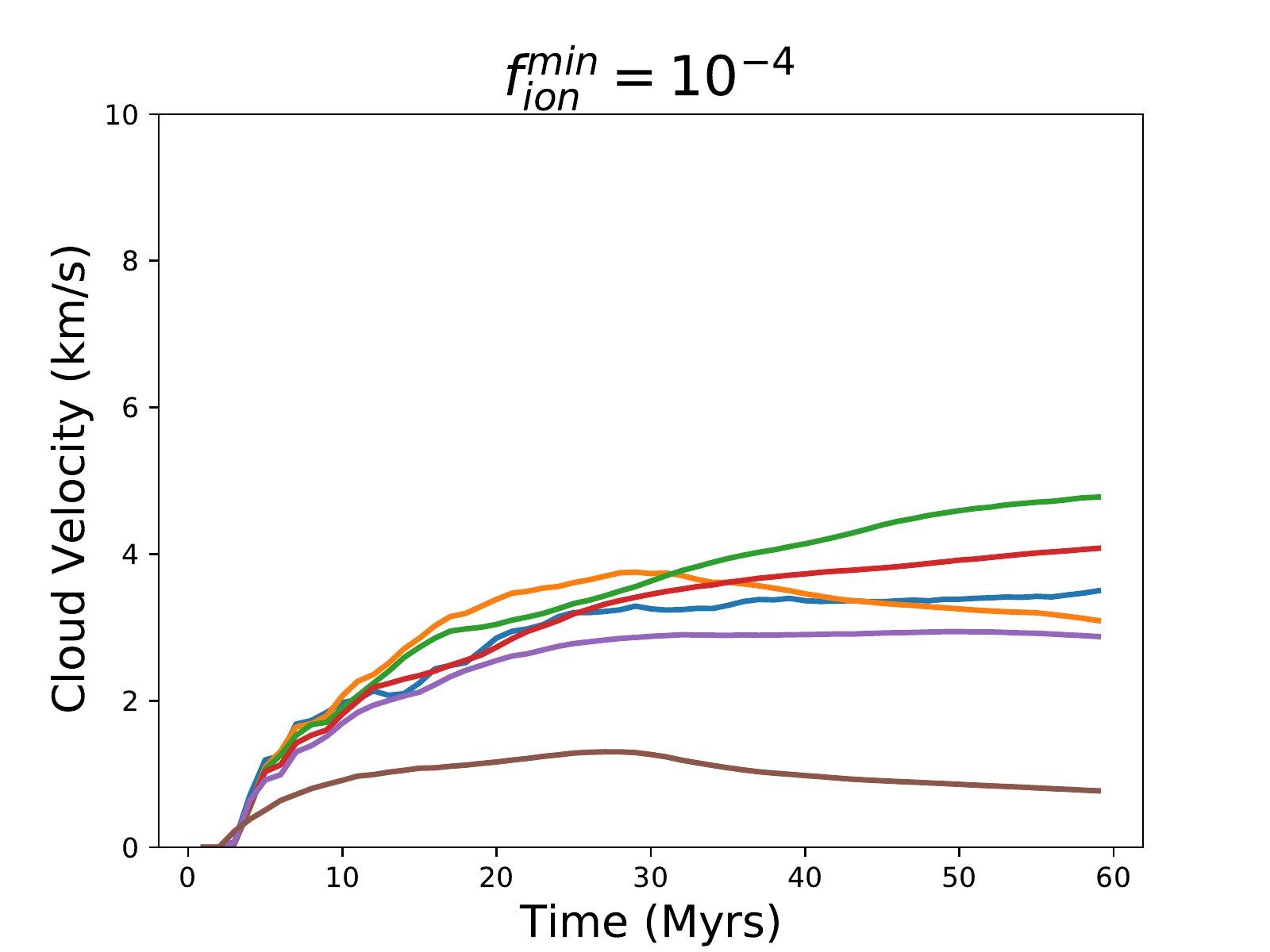}
\includegraphics[width = 0.32\textwidth]{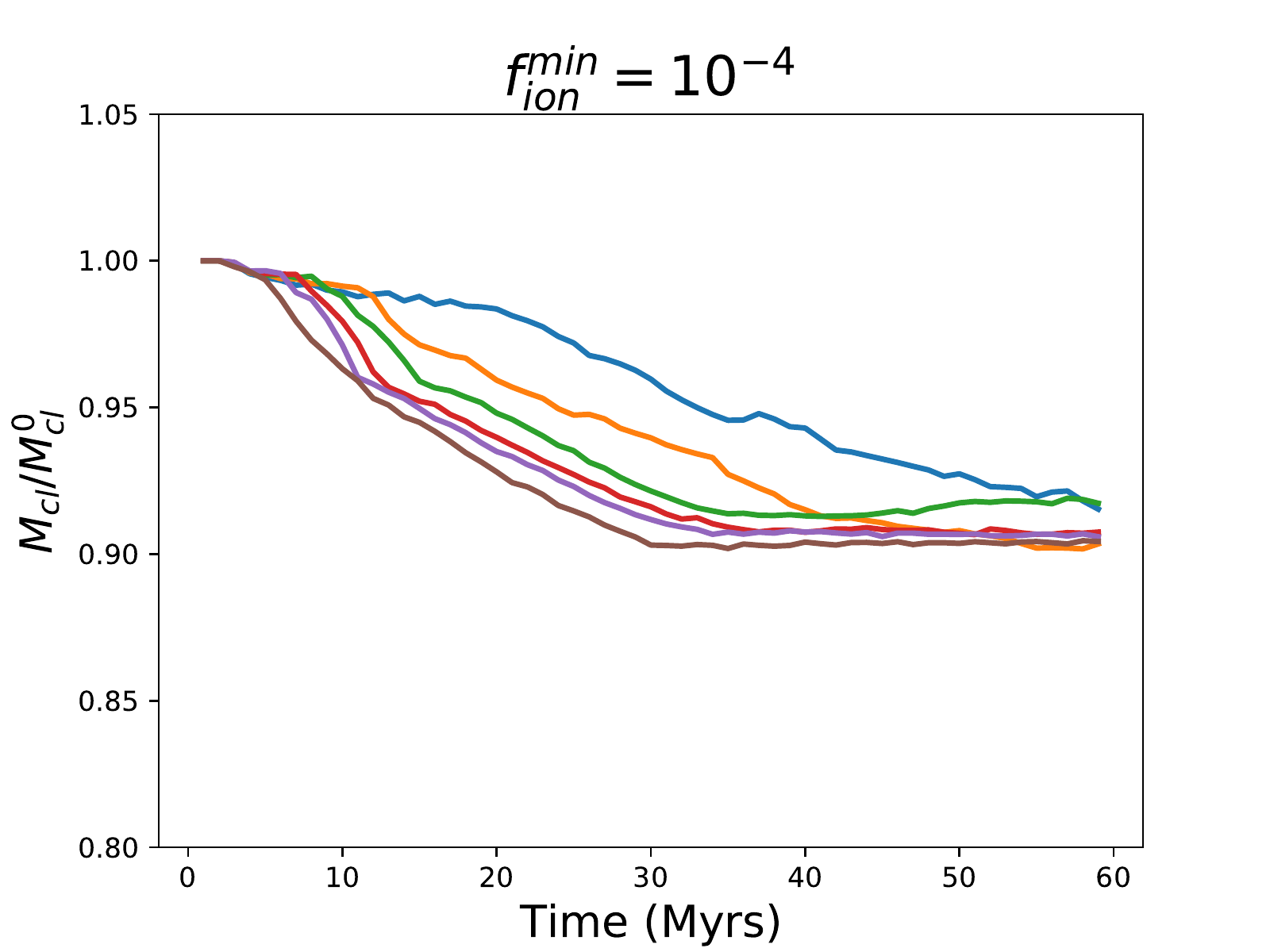}
\caption{Same as Figure \ref{2DCloud_Acceleration} but with $f_{ion}^{min} = 10^{-4}$. Cloud acceleration is decreased by a factor $\approx 2$ when ionization-dependent transport is included, and the greatest acceleration now favors somewhat lower magnetic field strengths.}
\label{2DCloud_Acceleration_fast}
\end{figure*}

%\begin{figure}
%\centering
%\includegraphics[width = 0.45\textwidth]{TotalMomentum_2DClouds.pdf}
%\caption{Momentum measured in the single cloud, 2D simulation in a $y = \pm 50$ pc, x = 0 to 2 kpc region. \textcolor{olive}{This figure will get replaced with plots of cloud mass, velocity, and momentum once we understand the results.}}
%\label{fig:2D_Momentum}
%\end{figure}

Figure \ref{fig:2D_ecr} shows snapshots from a subset of our 2D simulations with magnetic field strengths of 1 $\mu$G (bottom row) and 5 $\mu$G (top row). The cloud properties are the same as in \S \ref{1DSection}, but the extra dimensionality means that not all field lines (and hence, cosmic rays) enter the cloud, and field lines are allowed to evolve. For varying magnetic field strengths (top vs bottom rows) and varying ion fraction (left vs right columns), we see differences in how cosmic rays penetrate and pressurize the cloud and in the cloud morphology (shown by white and yellow contours of density 1 and 10 cm$^{-3}$, respectively). 

In the $B = 5 \mu$G cases, magnetic field line warping is negligible due to strong magnetic tension, and cosmic rays propagate straight across the cloud. Assuming full ionization, $v_{st}$ decreases sharply in the cloud, inducing a bottleneck on the front edge and a cosmic ray shadow behind the cloud. This shadow fills in very slowly since the collisional loss time in the cloud is significant compared to the transport time. The cloud morphology is very similar to what is seen in \cite{Wiener2019, Bruggen2020}, where the top and bottom cloud edges with the lowest gas column get pushed and form tails behind the cloud, followed eventually by full-cloud acceleration to the right. When ionization-dependent transport is included, the formation of these tails is not as dramatic since the cosmic ray pressure profile becomes flatter in the cloud and therefore doesn't push on the gas as much. The cosmic ray pressure behind the cloud is clearly larger, as well, since the cosmic ray transport time is shorter. A second bottleneck, as seen in the 1D simulations, is faintly visible on the back side of the cloud where the transport speeds drops precipitously to the gas Alfv{\'e}n speed, $v_{A}$ instead of $v_{A}^{ion}$. 

With a magnetic field strength of $B = 1 \mu$G, corresponding to a plasma beta of 75 instead of 3, the magnetic field topology plays a more significant role. Before the cosmic ray front even reaches the cloud, the preceding acoustic wave disturbs the magnetic field and warps it around the cloud. When cosmic rays approach the cloud, they must follow the stretched field lines that are draping around the cloud and therefore take longer to enter. For field lines that do penetrate the cloud, regardless of transport, the slower Alfv{\'e}n speed outside the cloud leads to a build-up of cosmic ray pressure and a severe bottleneck on the front edge. 

%One can also imagine that, if multiple clouds were placed close together, as is the case in our simulations of \S \ref{mainSims}, these bowed field lines would compress between clouds, increasing the Alfv{\'e}n velocity and leading to faster cosmic ray transport. Cosmic rays on field lines that don't initially thread a cloud, then, would be boosted through the gaps in the ISM, forming a picture akin to photons squeaking through under-densities in the ISM. 
\subsubsection{Cloud Acceleration: 1D vs 2D vs 3D} 
In each simulation, the cloud is eventually accelerated but to different extents depending on transport treatment, magnetic field strength, and dimensionality. Figures \ref{1D_vs_2D} and \ref{1D_vs_2D_vaion} compare profiles down the cloud midplane of cosmic ray energy density and gas density after $t = 60$ Myrs. Solid lines show the 1D simulations, dashed lines show the 2D simulations, and in two cases, we ran pc-resolution 3D simulations without ionization-dependent transport, shown by the dot-dashed lines. 

Let's first analyze the 1D results without ionization-dependent transport (the solid lines in Figure \ref{1D_vs_2D}), which are the most straightforward to interpret. Cloud acceleration depends on the magnitude and duration of the cosmic ray pressure gradient. Keeping in mind that the cosmic ray pulse only lasts for a finite 30 Myrs, at which point the upstream pressure will fade at a rate proportional to the Alfv{\'e}n speed, the cloud gets a push only until the cosmic ray pressures upstream and downstream equilibrate. At late times, when the pressure has equilibrated, the cosmic ray energy density slowly drops due to the collisional energy sink inside the cloud and also cosmic ray escape out the right box edge, which occurs because we enforce a cosmic ray gradient at that boundary. For a higher Alfv{\'e}n speed (higher magnetic field strength), cosmic rays build up a short-lasting and relatively low pressure upstream, and since they travel more quickly through the cloud, they lose less energy to collisions and emerge with a non-negligible pressure downstream. So the cosmic ray pressure gradient is small and also short-lived, leading to the least cloud acceleration for the B = 10 $\mu$G simulation. As seen in Figure \ref{1D_vs_2D}, the distance the cloud travels monotonically increases for decreasing field strength. 

In 2D, the magnetic field topology is a complicating factor, as we see the low and high magnetic field cases both give very little cloud acceleration. For high magnetic field strength, field line warping is negligible, and the setup is effectively 1D. Again, the pressure gradient is relatively small, and the pressures equilibrate quickly to give very little cloud acceleration. On the contrary, for low magnetic field strength, the field lines warp so much that it allows cosmic rays to move around the cloud, taking pressure from the upstream, while cosmic rays instead converge and build up behind the cloud. The cosmic ray pressure drop across the cloud is small, which again leads to relatively little cloud acceleration. 

Intermediate magnetic field strengths B $\approx 3-5 \mu$G ($\beta \approx 4.5-1.6$) give a sweet spot for cloud acceleration. In these cases, field lines don't warp as much, so the upstream pressure is higher, while the cosmic ray transit time through the cloud is intermediate (of order 5-10 Myrs and comparable to the collisional loss time), so the downstream pressure is low. This creates a sizeable and fairly long-lasting pressure gradient; however, note that the 3D simulations show flat cosmic ray pressure profiles after 60 Myrs and clouds that have better maintained their shape and moved less distance. While the differences between 2D and 3D are small compared to the differences between 1D and 2D, clearly the extra dimensionality plays a role, motivating a larger sample of 3D simulations that we will run in the future. 

\subsubsection{Effects of Ionization-Dependent Transport}
Figure \ref{1D_vs_2D_vaion} shows the same 1D vs 2D comparison as before, but we now turn on ionization-dependent transport with $f_{ion}^{min} = 10^{-4}$. Again, for the 1D simulations, the cloud distance increases monotonically for decreasing magnetic field strength, as the same physical picture of pressure equilibration applies. Note, however, that for a given magnetic field strength, the fully ionized case always gives slightly greater cloud acceleration than when partially neutral gas is accounted for. This difference is most dramatic when the magnetic field strength is high and ionization-dependent transport through the cloud leads to an especially quick pressure equilibration; for low magnetic field strengths, the bottleneck at the cloud interface is severe and results in a pressure gradient that pushes the cloud to nearly the same extent regardless of $f_{ion}^{min}.$ While the cosmic ray profiles are markedly different inside the clouds (steep for $f_{ion}^{min} = 1.0$ and flat for $f_{ion}^{min} = 10^{-4}$), the total pressure drop from upstream to downstream is comparably large in both cases, and it is the \emph{total} cosmic ray pressure difference across the cloud, regardless of the smaller-scale pressure profile differences, that determines the force. 

In 2D, efficient cloud acceleration again disfavors high magnetic field strengths, but clearly ionization-dependent transport has suppressed cloud acceleration in the intermediate field regime as well, as the density peaks shift to at most 200 pc after 60 Myrs. In the fully ionized case, intermediate field strengths led to a sweet spot with only small field line warping and moderate transport speed across the cloud; combined this created a large pressure difference. Ionization-dependent transport, however, equilibrates the pressure on a shorter timescale, breaking down the sweet spot. Low magnetic field strengths allow for a greater build-up of cosmic ray pressure upstream, but cosmic ray flow around the cloud once again makes the pressure drop across the cloud almost negligible. 

Consistent with the morphological differences in Figure \ref{fig:2D_ecr}, ionization-dependent transport keeps the cloud fairly compact, since there is no pressure gradient within the cloud, whereas the $f_{ion}^{min} = 1.0$ density profiles in Figure \ref{1D_vs_2D} show the clouds being stretched significantly, which decreases the average cloud density. While none of our ionization-dependent transport simulations resulted in high cloud velocities, it's intriguing to wonder whether ionization-dependent transport and a larger cosmic ray flux could lead to significant acceleration of molecular gas while keeping it intact instead of stretched apart. This will be addressed with future simulations.  

Our main conclusion is that cosmic ray acceleration of partially neutral clouds is less efficient than acceleration of fully ionized clouds, but the differences in imparted momentum are small (less than a factor of 2). This is quantified in Figures \ref{2DCloud_Acceleration} and \ref{2DCloud_Acceleration_fast}, which show the cloud momentum, velocity, and change in mass, defining the cloud to be all gas with temperature T $<$ 2 x 10$^{4}$ K. As expected, the momentum and velocities with ionization-dependent transport included are lower by a factor $\approx 2$ compared to the fully ionized case. The cloud masses decrease over time in all cases, and they decrease most quickly for higher magnetic field strengths (higher Alfv{\'e}n speeds) that most quickly transform the cloud edges into long filaments that are sheared apart by the Kelvin-Helmholtz instability, consistent with findings from \cite{Bruggen2020}. For similar reasons, the mass decreases more for clouds that reach greater velocities, as the cloud crushing time $t_{cc} \sim \xi^{1/2} r_{cl}/v$ is then shorter, where $\xi$ is the cloud overdensity relative to the background, $r_{cl}$ is the cloud radius, and $v$ is the velocity of the ambient medium relative to the cloud \citep{Klein1994}. It's worth noting again that there is no radiative cooling in our simulations, which may suppress the ablation of gas at the cloud boundary and even facilitate cloud growth as the surrounding gas mixes and condenses onto the cloud \citep{Gronke2018, Gronke2020, Kanjilal2020, 2020arXiv201105240B}.

%\subsubsection{Comparison to Previous Studies}

%To our main point of assessing energy loss and using gamma-ray emission to discriminate between transport models, we interestingly find that allowing for fast transport within the cloud does not efficiently decrease collisional losses within the $\approx 100$ Myr simulation duration. This is because, although fast transport means the cosmic ray population should pick up less grammage as it moves through the cloud, that population has access to the entirety of the cloud's collision targets. Meanwhile, cosmic rays with slower transport that severely bottleneck are confined to a thinner boundary layer, as one can see in Figure \ref{fig:2D_ecr} and as we showed analytically in ...  In lieu of discussing this more here, where there is only one target for cosmic ray collisions, we'll discuss this in detail in the following section where we set up a clumpy, mock ISM with exponential density profile. 

%Energy loss with fast transport happens sooner since cosmic rays can ``leak" into the dense clouds, and over a finite time period, then, . As discussed in \cite{Hopkins2020}, this makes sense because the cosmic ray residence time in diffuse ionized gas is much longer than in  Instead, especially for simulations with denser clouds, we generally find a slight \emph{increase} in collisional energy loss \cite{Hopkins2020} argue that, since the SHOULD PROBABLY TALK MORE ABOUT THIS HERE. 

\section{A Multiphase Cosmic-Ray Obstacle Course}
\label{mainSims}

\begin{table*}
  \centering
  \caption{Simulation parameters for the multiple clump simulations of \S \ref{mainSims}. Below are the mean number density, $\bar{n}$, and mean temperature, $\bar{T}$, at the lower (y = 0) boundary; the scale height H over which the density drops but the gas pressure is initially constant at $P_{g} = 3.23 \times 10^{-12}$ dyne cm$^{-2}$; magnetic field strength, B, which is varied between 1 and 5 $\mu$G; cosmic ray flux at the bottom boundary, $F_{CR}^{0}$; duration of the cosmic ray pulse, 2D and 3D grid dimensions in kpc, which depend on the clump sizes controlled by the parameter L; and the resolution, which depends on the parameter L. The maximum speed of light parameter, $V_{m}$, capped streaming speed, and capped diffusivity are the same as in \S \ref{sec:singleCloud}.} \label{table2}
  \begin{tabularx}{\textwidth}{ccccc}
  \toprule
      \textbf{$\bar{n}$}, \textbf{$\bar{T}$} & \textbf{Scale Height (H)} & $\mathbf{B} = B \hat{y}$ & \textbf{F$_{CR}^{0}$} & \textbf{Pulse Duration} \\
      \hline
      1 cm$^{-3}$, 10$^{4}$ K  & 250 pc & 1,5 $\mu$G & $1.54 \times 10^{-5}$ erg cm$^{-2}$ s$^{-1}$ & 30 Myrs \\
      & & & \\
     \hline
     \multicolumn{3}{c}{\textbf{Grid Size}} & \multicolumn{2}{c}{\textbf{Resolution}} \\
     \hline
     2 x 4 kpc (2D; L = 5) & 1 x 4 kpc (2D; L = 2) & & 4 pc (2D; L = 5) & 2 pc (2D; L = 2) \\
     2 x 4 x 1 kpc (3D; L = 5) & 1 x 4 x 0.5 kpc (3D; L = 2) & & 4 pc (3D; L = 5) & 2 pc (3D; L = 2) \\ 
     \hline
  \end{tabularx}
\end{table*}

We now show a suite of simulations of cosmic ray propagation through multi-clump ``obstacle courses." Most simulations were run in 2D, as that was the most computationally feasible way to scan our parameter space. Our analysis in \S \ref{sec:61}, \ref{sec:62} focuses entirely on the 2D simulations; however, we do run a subset of 3D simulations and quantify their cosmic ray calorimetry in \S \ref{sec:63}. We intend to run more 3D simulations in the future to probe differences between 2D and 3D further. 

Compared to the single-cloud simulations of \S \ref{2DSection}, where the cosmic ray front moved horizontally from left to right, we now rotate the box to help the reader envision the scenario we probe: a cosmic ray front vertically escaping from a clumpy disk.  Table \ref{table2} contains the simulation parameters, grid sizes, and resolutions we explore. Table \ref{table3} lists explicitly which simulations we ran in 2D and 3D for different clump distributions, magnetic field strengths, and transport models.

The mean density at the bottom boundary in each simulation is $\bar{n} \approx 1 cm^{-3}$ and initially falls off exponentially with a scale height H = 0.25 kpc. The magnetic field is constant in the y-direction (vertical direction) with a value of either 1 or 5 $\mu G$, and the pressure is constant throughout the simulation box at $3.23 \times 10^{-12}$ dyne cm$^{-2}$. In each simulation, the domain extends to the y = 4 kpc top boundary, which is sufficiently far away from the clumpy ISM to alleviate worries about boundary effects. As in \S \ref{2DSection}, the boundary conditions perpendicular to the magnetic field direction are set to outflow for all quantities, the bottom boundary injects cosmic rays by setting the cosmic ray flux (again with $F_{CR}^{0} = 1.54 \times 10^{-5}$ erg cm$^{-2}$ s$^{-1}$ and following the time profile of Equation \ref{fluxprof}), and the top boundary is an outflow boundary except for the cosmic ray energy density, where we again impose the ghost value to be slightly smaller than the domain value to ensure that cosmic rays leave the box. 

On top of this tapered density profile, we impose a lognormal distribution of density perturbations \citep{Bruggen2013} with 20 Fourier components in each of the x-, y-, and (if in 3D) z-directions. As with the mean density, the perturbation amplitude also decreases exponentially with scale height 0.25 kpc in order for the density variance to attenuate above the disk into a smooth halo. The density profile is then:

\begin{equation}
    n = \bar{n} e^{-y/H} e^{\alpha f(x,y,z)}
\end{equation}
\begin{equation}
\begin{split}
    f(x,y,z) = e^{-y/H}(1-e^{-y}) \sum_{k_{x}, k_{y}, k_{z}} A_{i} sin(2 \pi k_{x} x/ L + \phi_{i}) \\ sin(2 \pi k_{y} y/ L + \theta_{i})sin(2 \pi k_{z} z/ L + \eta_{i})
\end{split}
\end{equation}
where x, y, z are in units of kpc, H = 0.25 kpc, and $k_{x}$, $k_{y}$, $k_{z}$ $\in [10, 30]$. $A_{i}$ and $\phi_{i}$, $\theta_{i}$, $\eta_{i}$ are randomly chosen from a uniform distribution between [0, 1] and [0, 2$\pi$], respectively. The parameter L controls the characteristic size of perturbations, while $\alpha$ sets the density contrast. The extra factor of $(1-e^{-y})$ in the perturbation equation excludes the clumps away from the bottom boundary. We found in test simulations that if a dense clump is on the boundary, the outflow boundary we impose will copy the gas density into the ghost cells and therefore draw in a large amount of artificial mass. Since, as we'll see, cosmic rays push on clumps differently depending on the transport model, this results in different total masses in the simulation box, which muddies our analysis of gamma-ray emission, gas momentum, etc.

\begin{figure}
\centering
\includegraphics[width = 0.45\textwidth]{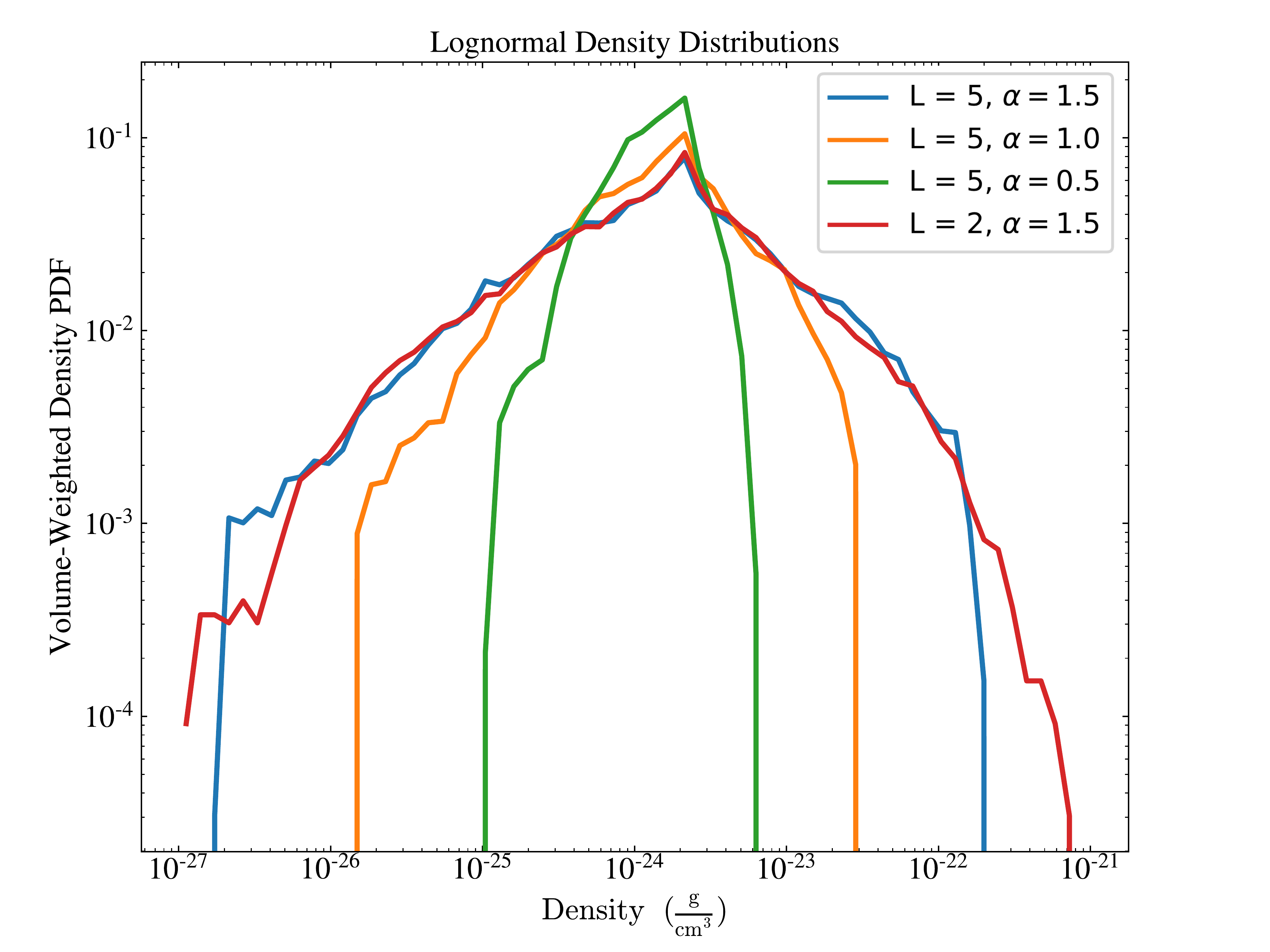}
\caption{Density probability distribution function (PDF) for each combination of (L,$\alpha$) we present.}
\label{fig:densityPDFs}
\end{figure}

Our clump setup varies, with (L, $\alpha$) between (5, 1.5) and (2, 0.5), giving different density probability distribution functions (PDFs) shown in Figure \ref{fig:densityPDFs}. For the gas pressure we initialize, at the mean density $\sim 1 cm^{-3}$, the temperature is $\sim 10^{4}$ K. For L = 5, this gives a porous ISM dominated by large ($\approx 50-100$ pc) over/under-densities. The simulation domain extends to $\pm 1$ kpc in the x-direction (horizontal direction) in this case. In simulations with L = 2, the box only extends to $\pm 500$ pc in the x-direction. With small density structures ($\approx 10-50$ pc) more typical of real ISM clouds, these domain widths are sufficient to present cosmic ray interactions with clouds of varying densities; however, they necessitate higher resolution to more properly resolve the cloud interfaces, which we've found to be important (\S \ref{1DSection}). For L = (5, 2) the resolution is uniformly (4, 2) pc until $y = 2$ kpc, at which point we make use of static mesh refinement to decrease resolution to (16, 8) pc and save computation time.

\begin{table}
  \centering
  \caption{Multiple clump simulations run (\S \ref{mainSims}). Most simulations for a given (L, $\alpha$, B) are run with $f_{ion}^{min} = 1.0$ and again with $f_{ion}^{min} = 10^{-4}$. For a subset of the 2D simulations, additional parallel diffusion is also included.}
  \begin{tabularx}{0.48\textwidth}{c|cccc}
  \toprule

%\begin{tabular}{c|c|c|c|c|c|c}
      \textbf{2D} & \textbf{L} & $\alpha$ & \textbf{B} ($\mu$G) & \textbf{Transport} \\
      \hline
      
      & 5 & 1.5 & 5 & $f_{ion}^{min} = 1.0$, $f_{ion}^{min} = 10^{-4}$, \\
      & & & & $f_{ion}^{min} = 1.0$ + $\kappa_{||} = 3 \times 10^{27}$, \\
      & & & & $f_{ion}^{min} = 1.0$ + $\kappa_{||} = 3 \times 10^{28}$, \\
      & & & & $f_{ion}^{min} = 10^{-4}$ + $\kappa_{||} = 3 \times 10^{27}$   \\
      & 5 & 1.5 & 1 & $f_{ion}^{min} = 1.0$, $f_{ion}^{min} = 10^{-4}$ \\
      & 5 & 1.0 & 5 & $f_{ion}^{min} = 1.0$, $f_{ion}^{min} = 10^{-4}$ \\
      & 5 & 0.5 & 5 & $f_{ion}^{min} = 1.0$, $f_{ion}^{min} = 10^{-4}$ \\
      & 2 & 1.5 & 5 & $f_{ion}^{min} = 1.0$, $f_{ion}^{min} = 10^{-4}$ \\
      & 2 & 1.5 & 1 & $f_{ion}^{min} = 1.0$, $f_{ion}^{min} = 10^{-4}$ \\
     \hline
      \textbf{3D} & \textbf{L} & $\alpha$ & \textbf{B} ($\mu$G) & \textbf{Transport} \\
      \hline
     
      & 5 & 1.5 & 5 & $f_{ion}^{min} = 1.0$, $f_{ion}^{min} = 10^{-4}$ \\
      & 2 & 1.5 & 5 & $f_{ion}^{min} = 1.0$ \\

    \hline
\end{tabularx}
\label{table3}
\end{table}

It's important to note that these simulations do not include cooling or gravity and are not meant to substitute for more realistic, stratified ISM simulations, though they do replicate the propagation of cosmic rays through layers of a fixed thickness and mean column density. Having a constant magnetic field strength out to many scale heights is also not realistic, so we focus our analysis to the first kpc above the midplane.

\subsection{Qualitative Differences}
\label{sec:61}

\begin{figure*}
\centering
\includegraphics[width = 0.98\textwidth]{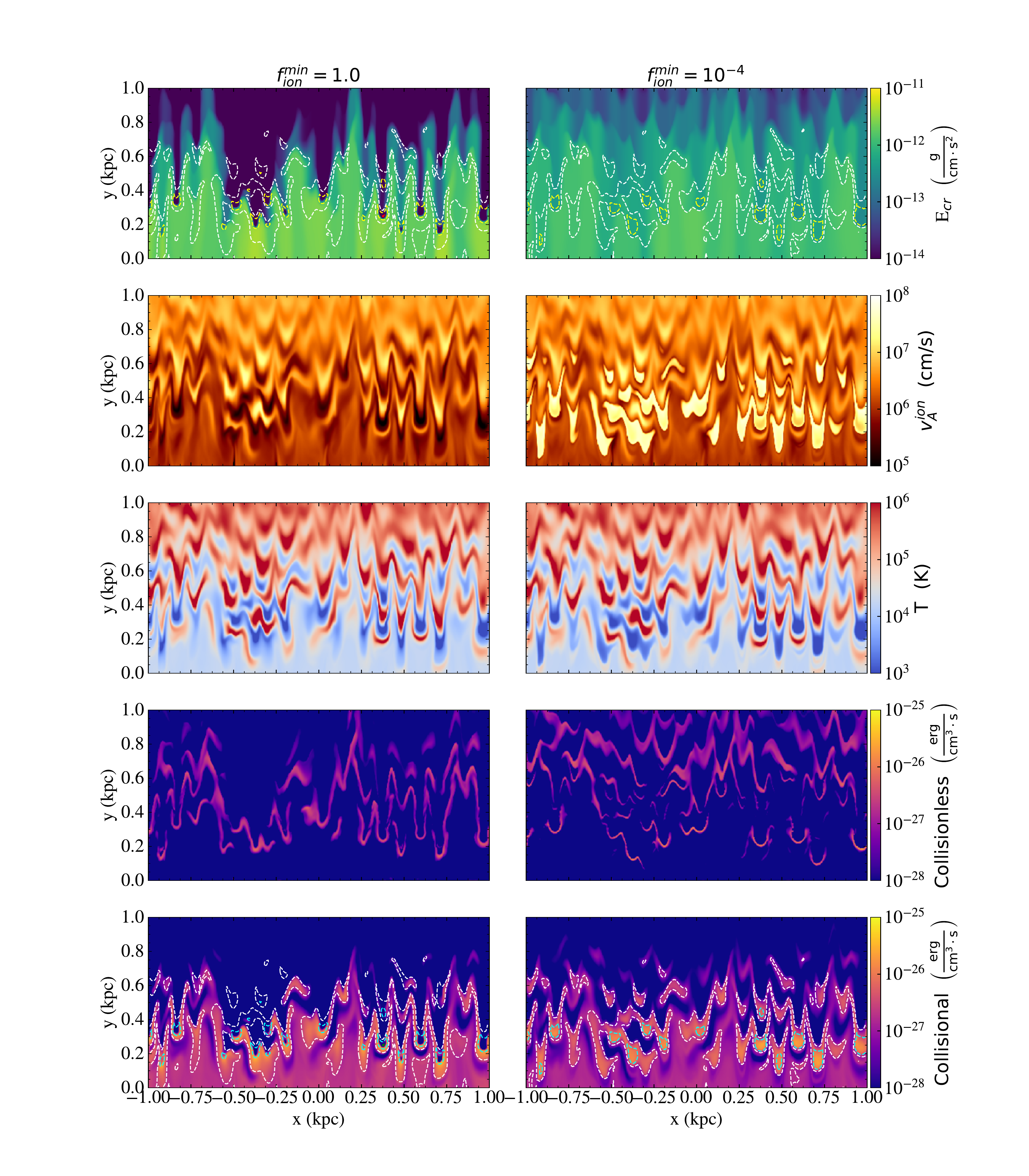}
\caption{Snapshots at $t = 40$ Myrs for the 2D ISM setup with B = 5 $\mu$G, mean density = 1.0 cm$^{-3}$, and (L, $\alpha$) = (5, 1.5). Simulation parameters are listed in Table \ref{table2}. \emph{Left panels:} Fully ionized assumption ($f_{ion}^{min} = 1.0$), \emph{Right panels:} Ionization-dependent transport included ($f_{ion}^{min} = 10^{-4}$). The top panel shows cosmic ray energy density and density contours of 1 and 10 cm$^{-3}$, which shows clear variations in how cosmic rays preferentially penetrate or flow around cold clouds. The second row shows $v_{A}^{ion}$, followed by gas temperature, collisionless energy loss rate ($|v_{A}^{ion} \cdot \nabla P_{CR}|$), and collisional ($\propto e_{CR} n_{gas}$) loss rate, again with dashed contours showing densities of 1 and 10 cm$^{-3}$. Collisionless heating at the cloud interfaces is apparent in both transport cases, while collisional energy loss is preferentially higher in dense gas when ionization-dependent transport is included. An animated version of this figure, showing the evolution of these panels from t = 1 - 100 Myrs, is available in the HTML version of this article.}
\label{fig:lognormal_L5_alpha1_5}
\end{figure*}

\begin{figure*}
\centering
\includegraphics[width = 0.98\textwidth]{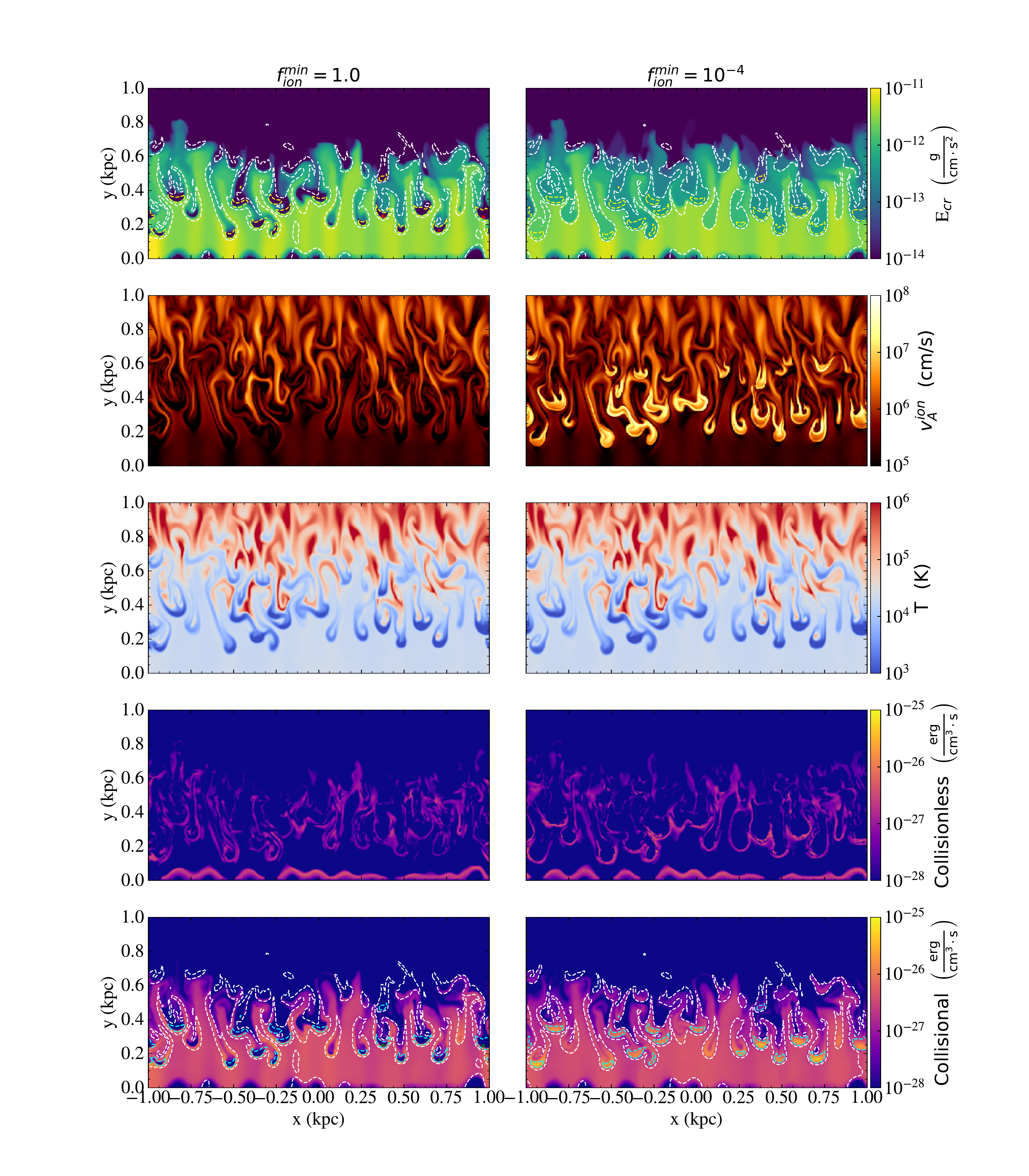}
\caption{Same as Figure \ref{fig:lognormal_L5_alpha1_5} but with B = 1 $\mu$G instead of 5 $\mu$G. Magnetic field lines now warp around the cold clumps, squeezing cosmic rays through the gaps in the ISM instead of funneling them through clouds. This effect dominates over differences in transport, as the left and right columns (varying $f_{ion}^{min}$) now show smaller differences. Also note that, compared to the B = 5 $\mu$G simulations shown in Figure \ref{fig:lognormal_L5_alpha1_5}, there is now a larger build-up of cosmic ray pressure that more effectively accelerates the gas column. This is especially true for the hot gas, which gets pushed out by the cosmic rays sweeping through the under-dense channels. An animated version of this figure, showing the evolution of these panels from t = 1 - 100 Myrs, is available in the HTML version of this article.}
\label{fig:lognormal_L5_alpha1_5_lowerB}
\end{figure*}

Figure \ref{fig:lognormal_L5_alpha1_5} shows snapshots at $t = 40$ Myrs of our 2D, B = 5 $\mu$G simulations with mean density = 1.0 cm$^{-3}$, (L, $\alpha$) = (5, 1.5), and $f_{ion}^{min} = 1.0$ (left) and $f_{ion}^{min} = 10^{-4}$ (right). Figure \ref{fig:lognormal_L5_alpha1_5_lowerB} shows the same setups but with B = 1 $\mu$G. The plots are oriented such that cosmic ray flux is input at the lower boundary, and magnetic field lines are vertical. The contours show densities of n = 1 cm$^{-3}$ and n = 10 cm$^{-3}$. The under-dense, inter-cloud region extends down to densities of n $\approx 10^{-3}$ cm$^{-3}$ (see Figure \ref{fig:densityPDFs}). Such a configuration, if entirely ionized, gives a range of Alfv{\'e}n velocities predominantly between $10^{5}$ and $10^{7}$ cm/s, but when neutral particles are accounted for, the lowest Alfv{\'e}n velocity is $\approx 10^{6}$ cm/s, with ion Alfv{\'e}n velocities in cloud interiors reaching beyond $10^{8}$ cm/s. 

This difference in propagation speed is clearly reflected in the top row of Figure \ref{fig:lognormal_L5_alpha1_5}, which shows that cosmic rays with ionization-dependent transport have propagated further into the domain and have reached the inner ``halo", where the average density has dropped. Cosmic rays from here will free-stream out of the simulation domain at high Alfv{\'e}n velocities since $v_{A} \propto \rho^{-1/2}$ for constant B. In the fully ionized case, the cosmic rays that have managed to propagate furthest are those that have avoided cold clouds and squeaked through the cracks. As in the 2D, single cloud simulations, for an initial plasma $\beta$ = 1.6, magnetic field line warping plays only a limited role, meaning most cosmic rays are eventually forced into the clouds (and additionally, accelerate the clouds). When accounting for ionization-dependent transport, cosmic rays leak into the dense cores and quickly reappear on the other side. As in Figure \ref{fig:2D_ecr}, there are sharp steps in cosmic ray energy at cloud edges, both at the front \emph{and} the back edges in the $f_{ion}^{min} = 10^{-4}$ case, as each interface induces a bottleneck. 

For a magnetic field strength of 1 $\mu$G (Figure \ref{fig:lognormal_L5_alpha1_5_lowerB}), magnetic field line warping dominantly determines how cosmic rays navigate the medium. Now, cosmic rays with either transport model reside primarily in the inter-cloud regions. The lower magnetic field strength also leads to a greater build-up of cosmic ray pressure that exerts a force on the medium. As in the single cloud simulations of \S \ref{2DSection}, such a lower field strength leads to only a small acceleration of the cold gas, but we see now that the hot gas is very efficiently pushed out by the cosmic rays that are forced to squeeze through the under-dense channels of the ISM.

\subsection{Influence on Clouds}
\label{sec:62}

\begin{figure}
\centering
\includegraphics[width = 0.43\textwidth]{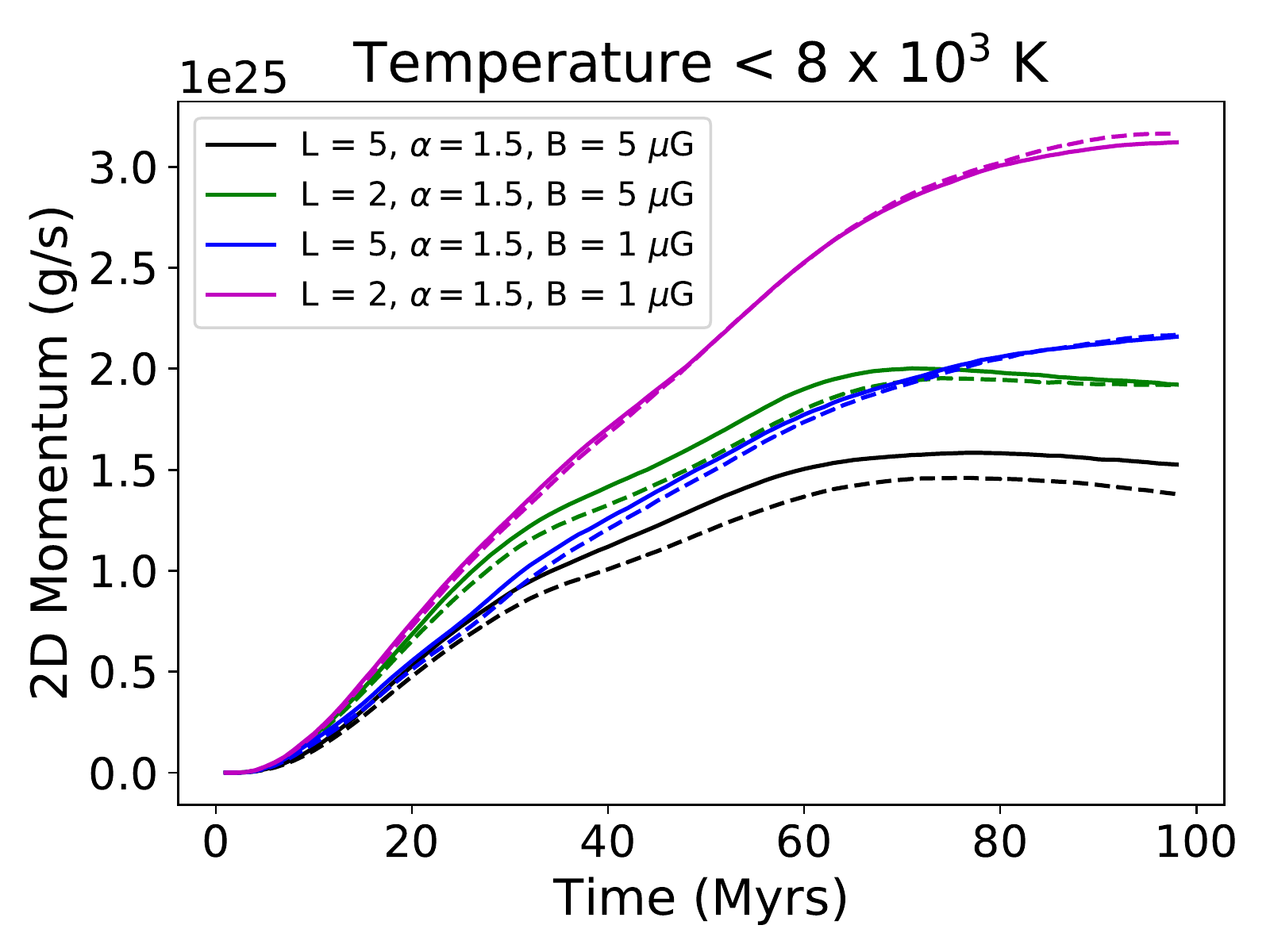}
\includegraphics[width = 0.43\textwidth]{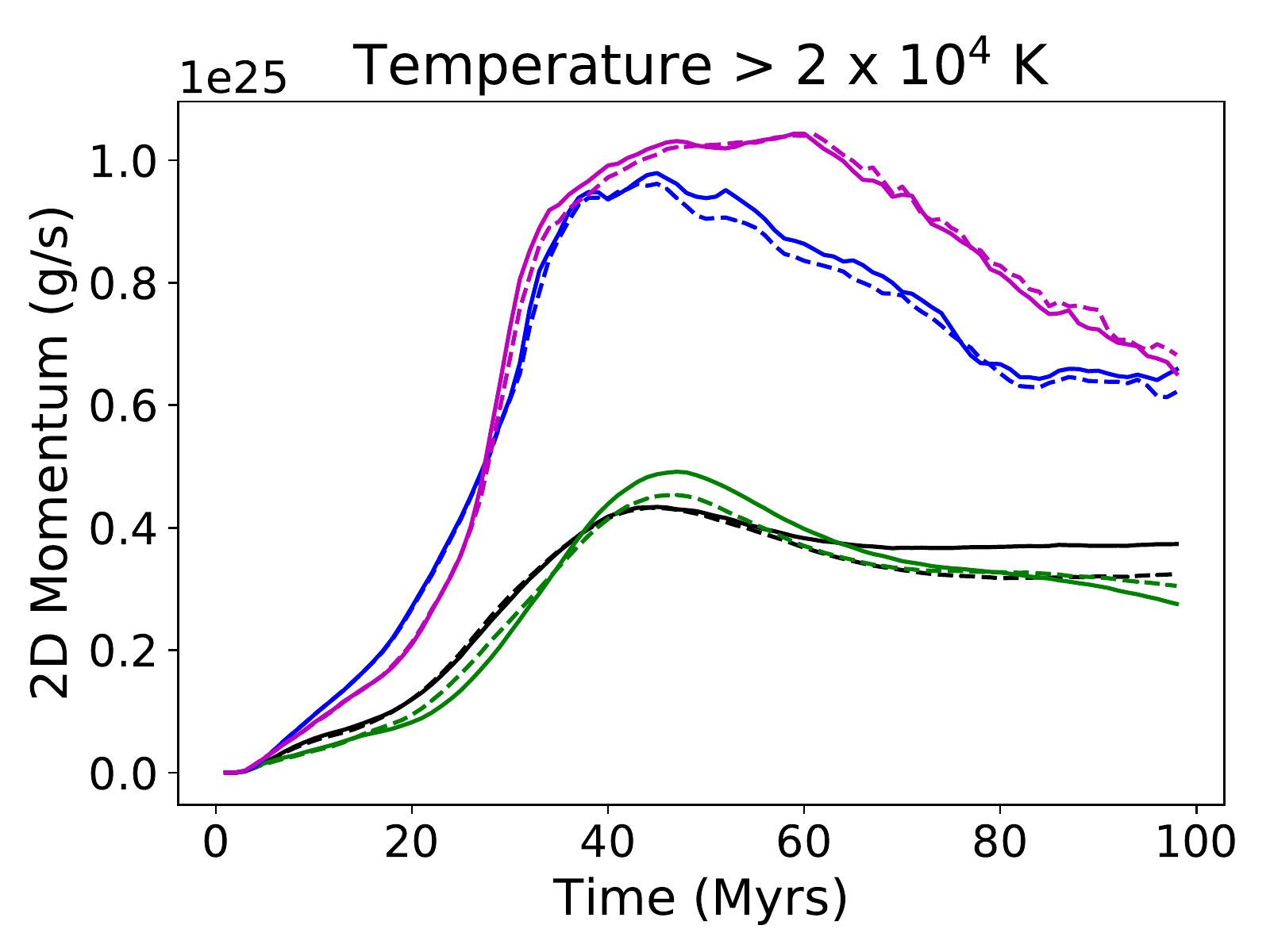}
\caption{Momentum measured in 2D clump simulations with varying L and $\alpha$ and varying transport, with $f_{ion}^{min} = 1.0$ (solid lines) and $f_{ion}^{min} = 10^{-4}$ (dashed lines). The top panel shows the momentum just in the cold gas (T $< 8 \times 10^{3}$ K), and the bottom panel shows the momentum in the hot gas (T $> 2 \times 10^{4}$ K). Note the change in y-axis reflecting that most of the momentum, in either transport model, is in the cold phase, but lower magnetic field strengths do increase the hot gas momentum quite substantially as cosmic rays squeeze out the gas between dense clouds. With a lower field strength, the cold momentum increases somewhat, as well, due to an increased cosmic ray pressure gradient as we saw in \S \ref{2DSection}. When neutral particles are accounted for, the momentum is lower, but only slightly, compared to the fully ionized case.}
\label{fig:lognormal_Momentum}
\end{figure}
 
\cite{Farber2018} probed the implications of cosmic ray decoupling for galactic wind driving and star formation feedback. They compared three transport models: cosmic ray advection, field-aligned diffusion with a single diffusion coefficient (one-$\kappa$ model), and field-aligned diffusion with a higher diffusion coefficient in cold gas (two-$\kappa$ model) to mimic the effects of ion-neutral damping. Comparing the one-$\kappa$ and two-$\kappa$ models, they found that boosting the diffusivity in the cold phase led to a faster, hotter outflow since the cold phase received less impulse from cosmic ray fronts. Changing $f_{ion}^{min}$, we find that this trend partially occurs. Figure \ref{fig:lognormal_Momentum} shows the momentum in the full simulation box as a function of time for our (L,$\alpha$) = (5, 1.5) ISM setup. The top panel shows the momentum regardless of gas temperature, while the bottom two panels split the momentum into ``cold" gas (T $< 2 \times 10^{4}$ K) and ``hot" gas (T $> 2 \times 10^{4}$ K). When ionization-dependent transport is included, slightly less momentum is imparted to both the cold and hot gas, but only on the level of tens of percent. As argued in \S \ref{2DSection}, we attribute this to the bottleneck and subsequent suppression of the diffusive flux at the cloud boundary, where large cosmic ray pressure gradients still drive waves and enact a force on the cloud. 

Regardless of whether ionization-dependent transport is included, the momentum in the cold phase exceeds the momentum in the hot phase, which is consistent with more complete simulations of supernova-driven outflows where mass-loading in the cold phase exceeds that in the hot phase (e.g. \citealt{Kim2020}). The time profiles of cold and hot momentum are noticeably different, though. The hot gas is quickly swept out by the preceding acoustic pulse and cosmic ray front that moves through. A short time after the cosmic ray source turns off at t = 30 Myrs, the cosmic ray pressure gradient diminishes, and the momentum slightly drops. The cold gas momentum rises a bit more slowly; cosmic ray bottlenecks take time to build up (on the order of the Alfv{\'e}n crossing time between the cloud and the source), and the cosmic ray pressure releases only after a long time since the cold clouds move relatively slowly. Pressure gradients on the cold gas then build to greater amplitudes and longer durations than on the hot gas, leading to a larger but more gradual acceleration of the clouds.

%The largest differences in cloud acceleration occur when an additional diffusion term is included or when the magnetic field strength is varied. First, as explained earlier in terms of energy loss, even a relativelly small diffusion coefficient of $3 \times 10^{27}$ cm$^{2}$ s$^{-1}$ can efficiently smooth cosmic ray pressure gradients at the cloud edge. This breaks down the formation of a bottleneck and allows cosmic rays to pass through the cloud with a higher diffusive flux, as they cannot efficiently generate waves to overcome ion-neutral damping. Clouds, then, are left relatively untouched; the total momentum in the simulation box clearly rises much slower. When not accounting for decoupling, this same lowered impulse to clouds is achieved when a higher diffusion coefficient is included -- in this case, the $f_{ion}^{min} = 1.0$, $\kappa_{||} = 3 \times 10^{29}$ and the $f_{ion}^{min} = 10^{-4}$, $\kappa_{||} = 3 \times 10^{27}$ curves are fairly similar. 

Rather than ionization-dependent transport leading to changes in cloud acceleration, the largest changes in momentum correlate with changing magnetic field strength. For the same injected cosmic ray flux, a lower magnetic field strength (slower transport) translates to a higher cosmic ray pressure gradient to drive the gas column outwards. Both the $f_{ion}^{min} = 1.0$ and $f_{ion}^{min} = 10^{-4}$ models show an increase in impulse. The cold gas momentum increases because cosmic ray bottlenecks are severe in either case at the cloud edges. The hot gas momentum increases because cosmic rays squeeze between the gaps in clouds, sweeping out the hot gas in front of them (see Figure \ref{fig:lognormal_L5_alpha1_5_lowerB}).

Interestingly, changes in momentum also correlate with the characteristic size of clumps controlled by our $L$ parameter. More momentum is imparted especially to the cold gas when $L$ is smaller; this makes sense as the limit $L \rightarrow 0$ is effectively a 1D simulation, which we saw in \S \ref{2DSection} was the most effective at accelerating cold clouds.

%%%%%%%%%%%%%%%%%%%%%%%%%%
\subsection{Quantifying Cosmic Ray Propagation and Energy Loss}
\label{sec:63}

\begin{figure*}
\centering
\includegraphics[width = 0.42\textwidth]{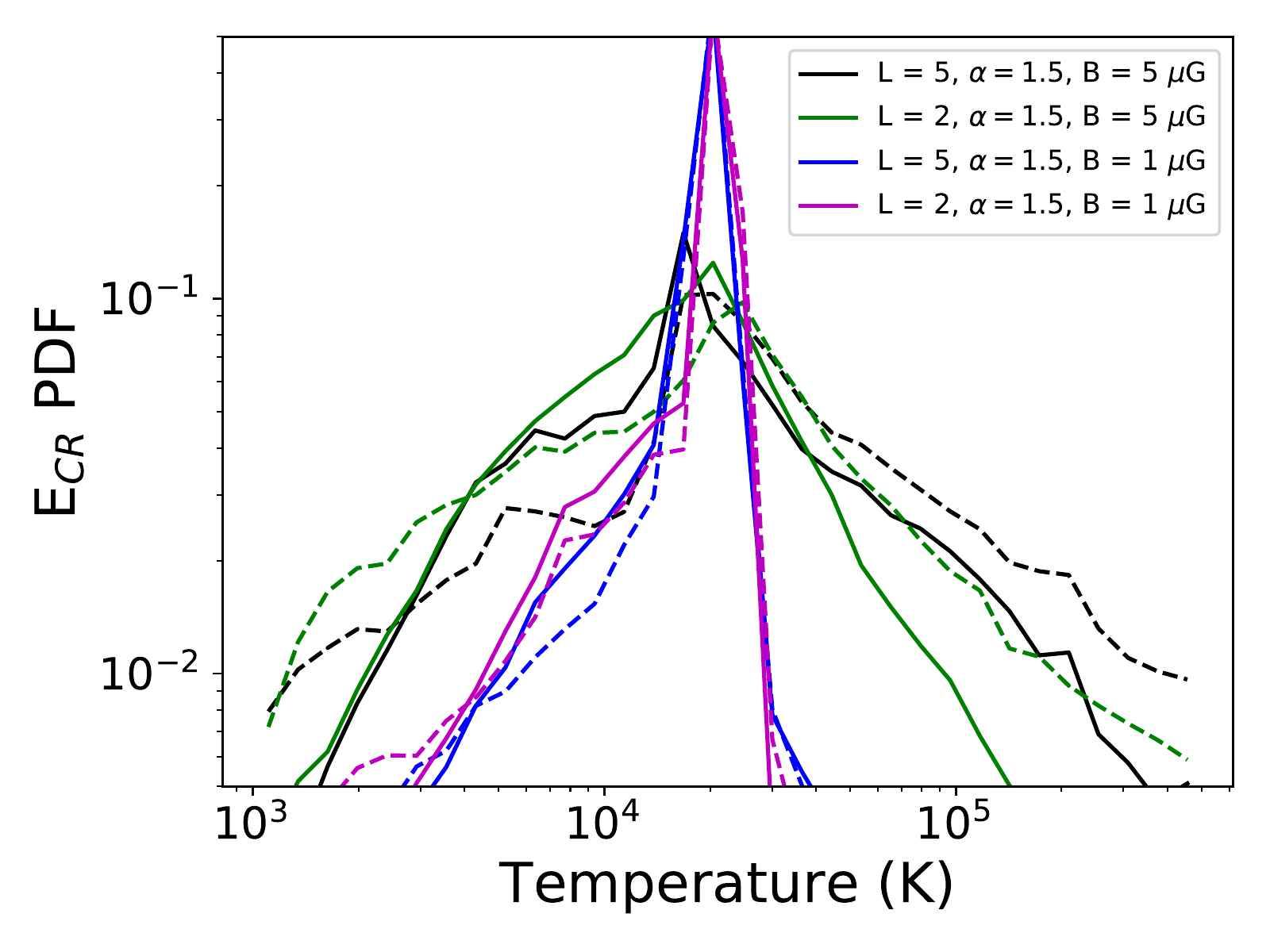}
\includegraphics[width = 0.42\textwidth]{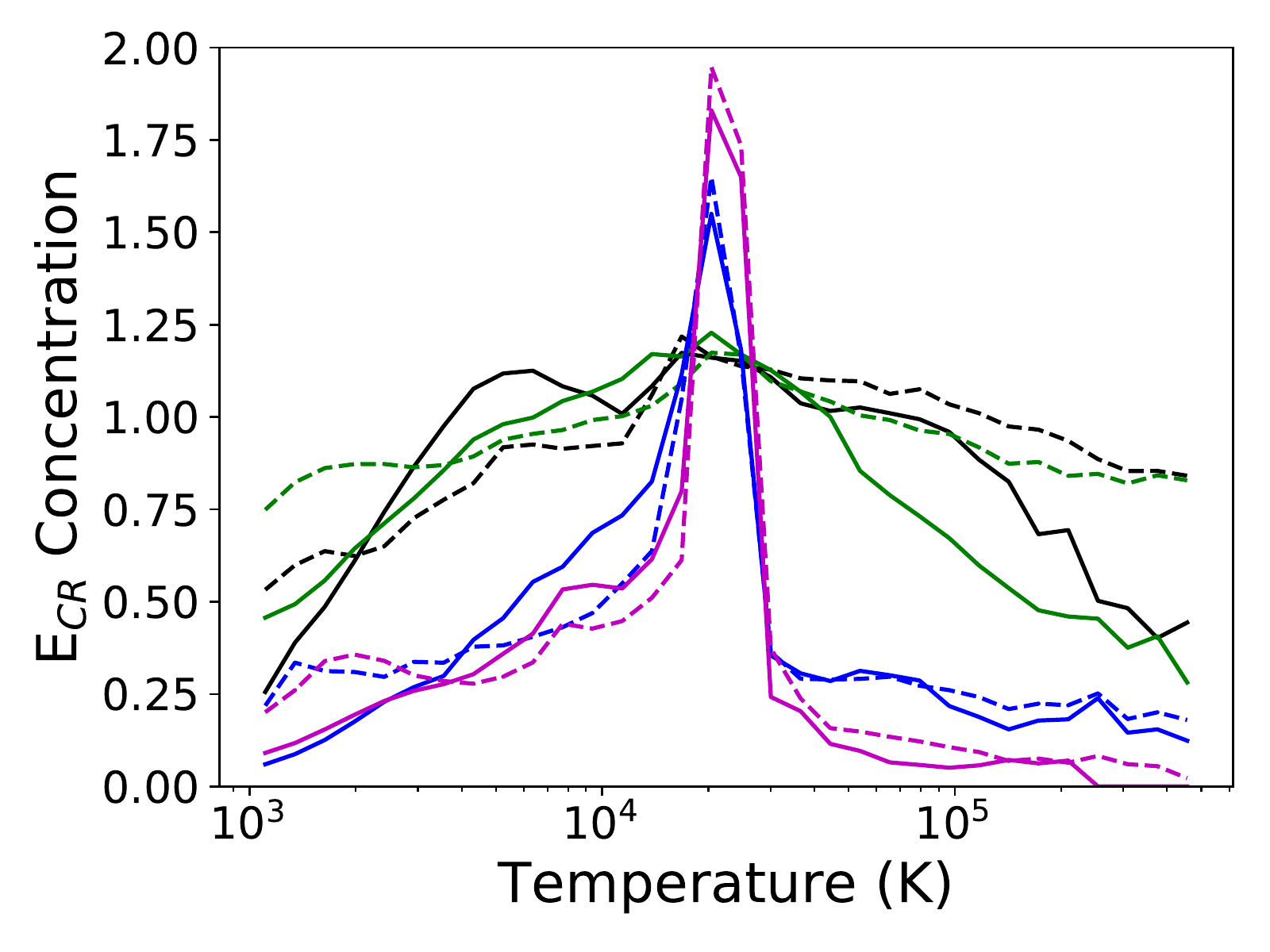}
\includegraphics[width = 0.42\textwidth]{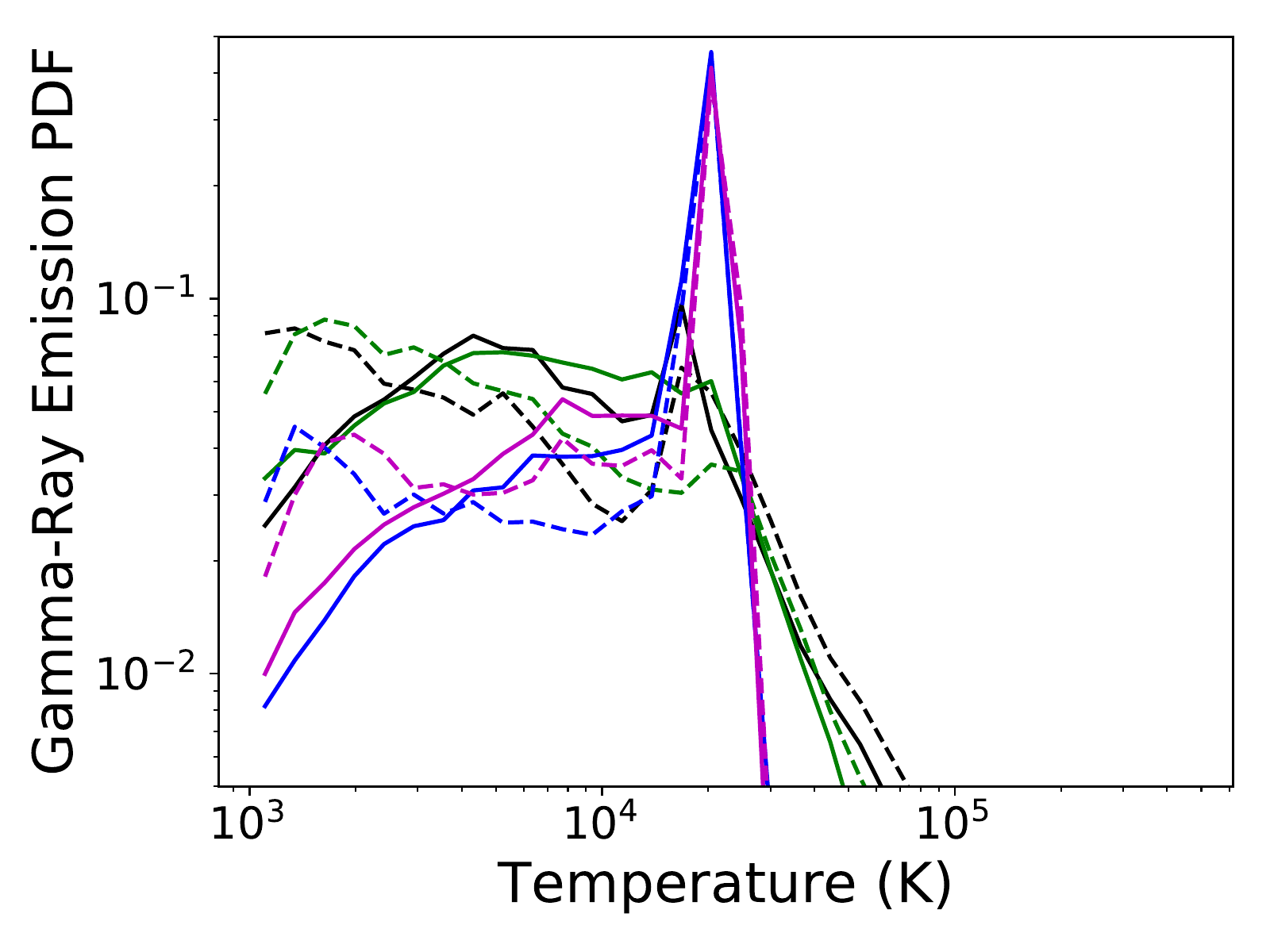}
\includegraphics[width = 0.42\textwidth]{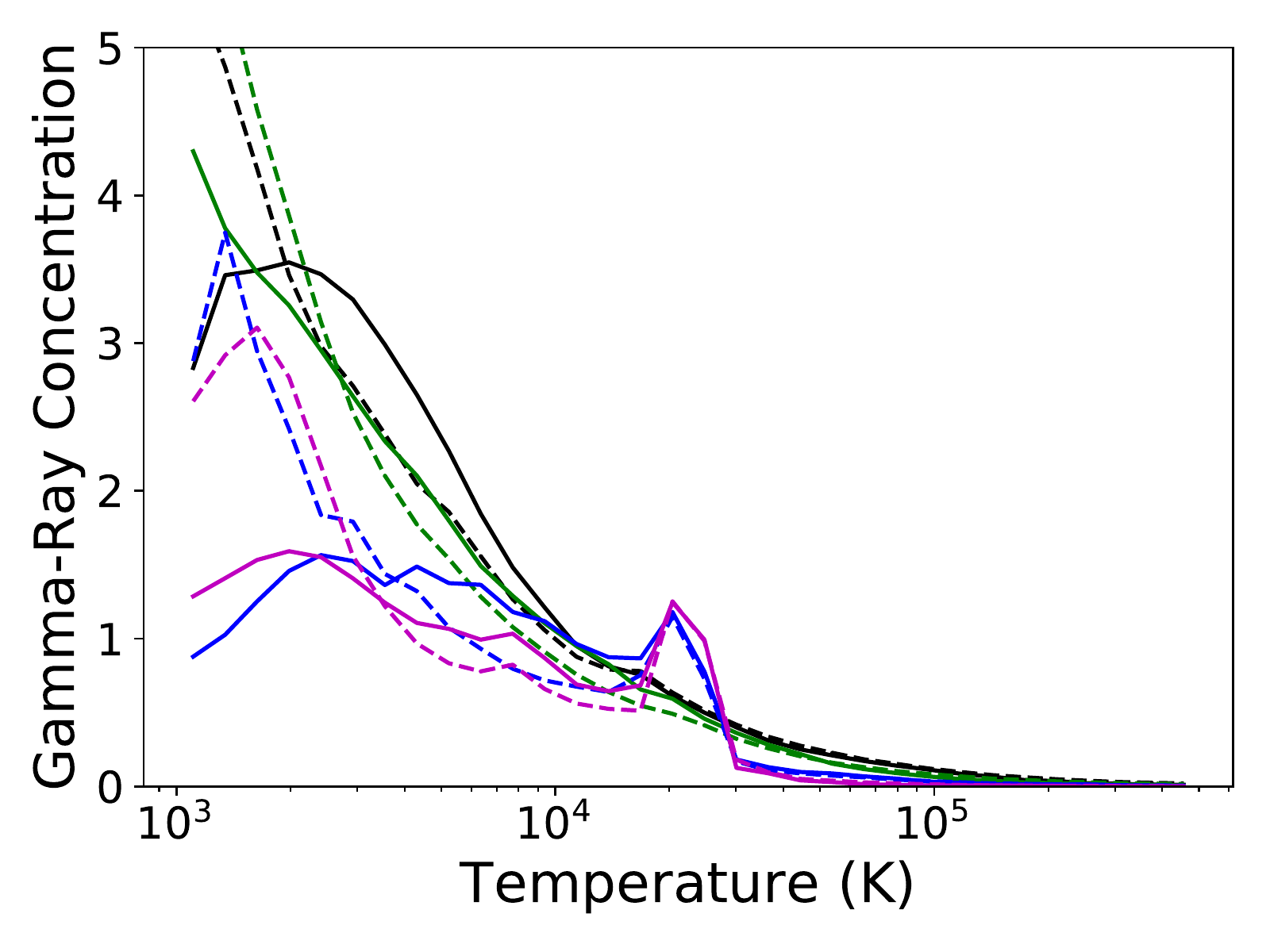}
\caption{All panels are for 2D simulations. The left column shows the probability distribution function (PDF) of cosmic ray energy density $E_{CR}$ (top) and gamma-ray emission $\sim E_{CR} n_{gas}$ (bottom) in 32 temperature bins ranging from T = $10^{3} - 5 \times 10^{5}$ K. The right column shows the concentrations, obtained by dividing the PDFs in the left column by the temperature PDF, which gives a comparison of how concentrated the quantities are compared to the fraction of the ISM at each temperature bin. If the fraction of energy or emission equals the volume fraction at that temperature bin, the concentration will be 1.0. Clearly there are differences depending on transport model, with $f_{ion}^{min} = 1.0$ shown by solid lines and $f_{ion}^{min} = 10^{-4}$ shown by dashed lines, as well as varying ISM parameters and magnetic field strengths (different colors). Cosmic rays in all cases preferentially sample the mean density and temperature $\approx 2 \times 10^{4}$, where the gas is fully ionized, Alfv\'en speeds are low, and cosmic ray residence times are long. Ionization-dependent transport leads to preferentially more occupancy in dense, cold clouds but less occupancy at transition temperatures $T \sim 8 \times 10^{3}$ K. This trade-off leads to roughly the same total gamma-ray luminosity, regardless of transport (\S \ref{sec:calorimetry}). With a lower magnetic field strength, field line warping allows cosmic rays to squeeze through the under-dense channels of the ISM, pushing out high temperature gas in the process. In those cases, cosmic rays are the most concentrated at the mean temperature.}
\label{fig:Ecr_Collisions_PDFs}
\end{figure*}

While shown qualitatively in Figures \ref{fig:lognormal_L5_alpha1_5} and \ref{fig:lognormal_L5_alpha1_5_lowerB}, how cosmic rays sample the ISM is shown more quantitatively in Figure \ref{fig:Ecr_Collisions_PDFs}. The left column shows PDFs of cosmic ray energy density (top) and gamma-ray emission (bottom), and the right column shows the concentration of cosmic ray energy and gamma-ray emission relative to the fraction of the volume occupied by each of the 32 temperature bins. We show this concentration because the temperature PDF changes depending on transport mode, magnetic field strength, and ISM parameters. So the most clear comparison between simulations is to divide the actual PDF by the temperature PDF; in this case, a concentration of 1.0 means that the fraction of cosmic ray energy or gamma-ray emission in a given temperature bin equals the volume fraction within that temperature bin. We restrict our analysis to the clumpiest part of the simulation, which is within 500 pc of the lower boundary.

In all simulations, most cosmic ray energy resides near the mean density and temperature, partially because that represents a maximum in the density PDF (Figure \ref{fig:densityPDFs}) and partially because that is close to the minimum in $v_{A}^{ion}$. This is consistent with previous work that found, using a Monte Carlo method, that cosmic rays in M82-like starburst galaxies would generally sample the mean density \citep{Boettcher2013}. At temperatures below T $\approx 10^{4}$ K, as cosmic rays encounter clumps, the PDFs and concentrations start to diverge. With ionization-dependent transport (dashed lines), there is a steep drop in cosmic ray occupancy near temperatures T $\approx 8 \times 10^{3}$ K, where $v_{A}^{ion}$ rapidly increases. At similar temperatures, cosmic rays with $f_{ion}^{min} = 1.0$ are actually overabundant. These differences make sense if we imagine a single cloud - cosmic ray front interaction: if the interface is assumed to be fully ionized, cosmic ray streaming speeds will be very low, forcing a build-up in pressure, while changing ionization would induce ionization-dependent transport and a drop in cosmic ray energy at the same temperature band. As we go to lower temperatures, the $f_{ion}^{min} = 1.0$ curves continue to drop as slow-moving cosmic rays are consumed by collisions, but the $f_{ion}^{min} = 10^{-4}$ curves drop more gradually, as cosmic rays uniformly fill cold cloud interiors. 

Lowering the magnetic field strength gives the most dramatic change. Cosmic rays streaming along the now-warped magnetic field lines preferentially squeeze through the under-dense channels between clouds. The cosmic ray and gamma-ray concentrations are both suppressed at low temperatures. Cosmic rays also do not occupy the high temperature space: this low-density gas gets pushed into the halo, forced out by the cosmic ray front that sweeps between clouds. In these low magnetic field cases, then, the great majority of cosmic rays lie within a narrow temperature range near the mean. 

%Beyond that density, cosmic rays fill that density space almost uniformly. If cosmic rays were streaming at the same speed, their PDF would follow that of the density, which slowly drops as density increases; however, because cosmic ray streaming speeds slightly decrease as density increases (when ion-neutral damping is overcome, and the transport closely follows $v_{A}^{ion}$, which appears true here; see Figure 1), a relatively flat PDF emerges. At the highest densities, the PDF falls as collisional losses become more important and because the highest density gas is only a small fraction of the volume. 

The bottom row of Figure \ref{fig:Ecr_Collisions_PDFs} shows the PDF of hadronic gamma-ray emission $\propto e_{CR} n_{\rm gas}$. The trends from the top row are clearly imprinted. For both transport models, gamma-ray emission is very concentrated near the mean density when B = 1 $\mu$G, but with B = 5 $\mu$G, a substantial fraction of the gamma-ray emission comes from cold clouds at temperatures $T < 10^{4}$ K. The low-temperature bins that dominate emission depend on transport model. At intermediate densities $n \approx 1 - 10$ cm$^{-3}$ (temperatures around $8 \times 10^{3}$ K), collisional energy losses are decreased considerably when ionization-dependent transport is included, with more energy loss occurring at the higher densities that fast-moving cosmic rays can leak into. In the fully ionized case, the highest densities are inaccessible to cosmic rays, as they lose energy instead in cloud interface layers ($T \approx 8 \times 10^{3}$ K) before penetrating into cloud cores. As we'll see, this trade-off still leads to very similar \emph{total} collisional loss rates (\S \ref{sec:calorimetry}).

The spatial differences in collisional and collisionless energy loss are apparent in the bottom two panels of Figure \ref{fig:lognormal_L5_alpha1_5}. In both columns, most of the energy losses occur near the cosmic ray source where the mean density is the highest. Compared to the fully ionized case, collisional losses are much more prevalent in dense gas, and collisionless energy loss not only occurs at the front cloud edge but also the back due to a second drop in $v_{A}^{ion}$ that induces another steep cosmic ray pressure gradient. 

%While this is just one representative configuration shown in Figures \ref{fig:Ecr_Collisions_PDFs} and \ref{fig:lognormal_L5_alpha1_5}, we expect that there will be a dependence on our chosen (L, $\alpha$) values, magnetic field strength, and 2D vs 3D. 

\subsection{Total Cosmic Ray Calorimetry}
\label{sec:calorimetry}

\begin{figure*}
\centering
\includegraphics[width = 0.95\textwidth]{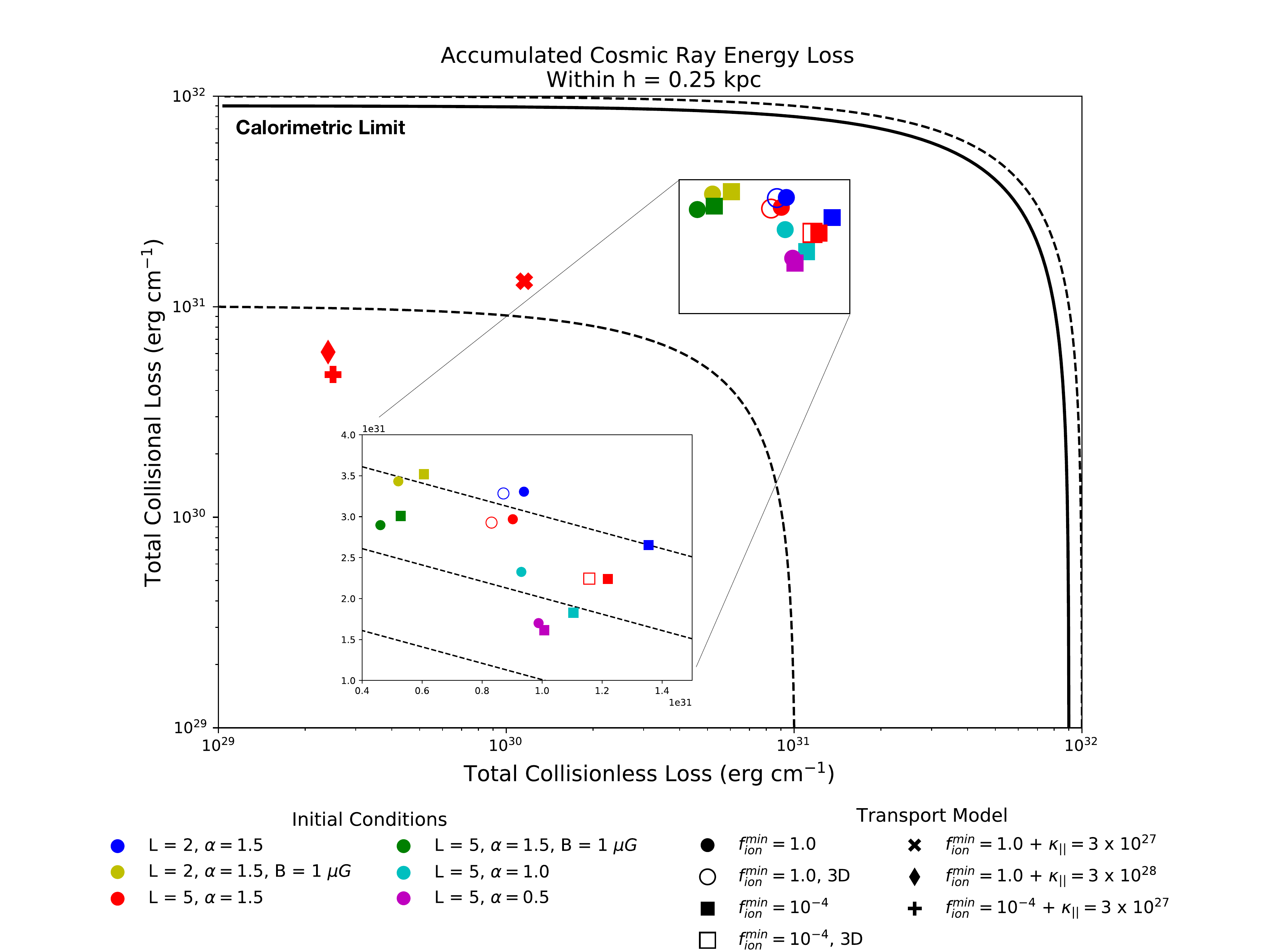}
\caption{Accumulated collisional and collisionless cosmic ray energy loss within one scale height (250 pc) for each of the mock ISM simulations. Different symbols represent changes in transport model, while color represents the ISM setup varying L, $\alpha$, and B. The black dashed lines, both in the large figure and smaller zoom-in, show contours of constant total energy loss, and the solid black line shows the calorimetric limit, i.e. the total cosmic ray energy injected into the box. Despite some significant differences in collisionless energy loss, especially when an additional, constant diffusion term is included, the collisional energy loss (and hence, diffuse gamma-ray emission) only varies within a factor of a few for most models. Changes in ISM setup and 2D vs 3D impart only small variations in total energy loss that are explained in the text.}
\label{fig:BigFigure}
\end{figure*}

Figure \ref{fig:BigFigure} compiles the total collisional and collisionless energy loss from each of the mock ISM simulations we ran. Each symbol represents a different transport model, including a few simulations where we included an additional, constant cosmic ray diffusion coefficient; this boosts cosmic ray escape most drastically in the diffuse medium and cloud interfaces where the residence times are otherwise very long. Each color is a different ISM configuration, varying L, $\alpha$, and B. Black dashed lines show contours of constant \emph{total} cosmic ray energy loss. 

What one can see most clearly is that, although collisionless energy loss can vary by orders of magnitude, collisional energy loss only varies within a factor of a few regardless of transport model. All models without additional diffusion are clustered within the outlined box, meaning that, although the collisions occur in very different places (see Figure \ref{fig:Ecr_Collisions_PDFs}), the total collisional energy loss is about the same.

%As postulated in \cite{Hopkins2020}, what this boils down to is that, in the fully ionized case, the lower mean density sampled by cosmic rays is offset by the long residence times in diffuse gas. With the effect of ionization on the Alfv{\'e}n speed accounted for, cosmic ray residence times are shorter, but cosmic rays access denser gas where the collision rate is higher. 

Zooming in on the clustered region of Figure \ref{fig:BigFigure}, we do see some variation depending on ISM configuration and transport model, but the changes are within a factor of a few. For instance, while a lower magnetic field strength on average leads to lower transport speeds and longer cosmic ray residence times in the ISM, the net collisional energy loss is comparable to that with a higher magnetic field strength. With weak magnetic fields, the collisionless energy loss $\propto v_{A}^{ion} \cdot \nabla P_{CR}$ decreases since $v_{A}^{ion} \propto B$ also decreases; the decrease is only by a factor of $\approx 2$, however, since slower transport speeds lead to a build-up in cosmic ray pressure and higher $\nabla P_{CR}$. 

Varying $\alpha$, which determines the density contrast, also makes a difference. With higher (lower) density contrast, the collisional energy loss $\propto n_{gas}$ increases (decreases) accordingly. Varying L has almost no consequence for collisional loss, however. The slight increase in collisions for L = 2 compared to L = 5 can be attributed to the slightly larger density range for L = 2 (see Figure \ref{fig:densityPDFs}), which follows the same explanation as varying $\alpha$.

We also ran three of our clump simulations in 3D, two of which had $f_{ion}^{min} = 1.0$ and one with $f_{ion}^{min} = 10^{-4}$. Each of these simulations had B = 5 $\mu$G, which as we saw in the single cloud simulations of \S \ref{2DSection}, shows only small differences between 2D and 3D. That is reflected in Figure \ref{fig:BigFigure}, as well, where the 2D and 3D energy losses are very similar. For larger cosmic ray fluxes or larger plasma $\beta$, the differences between 2D and 3D may become more important, which we will explore in future work. 

It's illuminating to also disentangle the differences in collisionless energy loss. If one were to draw a line diagonally across the plot from lower left to upper right, that would show equal collisional and collisionless energy loss; we see that in all simulations run, collisonal losses exceed collisionless losses, but this gap narrows when the density contrast (controlled by $\alpha$) decreases and when ionization-dependent transport is included. Comparing e.g. $L = 2$, $\alpha = 1.5$ simulations with different $f_{ion}^{min}$, we see that ionization-dependent transport decreases collisional energy loss but boosts collisionless energy loss, ending up with a similar \emph{total} energy loss. The increase in collisionless energy loss can be explained by the second bottleneck that occurs at each back cloud edge (see e.g. Figure \ref{fig:1D_energyloss}), therefore giving two interfaces that induce energy loss instead of just one. So ionization-dependent transport not only fails to decrease collisional losses substantially but can also offset this reduction by increasing collisionless energy loss -- the net result is a very similar cosmic ray calorimetry. 

We find larger differences in collisionless and collisional energy loss when we include an additional diffusion term. Under the assumption of full ionization, an additional diffusion coefficient of $3 \times 10^{28}$ cm$^{2}$ s$^{-1}$ significantly flattens the cosmic ray pressure gradient and leads to a decrease in collisionless energy loss by a factor of tens. Interestingly, with $f_{ion}^{min} = 10^{-4}$, the same drop in energy loss can be achieved with a smaller diffusion coefficient of $3 \times 10^{27}$ cm$^{2}$ s$^{-1}$. Without this additional diffusion, we saw that cosmic rays still bottleneck at the cloud edges, which severely decreases the diffusive flux through the clouds. This bottleneck is effectively wiped out when even a small constant diffusivity is added, as it smooths the cosmic ray pressure gradient and prevents the generation of confining waves; this maintains a large diffusivity through the partially neutral cloud interior, and transport no longer simply follows the ion Alfv{\'e}n speed. The outcome is that cosmic rays do not sense the cloud and effectively free-stream through it. While this localized faster transport decreases the collisionless energy loss by orders of magnitude, the decrease in collisional loss is only a factor of a few since collisions primarily come from cosmic rays with long residence times in the warm ionized gas (see e.g. Figure \ref{fig:Ecr_Collisions_PDFs}).

This finding is consistent with previous arguments in \cite{Chan2019} and \cite{Hopkins2020}, who compared full-galaxy simulations with varying diffusive and streaming transport with gamma-ray data. They found that, especially for dwarf galaxies, diffuse gamma-ray emission is far over-produced (by a factor of ten or more) compared to observations of Local Group dwarf galaxies unless the cosmic ray escape time is significantly reduced; despite strong outflows generated in these dwarf galaxies that would naively lead to fast advective escape, an additional cosmic ray diffusion coefficient of $3 \times 10^{29}$ cm$^{2}$ s$^{-1}$ is needed to lower emission to reasonable levels. 

More specifically, \cite{Hopkins2020} found that a significant fraction of gamma-ray emission was coming from the diffuse medium, where long cosmic ray residence times compensated for the low gas density to enhance collisional losses $\propto e_{CR} n_{gas}$. The simulations that best fit observed gamma-ray and cosmic ray grammage data, then, had to allow for fast cosmic ray transport in the \emph{diffuse} medium, while fast transport in cold, dense gas (where one would naively think most collisions occur) was not sufficient. Our study seems to corroborate this result, as we find collisional energy loss (hence, gamma-ray luminosity) to be fairly degenerate as we vary ISM setups and transport models. The largest decrease in collisions comes when an extra diffusion term is present -- an additional simulation with $\kappa_{||} = 3 \times 10^{29}$ cm$^{2}$ s$^{-1}$ (not shown in Figure \ref{fig:BigFigure}) continues the trend towards lower collisional energy loss.

\section{Discussion}
\label{discussion}

\subsection{Comparison to Previous Bottleneck Simulations}
While we've only simulated cosmic ray to thermal pressure ratios of at most a few, therefore accelerating clouds to only a few km/s, we can already gleam some of the implications for galactic wind driving and compare to the broader literature on cloud ``wind-tunnel" simulations and cosmic ray bottlenecks. First, we have extended the work of \cite{Wiener2017, Wiener2019} with a faster, more accurate method of cosmic ray transport \citep{JiangCRModule} in the Athena++ code. This speed-up allowed us to, amongst other things, further explore the parameter space of magnetic field strengths and elucidate the implications of field line warping for cloud acceleration. 

Our simulations without ionization-dependent transport are also similar to another recent simulation suite \citep{Bruggen2020} but with some differences. \cite{Bruggen2020} simulate a cosmic ray front with ratio of cosmic ray pressure to thermal pressure ranging from a few to a few tens, impinging upon a warm, fully ionized halo cloud with density $10^{-26}$ g cm$^{-3}$ and radius 100 pc, much larger and lower density than our representative ISM cloud. Unlike our simulations, which define the cosmic ray flux at the boundary for a finite time, they maintain a constant cosmic ray energy density at the boundary for the entirety of their simulation. While we include cosmic ray heating and collisional losses (which would be negligible for a diffuse halo cloud), they do not include heating but do include radiative cooling.

These differences in implementation and setup make it hard to draw fair comparisons. One differing result, however, is that we find a clear dependence of cloud acceleration on magnetic field strength, while they find only a small dependence (see e.g. Figure 8 of \citealt{Bruggen2020}). This is likely because 1) they hold the boundary cosmic ray energy density fixed, while our upstream cosmic ray pressure depends on the Alfv{\'e}n speed, and 2) they implement a switch that constrains cosmic rays to accelerate only the cells with density greater than 10$\%$ above the ambient density. If implemented in our simulations, these differences would change our results, as the amplitude and duration of the upstream cosmic ray pressure gradient would have no magnetic field dependence, and the acoustic pulse that precedes the cosmic ray front wouldn't be present. As in \cite{Wiener2017, Wiener2019}, we find that, when $c_{s} > v_{A}$, this acoustic pulse plays an important role in ``prepping" the cloud for the cosmic ray front: it begins to accelerate the cloud, sometimes pushing it tens of pc before the cosmic ray front reaches, and it affects the path of streaming cosmic rays by warping the magnetic field lines around the cloud. In our simulations with multiple clumps, especially those with low magnetic field strength ($c_{s} >> v_{A}$), these effects are evidently consequential, as the cosmic ray front squeezes between the clouds and almost entirely evacuates the hot, inter-cloud gas from the disk (see Figure \ref{fig:lognormal_L5_alpha1_5_lowerB}). The role and evolution of this ambient gas are clearly interesting and should be explored in future work.

The main purpose of our paper is to consider the additional transport effects in denser, partially neutral clouds that are most appropriate for the ISM but may also be applicable to molecular gas expelled into galaxy halos.
\cite{Everett2011} found that, since the cosmic ray flux quickly becomes diffusive and flattens the cosmic ray pressure gradient (both upstream and inside the cloud), the cosmic rays can't exert pressure forces and accelerate the cloud. Instead, on the upstream of the cloud, they can only exert a pressure difference (similar to ram pressure).

This picture changes if ion-neutral damping is not very strong, which in our simulations, occurs in the cloud interfaces. In those regions, cosmic ray pressure gradients impart momentum to the cloud with an amplitude and duration dependent on the magnetic field strength. As shown in \cite{Wiener2017, Wiener2019, Bruggen2020}, this bottleneck can lead to cloud acceleration possibly to hundreds of km/s in galaxy halos if the cloud can stay intact, which may be reinforced by radiative cooling. 

Including ionization-dependent transport, the cloud acceleration we find lies somewhere between these fully ionized bottleneck simulations and the expectations from \cite{Everett2011}, with cosmic rays still able to exert forces at cloud interfaces, rather than cloud interiors. The net differences in cloud momenta considering fully ionized transport vs ionization-dependent transport in partially neutral gas differ by less than a factor of 2. While we have only explored cosmic ray to thermal pressure ratios of order unity, we expect cosmic ray confinement to be even stronger at higher ratios, where cosmic ray pressure gradients can more effectively excite confining waves. These results suggest that cosmic rays, especially near cosmic ray sources where the pressure gradient is large, can still accelerate multiphase gas directly out of the ISM, despite not having a pressure gradient in the cloud interior. This deserves future attention and a more galactic wind-focused study than the one we've presented here.

\subsection{Limitations}
This work focuses on the interplay between cosmic rays and cold clouds, taking a critical look at the roles of bottlenecks induced upstream and elevated transport speeds in partially neutral clouds. We focus on the self-confinement model of cosmic ray streaming, with advective flux modified by varying ion Alfv{\'e}n speed and diffusive flux resulting solely from ion-neutral damping. We have not, until this point in the paper, explored the role of other damping mechanisms or instabilities that suppress or enhance confinement. While we mainly leave this to future work, we did re-run a subset of our 1D cloud simulations including nonlinear Landau damping (NLLD), which results from thermal particles taking energy from interacting Alfv{\'e}n waves. Following \cite{Hopkins2020}, the damping rate can be written as 

\begin{equation}
    \Gamma_{\rm NLLD} = \left[\frac{(\gamma_{CR} - 1)\pi^{1/2}}{8} \left(\frac{c_{s}v_{A}^{ion}}{r_{L}l_{CR}} \right) \left(\frac{e_{CR}}{e_{B}} \right) \right]^{1/2}
\end{equation}
This is proportional to the cosmic ray pressure gradient, suggesting that waves may be efficiently damped in the cloud interface region where a steep pressure gradient forms, but the diffusive flux is inversely proportional: $\kappa_{||} \propto (\nabla P_{CR})^{-1/2}$ \citep{Loewenstein1991}. Because of this, our simulations show only a small additional diffusion coefficient $< 10^{25} cm^{2}/s$ in the cloud interface, and the resulting cosmic ray and cloud evolution is almost exactly the same whether we account for NLLD or not. 

%This paper seeks to partially address two fundamental questions: how do cosmic rays navigate the multiphase ISM, and can observations constrain cosmic ray propagation and inform the development of new, more complete models of cosmic ray transport? While we're compelled by our results, this work is not without limitations. 

Another limitation, which we plan to explore in future work, is that these simulations do not include cooling or conduction. These play important roles in cloud survival during acceleration \citep{Bruggen2016THECONDUCTION, Gronke2018}, and they also set the thermodynamic state of the gas and the width of intermediate temperature transition regions, or cloud envelopes. Note that a typical Fields length in $10^{4}$ K gas is less than 0.1 pc, much smaller than our fiducial cloud setup with a 5 pc interface. Heating from cosmic ray streaming, while it noticeably heats the interface in our simulations, is not likely to set the interface width in real, cooling clouds. From our simulations, we find collisionless loss rates (hence, heating rates) less than $10^{-25}$ erg cm$^{-3}$ s$^{-1}$, subdominant compared to the expected radiative cooling and conductive heating rates if those were taken into account (see also Figure 9 of \citealt{Everett2011}). 

Models and observations of molecular clouds exposed to interstellar radiation fields, however, do show a gradual transition between hot and cold phases, with warm, ionized envelopes surrounding the cold cores \citep{Goldsmith2005}. Indeed, a growing literature on cosmic ray penetration into molecular clouds has come to similar conclusions about the trapping of cosmic rays by either extrinsic or self-generated turbulence in interface regions \citep{Morlino2015, Schlickeiser2016, Dogiel2018, Ivlev2018, Inoue2019}, especially near cosmic ray sources where it appears that cosmic rays can be confined even in the presence of neutral particles \citep{Nava2016, Brahimi2020}. 

%In this case, collisions within the dense molecular cloud can be frequent enough to instigate a pressure gradient and the generation of confining waves. We compare to these works more directly in the Appendix but note here some of the recent observational evidence that may support a picture of cosmic ray bottlenecks at cloud interfaces.

\subsection{Further Observational Constraints}
Constraining these theoretical results with observations is important and already underway to truly diagnose cosmic ray transport in dense, cold clouds. \cite{Silsbee2019} considered whether the low-energy (less than GeV) cosmic ray spectrum from the Voyager probe would be better fit by diffusive or free-streaming propagation of cosmic rays penetrating through a large column density into the Local Bubble. While confinement will be stronger at these energies than for GeV cosmic rays, they find that diffusive propagation is a better fit to the spectrum than free-streaming, which would additionally require an unreasonably high column density to fit the Voyager data. 

\cite{Fujita2020} model free-streaming and diffusive propagation (due to preexisting MHD turbulence) in molecular clouds, obtaining profiles of the ionization rate, 6.4 keV line flux, and gamma-ray emission. While there are only a few observed supernova remnants for which all three of these diagnostics are available, the qualitative comparison in their Discussion section finds that both diffusive and free-streaming propagation are likely realized. 

\cite{Joubaud2020} studied a diffuse cloud near the Orion-Eridanus superbubble that was observed to have a 34$\%$ lower cosmic ray flux compared to the local flux estimate. As the authors note, the estimates of \cite{Everett2011} suggest only a slight drop in cosmic ray flux within cold clouds and would require a much higher than observed magnetic field strength to account for such a 30$\%$ drop. Our simulations assume cloud parameters closer to the Eridu estimates ($n \approx 7$ cm$^{-3}$ and $B \approx 5 \mu G$) and find that the cosmic ray pressure drops by a factor $\approx 2$ if the cloud is partially ionized and steadily drops much more if the cloud is fully ionized. If the cloud lies along a magnetic flux tube with cosmic rays streaming into the halo, as speculated by \cite{Joubaud2020}, then the cloud may even be accelerating due to a cosmic ray bottleneck. In any case, if a warm ionized envelope exists, in which the Alfv{\'e}n speed drops and reinforces the cosmic ray pressure gradient, it's not unreasonable, given our results, for the cloud-interior cosmic ray pressure to be 30$\%$ lower than in the surrounding medium. More observational studies in this vein will be very helpful in further constraining cosmic ray - cloud interactions.

\section{Conclusions}
\label{conclusions}
Many MHD + cosmic ray simulations, aimed at modeling galaxy evolution, now take into account the self-confinement of GeV-energy cosmic rays through frequent scattering off cosmic ray - generated Alfv{\'e}n waves. While this kinetic process occurs at scales of order the cosmic ray gyroradius, many orders of magnitude below the current resolution limit of galaxy evolution simulations, fluid theories of this streaming transport have been developed and, with the advent of various numerical techniques, implemented in MHD codes for use in a galaxy-scale context. With these tools now available, our understanding of cosmic ray influence in the interstellar, circumgalactic, and intracluster medium has rapidly progressed; however, as of this writing, two important trends are evident in most published simulations: 1) it's assumed that GeV-energy cosmic ray transport proceeds as if the background thermal gas is everywhere fully ionized. In reality, the waves the cosmic rays resonate with propagate only in the ions, and streaming is therefore at the ion Alfv{\'e}n speed $v_{A}^{ion} = v_{A}/\sqrt{f_{ion}}$. Also, ion-neutral friction can damp the confining waves, thereby decoupling cosmic rays and greatly increasing their diffusive flux; 2) partially neutral ISM clouds and other density irregularities that induce decoupling and nonlinear cosmic transport are frequently under-resolved. If we imagine the multiphase ISM as an obstacle course of coupled vs decoupled regimes, further complicated by hadronic and Coulomb collisions between cosmic rays and ambient gas, then how do cosmic rays navigate this medium, and can we constrain this transport with observations? The answers have implications for ISM dynamics, interpretation of observations, and the acceleration of cold interstellar gas in galactic winds and fountains. 

In this paper, we presented a suite of high-resolution, idealized simulations of cosmic ray fronts navigating the multiphase ISM. We ran each simulation twice: first, assuming the medium is fully ionized and, second, accounting for changes in ionization that induce ``ionization-dependent transport". We began with 1D simulations of an energetically significant cosmic ray front hitting a single, partially neutral cloud with radius and density typical of ISM conditions. We then explored how cosmic ray transport, cosmic ray energy loss, and cloud acceleration and morphology varied in multiple dimensions and varying magnetic field strength. We then ran the same cosmic ray fronts through mock ISM setups: slabs of lognormally-distributed multi-phase clumps on top of an exponentially decaying mean density. We assessed how cosmic rays sampled the ISM and how ionization-dependent transport affected cloud acceleration and collisional and collisionless energy loss.

Our main findings were: 

\begin{itemize}
\item Cloud interfaces can play a crucial role in setting the cosmic ray propagation through the cloud interior. The drop in Alfv{\'e}n speed while the gas is still fully ionized leads to a bottleneck in the interface; this reinforces the cosmic ray pressure gradient that generates confining waves and locks cosmic rays to the wave frame. The effects of ion-neutral damping are thereby suppressed, and the advective flux rather than diffusive flux generally dominates transport ($v_{st} \approx v_{A}^{ion}$). 

\item The stair-step structure that develops at cloud interfaces facilitates collisionless energy transfer from the cosmic rays to waves $\propto v_{A}^{ion} \nabla P_{CR}$. Unlike for a fully ionized cloud where a single cosmic ray gradient forms \citep{Wiener2017, Wiener2019}, there's now a bottleneck at both the front \emph{and} back cloud edges. In our mock ISM simulations, this extra energy loss partially offsets a reduction in collisional energy loss, keeping the total cosmic ray calorimetry similar to when full ionization is assumed. The additional gas heating may also have implications for ion abundances and kinematics at cloud interfaces, though we estimate that the role of heating would be secondary to cooling and conduction. We plan to explore this further by more self-consistently modeling interface properties rather than leaving the interface width as a free parameter.   

\item Cosmic rays sample the ISM differently when varying ionization is accounted for. Ionization-dependent transport allows cosmic rays to leak into dense cloud interiors while, assuming full ionization, cosmic rays are confined to thin cloud boundary layers. These differences are imprinted on gamma-ray emission maps. 

\item Total collisional losses are very similar for the two transport models. With ionization-dependent transport, short cosmic ray residence times in clouds are offset by the denser gas they access. So while gamma-ray emission maps with the two transport models look very different, the gamma-ray luminosity is dominantly set by the cosmic ray transport in the warm ionized medium, which is the same in each transport model. This result is consistent with \cite{Hopkins2020}. 

\item When ionization is accounted for, cloud acceleration is decreased since ionization-dependent transport in cloud interiors flattens the cosmic ray pressure profile. The change is not as large as one might expect, however, as our simulations exhibit a factor $< 2$ difference in cold gas momentum. Cosmic rays still bottleneck at the cloud edges, leading to pressure gradient forces that push the cloud. This suggests that, contrary to previous expectation, ionization-dependent transport does not greatly inhibit cosmic rays from accelerating cold, molecular gas in galactic outflows. The magnetic field topology, which can be warped upstream by strong cosmic ray fronts, may be a greater inhibitor of direct cosmic ray acceleration of clouds. We will explore these aspects more thoroughly in a future study.

\end{itemize}

Clearly, cosmic ray propagation through the multiphase ISM is a very rich problem with fundamental implications for ISM dynamics and feedback. To understand how cosmic rays shape their environments, we need to continue applying first-principle cosmic ray transport models to astrophysical problems, test the outcomes against observations, and use those constraints to create more complete models of cosmic ray transport. Simulations that employ streaming at the gas Alfv{\'e}n speed will drastically underestimate transport speeds in partially neutral clouds, thereby overestimating cloud acceleration. On the other hand, constant field-aligned diffusion models have no dependence on local gas conditions, thereby precluding bottleneck formation at cloud interfaces, and they omit the energy transfer from cosmic rays to thermal gas, fundamentally changing cosmic ray influence on galaxies \citep{2017MNRAS.467..906W}. ``Two-$\kappa$" prescriptions (e.g. \citealt{Farber2018}), while tying the transport speed to local gas temperature, again prohibit bottlenecks at cloud interfaces and may overestimate the effective transport speed depending on the choice of $\kappa$. These models, then, likely underestimate the dynamical influence of cosmic rays on partially neutral gas. 

In light of our results, we instead advocate that galaxy evolution simulations should, to zeroth order, model streaming at the ion Alfv{\'e}n speed with a temperature-based ion fraction. This method will more fully capture the generation of bottlenecks and steep pressure gradients at cloud interfaces, as well as ionization-dependent transport in cloud interiors that is robust, at least in these simulations, to changes in ionization fraction. More detailed modeling of ionization from local sources and lower energy cosmic rays could alter this picture and should be a next step.

%In light of our results, we advocate that galaxy evolution simulations should at least prescribe a temperature-based ion fraction and model streaming at the ion Alfv{\'e}n speed. Rather than streaming at the Alfv{\'e}n speed, constant field-aligned diffusion, or ``two-$\kappa$" models depending on gas temperature, this method will more fully capture the generation of bottlenecks and steep pressure gradients at cloud interfaces, as well as ionization-dependent transport in cloud interiors that is robust, at least in these simulations, to changes in ionization fraction. More detailed modeling of ionization from local sources and lower energy cosmic rays could alter this picture and should be a next step.

%Simulations with more realistic ISM setups should further illuminate the regimes of cosmic ray confinement and whether novel models of cosmic ray transport, taking into account other cosmic ray instabilities and wave damping processes, are really needed to match observations. 

\section*{Acknowledgments}
The authors are greatly indebted to Yan-Fei Jiang for providing us with his two-moment method of cosmic ray transport in Athena++. We also thank the anonymous referee whose comments greatly improved the quality and clarity of this manuscript, as well as Josh Wiener, Mateusz Ruszkowski, Ryan Farber, Peng Oh, and Max Gronke for stimulating discussions. This research was supported in part by the National Science Foundation under Grant No. NSF PHY-1748958 and by the Gordon and Betty Moore Foundation through Grant No. GBMF7392. EGZ is happy to acknowledge support from NSF Grant AST 2007323. This work made extensive use of the \emph{yt} \citep{ytPaper} software package, a toolkit for analyzing and visualizing quantitative data. Computations were performed using the Extreme Science and Engineering Discovery Environment (XSEDE), which is supported by National Science Foundation grant number ACI-1548562 \citep{xsede}. Specifically, our computing resources stemmed from allocation TG-AST190019 on the Stampede2 supercomputer.

\appendix

\subsection{Effect of Resolution}
\label{append1}

\begin{figure*}[h]
\centering
\includegraphics[width = 0.98\textwidth]{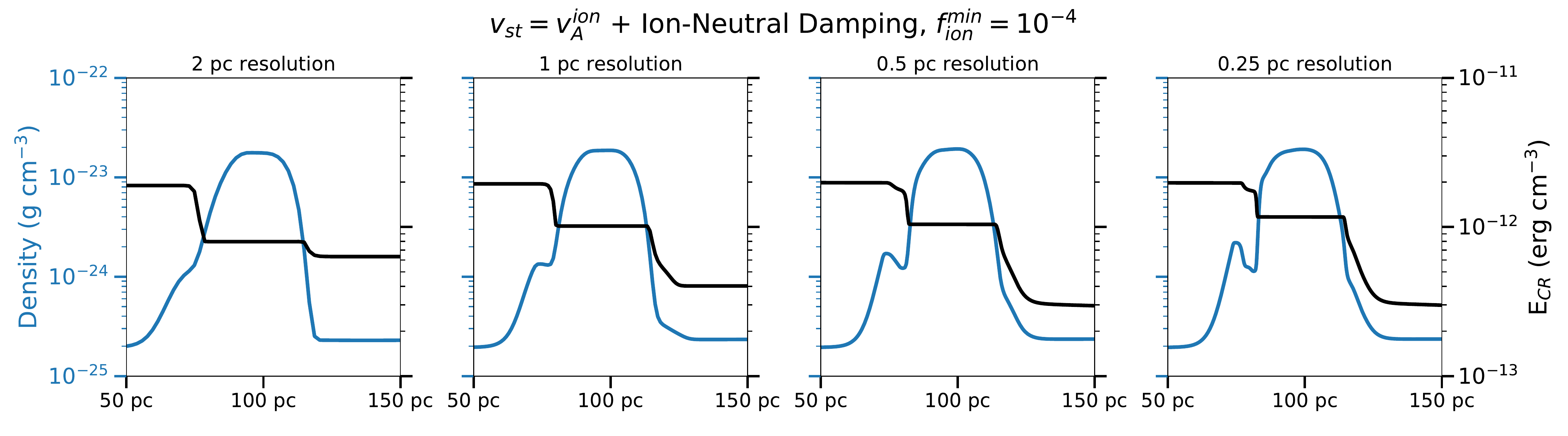}
\caption{Density and cosmic ray energy density for the fiducial cosmic ray front impinging on a cloud of radius 10 pc and interface width 5 pc. Each panel corresponds to a different resolution. When the interface region is well-resolved ($\approx 10$ cells per interface), the results are well-converged. With poorer resolution, the cosmic ray front doesn't respond to the drop in Alfv{\'e}n speed at the interface, therefore barreling through the cloud and appearing on the other side with a larger cosmic ray energy. The spike in density on the front cloud edge, caused by cosmic ray heating, is also not as noticeable at low resolution.}
\label{fig:1D_change_res}
\end{figure*}

In this section, we carry out a small resolution study, primarily in 1D, to test convergence and inform larger-scale simulations of galaxy evolution, which may under-resolve cosmic ray interactions with multiphase structures. Figure \ref{fig:1D_change_res} shows our fiducial 1D cloud of radius 10 pc and interface width 5 pc simulated at resolutions of 2, 1, 0.5, and 0.25 pc. As noted in Section \ref{1DSection}, not resolving the cloud interface is akin to having a physically smaller interface width. Lower resolution leads to a smaller total drop in cosmic ray energy through the cloud, less heating at the front edge, and different cloud morphologies (note the disappearance of the upstream ``thumb" caused by heating). Convergence is achieved at a resolution of 0.5 pc, which resolves the 5 pc - thick interface by a sufficient number of cells. Similarly, our simulations with a 10 pc interface width reached convergence at 1 pc. 

%\textcolor{red}{Any idea why 0.5pc? Have you checked to see whether the choice of $V_m$ has anything to do with it?}

\begin{figure*}[h]
\centering
\includegraphics[width = 0.4\textwidth]{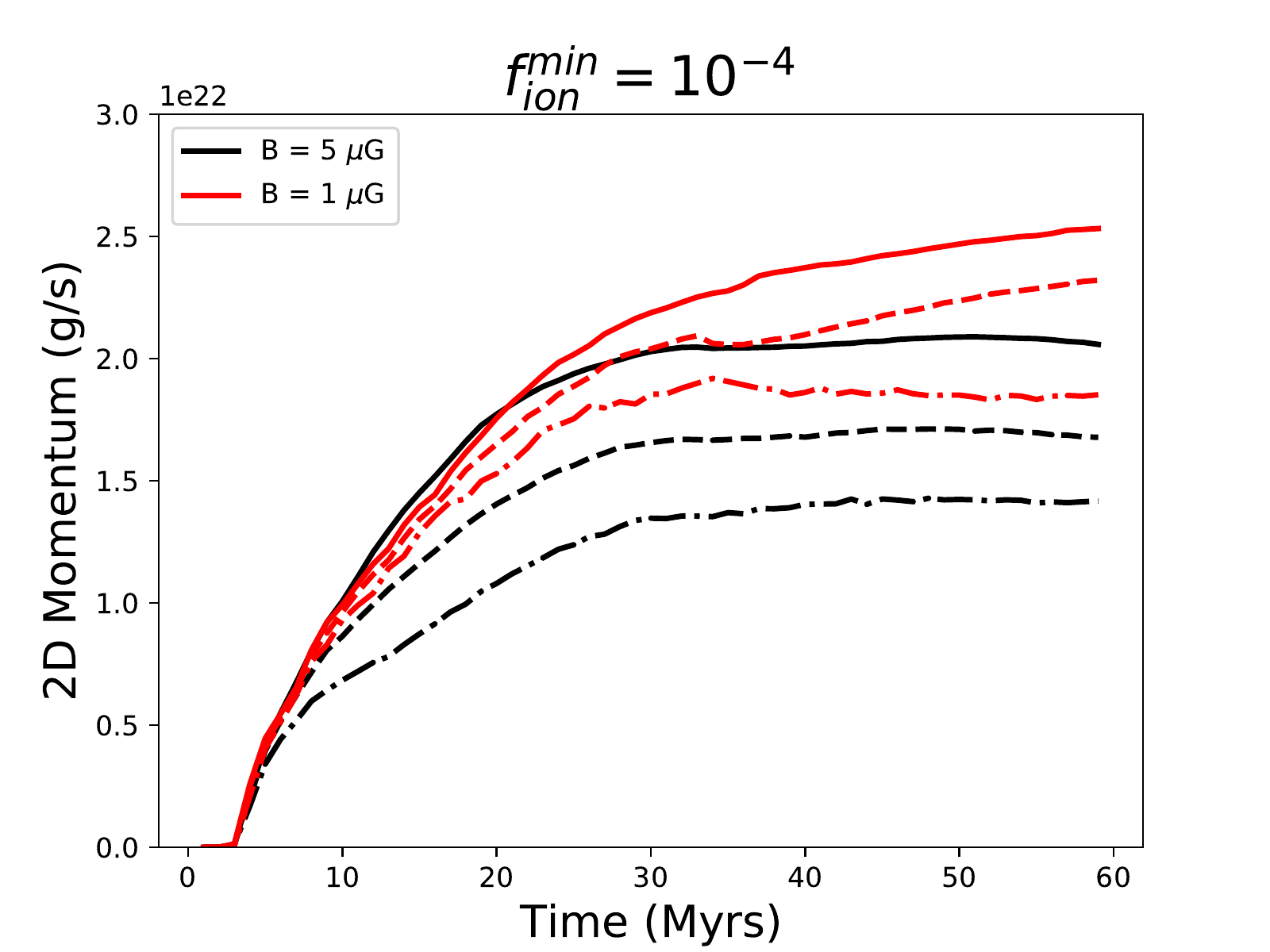}
\includegraphics[width = 0.4\textwidth]{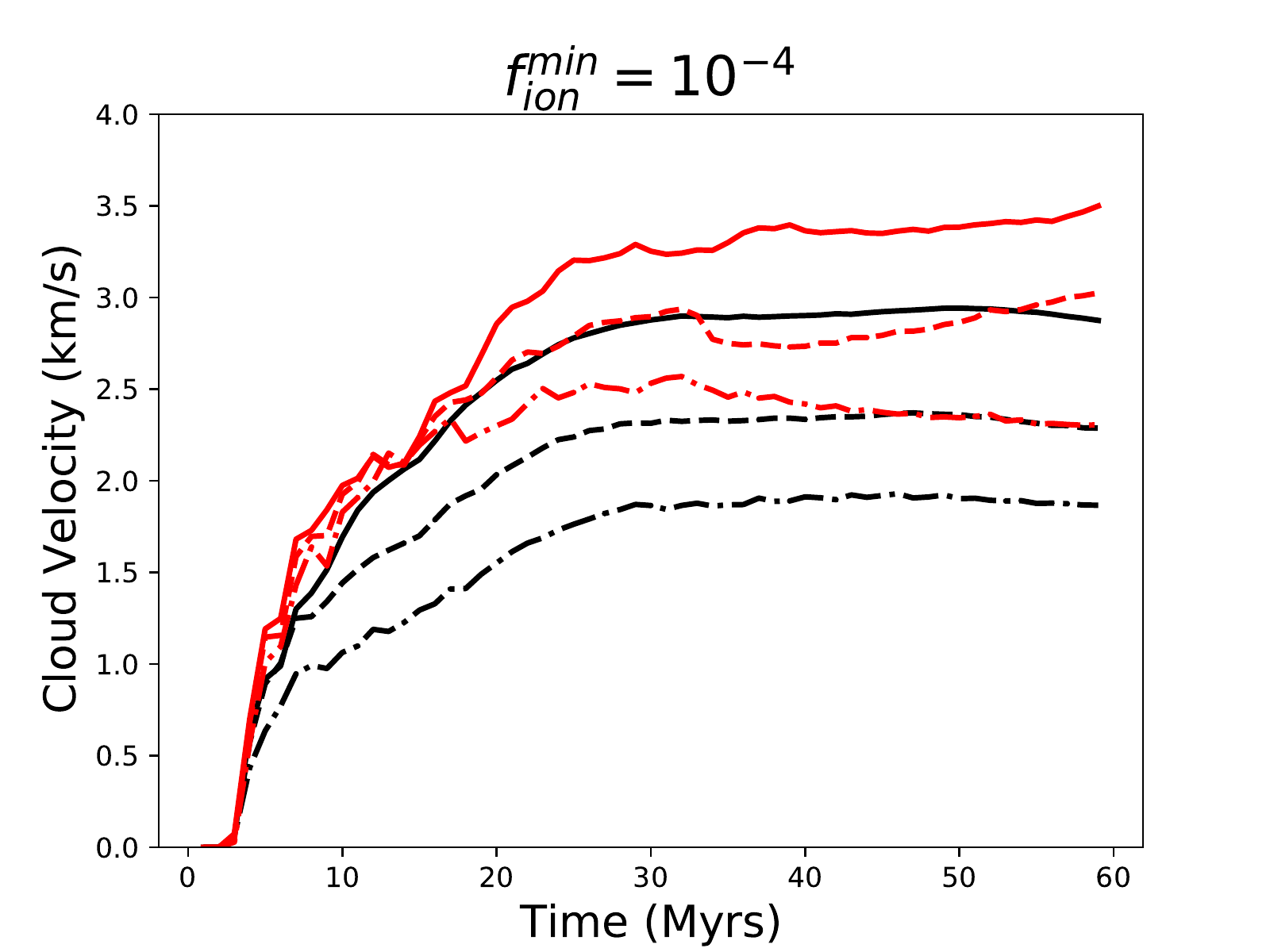}
\caption{Cloud momentum (left) and velocity (right) at varying resolutions of 1 pc (solid lines), 2 pc (dashed lines), and 4 pc (dot-dashed lines) for the 2D cloud simulation of \S \ref{2DSection} with B = 5 $\mu$G (black lines) and B = 1 $\mu$G (red lines). Deteriorating resolution leads to less cloud acceleration, since cosmic ray pressure gradients are smoothed at the unresolved cloud boundary. Even at the fiducial 1 pc resolution, we don't appear to be reaching convergence, as expected from Figure \ref{fig:1D_change_res}.}
\label{fig:2DCloud_resStudy}
\end{figure*}

Because the total pressure drop determines the force on the cloud, we expect that under-resolved simulations will underestimate cloud acceleration. To test this, we ran our 2D cloud simulations (with $f_{ion}^{min} = 10^{-4})$ at resolutions of 4, 2, and 1 pc. The resulting cloud momenta and velocities are shown in Figure \ref{fig:2DCloud_resStudy}. As expected, deteriorating resolution decreases cloud acceleration and, hence, the perceived effectiveness of cosmic ray-driven feedback. 

While this pc or sub-pc resolution requirement is initially daunting, it's worth noting that, if transport is dominated by the advective rather than diffusive flux, such problems become more computationally tractable. Our choice to set a high $V_{m} = 10^{10}$ cm/s, advective cap of $10^{9}$ cm/s, and diffusive cap of $3 \times 10^{30}$ cm$^{2}$/s in the first place was a pre-emptive measure anticipating that large diffusive fluxes, in particular, would need to be accommodated. As we found, that is rarely the case, and transport speeds are generally limited by $v_{A}^{ion}$. We found only small differences, then, when $V_{m}$ and the diffusive and advective caps are each \emph{decreased} by a factor of 5-10, which should provide a glimmer of hope for convergence of future large-scale simulations with ionization-dependent transport.

\begin{figure*}
\centering
\includegraphics[width = 0.4\textwidth]{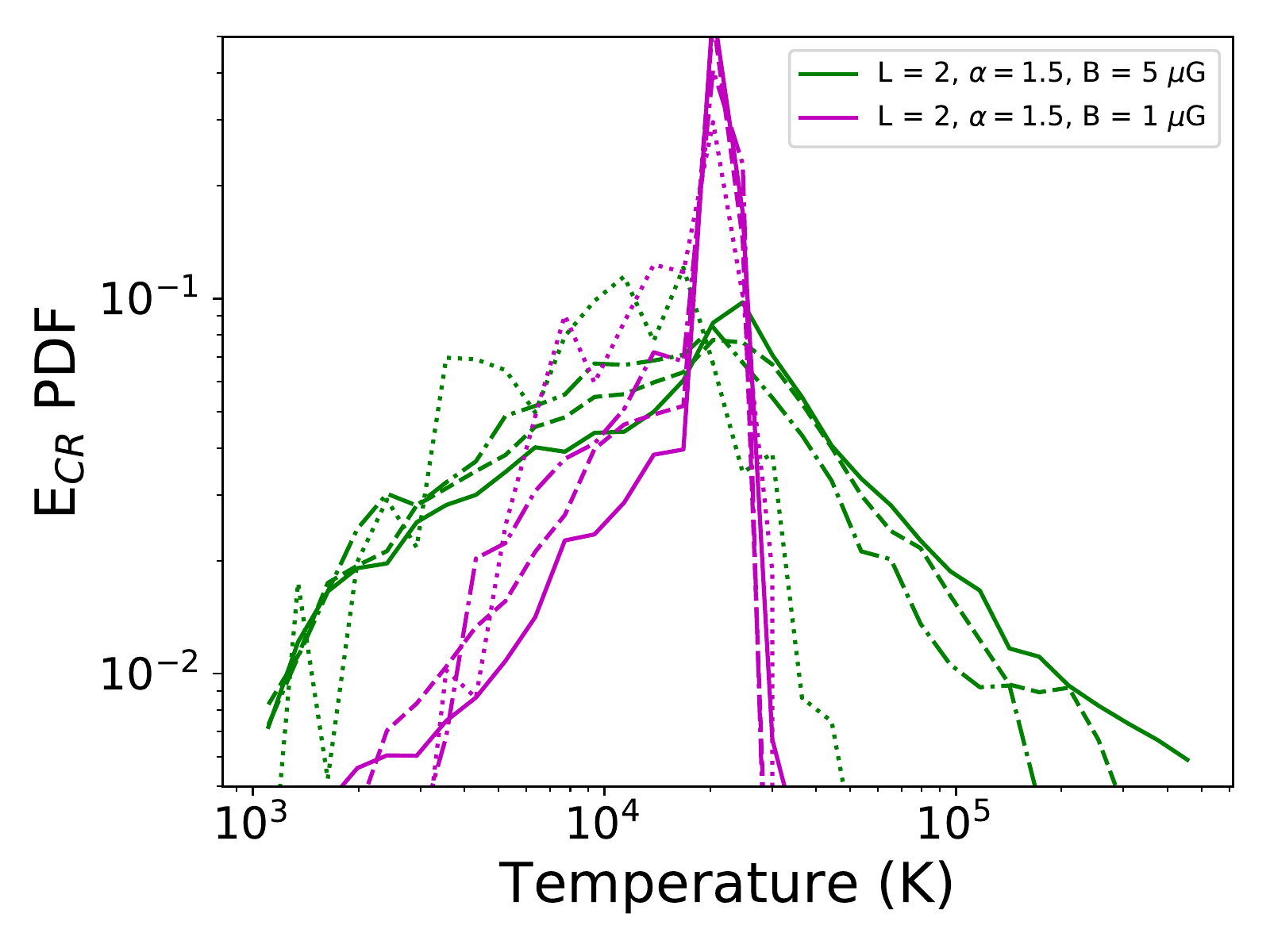}
\includegraphics[width = 0.4\textwidth]{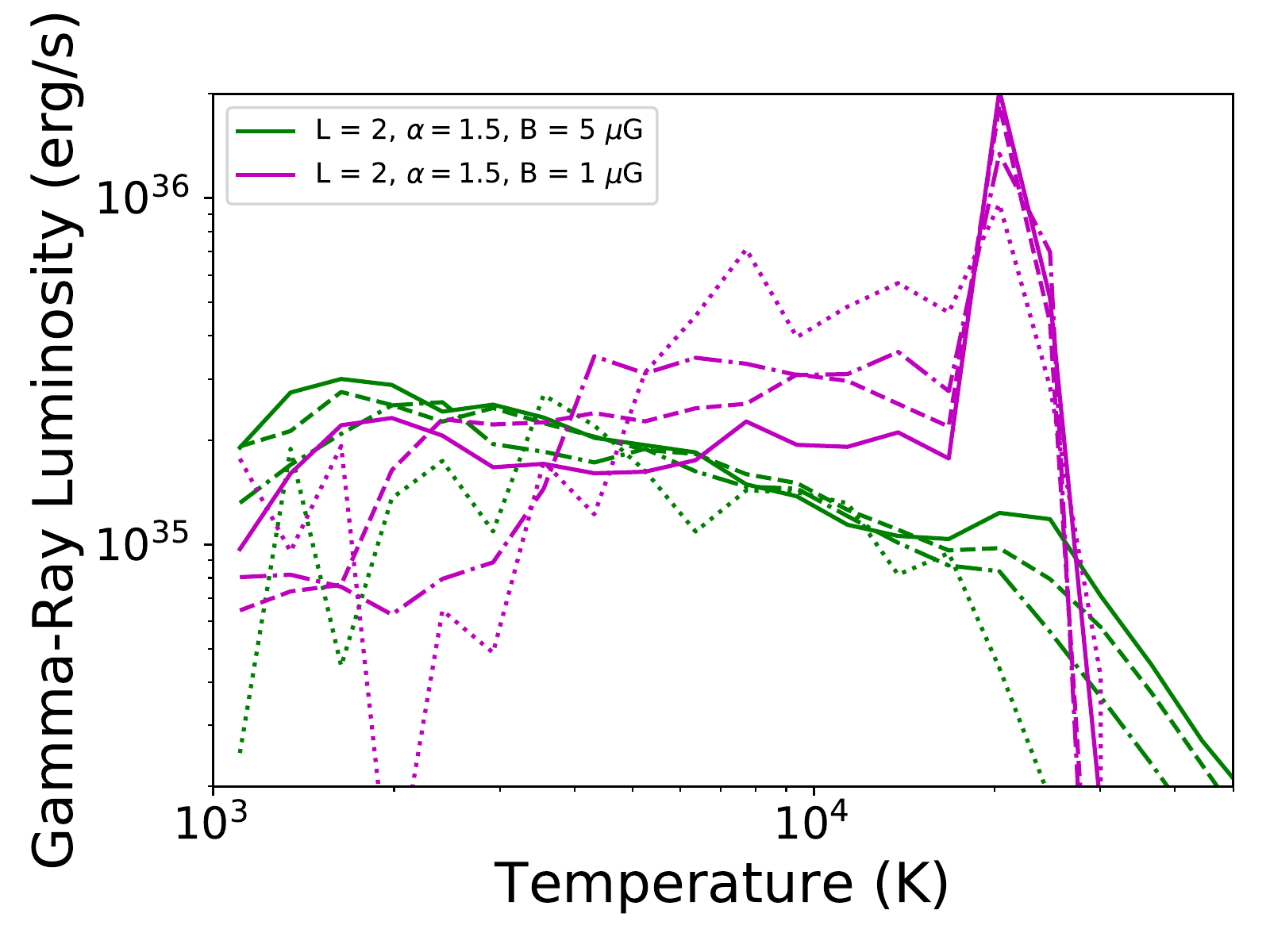}
\caption{Probability distribution function (see also Figure \ref{fig:Ecr_Collisions_PDFs}) of cosmic ray energy density (left) and histogram of gamma-ray luminosity (right) at varying resolutions of 2 pc (solid lines), 4 pc (dashed lines), 8 pc (dot-dashed lines), and 32 pc (dotted lines). For either magnetic field strength tested, lowering resolution leads to higher cosmic ray occupancy in cold gas, especially at interface temperatures $T \approx 10^{4}$ K. This leads to slightly more gamma-ray emission coming from cold clumps. This is especially noticeable in the B = 1 $\mu$G simulations, though the majority of emission still comes from the $T \approx 2 \times 10^{4}$ K band. Note the change in x-axis limits in the right panel to focus on the lower temperature region where most gamma-ray emission comes from. }
\label{fig:PDFs_resStudy}
\end{figure*}

Finally, we test how cosmic rays sample the ISM and transfer energy in our clumpy simulations with lower resolutions. We again focus on ionization-dependent transport with $f_{ion}^{min} = 10^{-4}$, and we test resolutions of 32, 8, 4, and 2 pc in our (L, $\alpha$) = (2, 1.5) setups. The resulting cosmic ray energy density PDFs and gamma-ray luminosity histograms are shown in Figure \ref{fig:PDFs_resStudy}. Lowering the resolution smooths cloud interfaces, allowing cosmic rays to leak more easily into cold, dense gas. This trend is evident in both the B = 5 $\mu$G and B = 1 $\mu$G simulations. This leads to more gamma-ray emission coming from the dense, cold gas, especially in the interface temperature range near $8 \times 10^{3}$ K. 

The total gamma-ray luminosities over all temperature bins are relatively unchanged when the resolution is varied, though, and even appear to decrease in the B = 5 $\mu$G case with 32 pc resolution. This is consistent, also, with our findings of \S \ref{mainSims} where we changed the smoothness of the clumpy ISM via the $\alpha$ parameter. In Figure \ref{fig:BigFigure}, one can see that lower $\alpha$, akin to lower resolution that would smooth out density perturbations, actually leads to a decrease in collisional losses. Poor resolution, then, appears not to be the dominant factor that grossly overestimates gamma-ray emission in large-scale simulations, but more high-resolution simulations of cosmic rays in different environments (e.g. superbubbles vs mean ISM) and for a range of mean gas densities are needed. 

%this suggests that the resolution requirement to resolve cosmic ray influence on multiphase gas is of order 1 pc or less assuming that a typical cold cloud in the ISM is similar to our fiducial 10 pc radius cloud. This is daunting given that cosmic ray propagation speeds are close to $10^{9}$ cm/s in cloud cores, therefore lowering the stable timestep. Resolved, 3D, galaxy-scale simulations are then likely to be quite computationally expensive. \textcolor{olive}{Still finishing this part up.}

\subsection{Validity of a Fluid Cosmic Ray Model}
A possible limitation of this work concerns the validity of a fluid model for cosmic rays. This assumption rests upon the cosmic ray mean free path $\lambda_{mfp}$ being shorter than other length scales of interest, namely the cosmic ray scale length $l_{CR}$ and is, in fact, baked into the diffusivity we've implemented: Assuming a short mean free path and balancing scattering against acceleration down the pressure gradient leads to a scattering rate $\nu\sim c^2/(l_{CR}v_s)$, where $v_s$ is the streaming velocity at which wave damping balances wave growth. This implies that the ratio of mean free path to scale length $\lambda_{mfp}/l_{CR}\sim v_s/c < 1$.  So even at cloud interfaces where the diffusivity is initially very large and the scale length becomes very short as the cosmic ray front approaches, $\lambda_{mfp} < l_{CR}$ by design.  %So it's natural to question whether the mean free path, which is initially very large inside the cloud, is ever greater than the cosmic ray scale length, which suggests we are really in a kinetic rather than a fluid regime. 

\begin{figure}
\centering
\includegraphics[width = 0.48\textwidth]{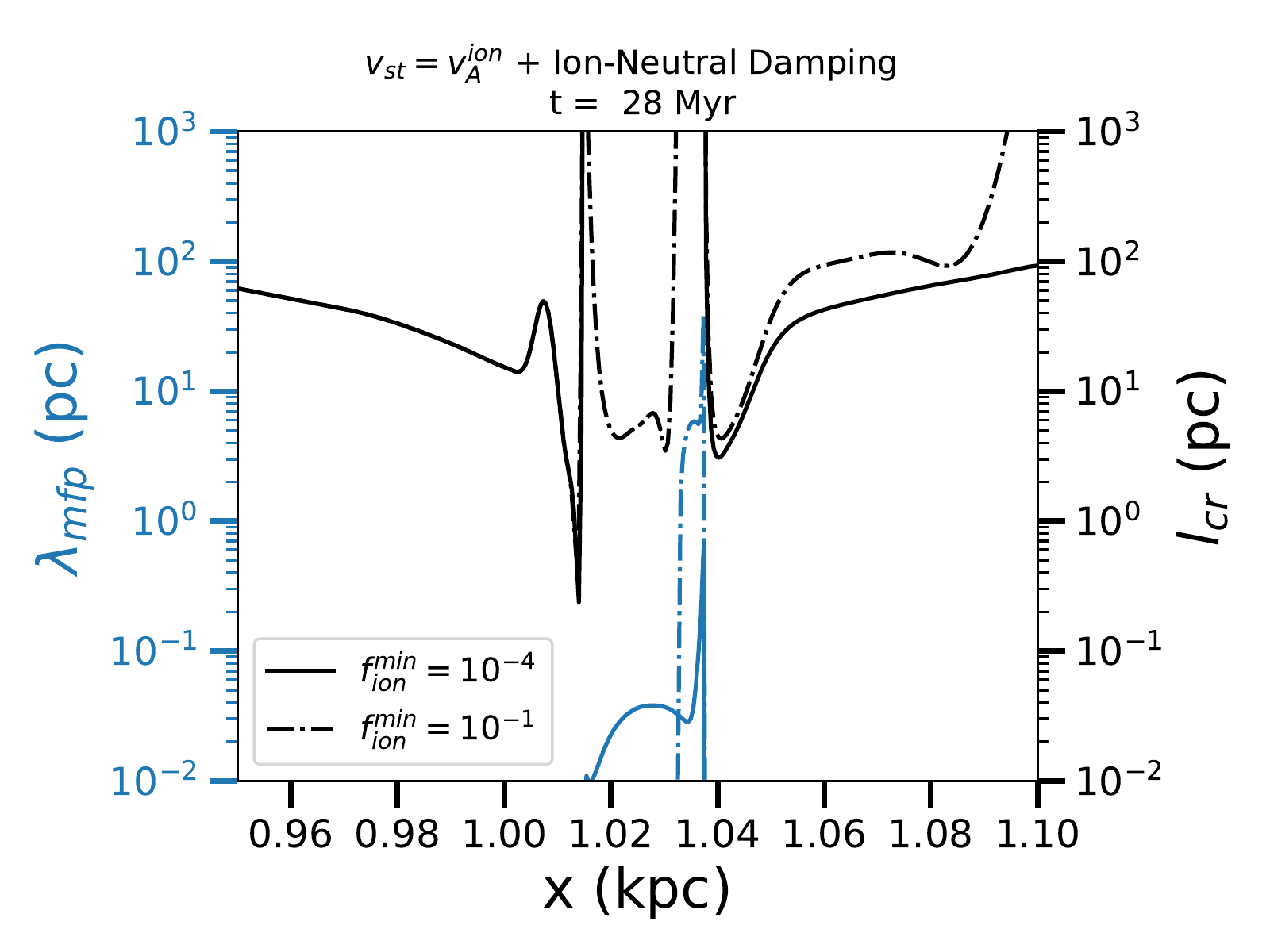}
\caption{The cosmic ray mean free path $\lambda_{mfp} = \kappa_{||}/c$ vs the cosmic ray scale length $l_{CR}$ for the 1D simulations of \S \ref{1DSection}.}
\label{fig:1D_mfp_lcr}
\end{figure}

In Figure \ref{fig:1D_mfp_lcr}, we plot the mean free path $\lambda_{mfp} = \kappa/c$ versus $l_{cr} = P_{CR}/\nabla P_{CR}$ at a time of 28 Myrs when the cosmic ray front is just hitting the cloud we presented in Section \ref{1DSection}. At every snapshot we output to data, the mean free path is below the cosmic ray scale length, but clearly, as in this snapshot, there are locations upstream of the cosmic ray front where the mean free path is greater than the scale length a few cells away from it. 

%So while, at every point $\lambda_{mfp} < l_{CR}$, it's unclear whether we are truly in a fluid regime. 

%\textcolor{red}{I have a possible insight into this. \textit{Assuming} a short mean free path and balancing scattering against acceleration down the pressure gradient leads to $\nu\sim c^2/(Lv_s)$, where $v_s$ is the streaming velocity at which wave damping balances wave growth. This implies that the ratio of mean free path to scale length $\lambda/L\sim v_S/c$. Since your $\kappa$ are derived on that basis I think the small $\lambda/L$ criterion is baked in.}

\subsection{Cosmic Ray Exclusion via Collisional Losses}

One of the main findings of our work is that the interface between hot and cold gas plays a crucial role in cosmic ray propagation and dynamical effects on the cloud. Our result is broadly consistent with a growing literature on cosmic ray penetration into molecular clouds, the history of which is quite rich (see e.g. \citealt{Skilling1976} and \citealt{Cesarsky1978} for seminal papers) and outlined in \cite{Everett2011}. Specifically, recent works have noted that cosmic ray penetration into molecular clouds is highly nonlinear and self-modulated by the excitation of waves in the cloud interface (see the Discussion section). While we have focused on a cosmic ray front moving with a preferred direction toward a cloud, it may be that molecular clouds sit in a ``sea" of cosmic rays -- an initially flat cosmic ray energy profile. In this case, with no pre-existing pressure gradient, the cosmic ray profile is modulated by collisional loss rates in the dense molecular cloud. The resulting pressure difference sucks fresh cosmic rays into the cloud, and how much those exterior cosmic rays can pressurize the cloud interior depends on the transport through the cloud interface.  
%\textcolor{red}{This is actually related to old work by Skilling \& Strong (1976) and Cesarsky \& V\"olk (1978), which I think John Everett \& I cited but you might want to comment.}

%This toy problem not only sheds light on cosmic ray diffusive and advective transport but also partially addresses whether Coulomb and hadronic collisions, themselves, can instigate cosmic ray wave zones. \cite{Skilling1976} proposed that, since low-energy cosmic ray protons (E $<$ 300 MeV) would be destroyed by collisions within clouds, the cosmic ray flux entering a cloud would be greater than the flux leaving it, setting up a cosmic ray pressure gradient and the generation of waves through the streaming instability. This would then limit the cosmic ray drift speed to the Alfv{\'e}n speed, decreasing the cosmic ray flow into the cloud. \cite{Cesarsky1978}, however, found that this would only happen for energies $<$ 50 MeV because the compression of the magnetic field on the cloud interface would raise the cosmic ray flux into the cloud. For cosmic rays of GeV energies, which makes up the peak of the cosmic ray energy spectrum, the outcome is still unclear. \textcolor{olive}{Mention and compare to Morlino and Gabici 2015, Schlickheiser+ 2016, various papers about self-modulation by Dogiel, Ivlev, Chernyshov, Fujita+ 2020 (diffusive and free-streaming both seem plausible in that analysis)}. 

We model this scenario for two different clouds each of radius 50 pc, with varying interface widths of 25, 5, and 0.5 pc, and with densities of 10 and 100 cm$^{-3}$. We use a uniform resolution of 0.5 pc for simulations with interface widths 25 and 5 pc, and we increase the resolution to 0.05 pc for the 0.5 pc interface run in order to keep the interface resolved. Initially, the cosmic ray energy density is uniform everywhere, but it drops within the cloud as collisions decrease the cosmic ray pressure. We use inflow/outflow boundaries to allow for a steady influx of cosmic rays into the simulation and back into the cloud. 

\begin{figure*}
\centering
\includegraphics[width = 0.46\textwidth]{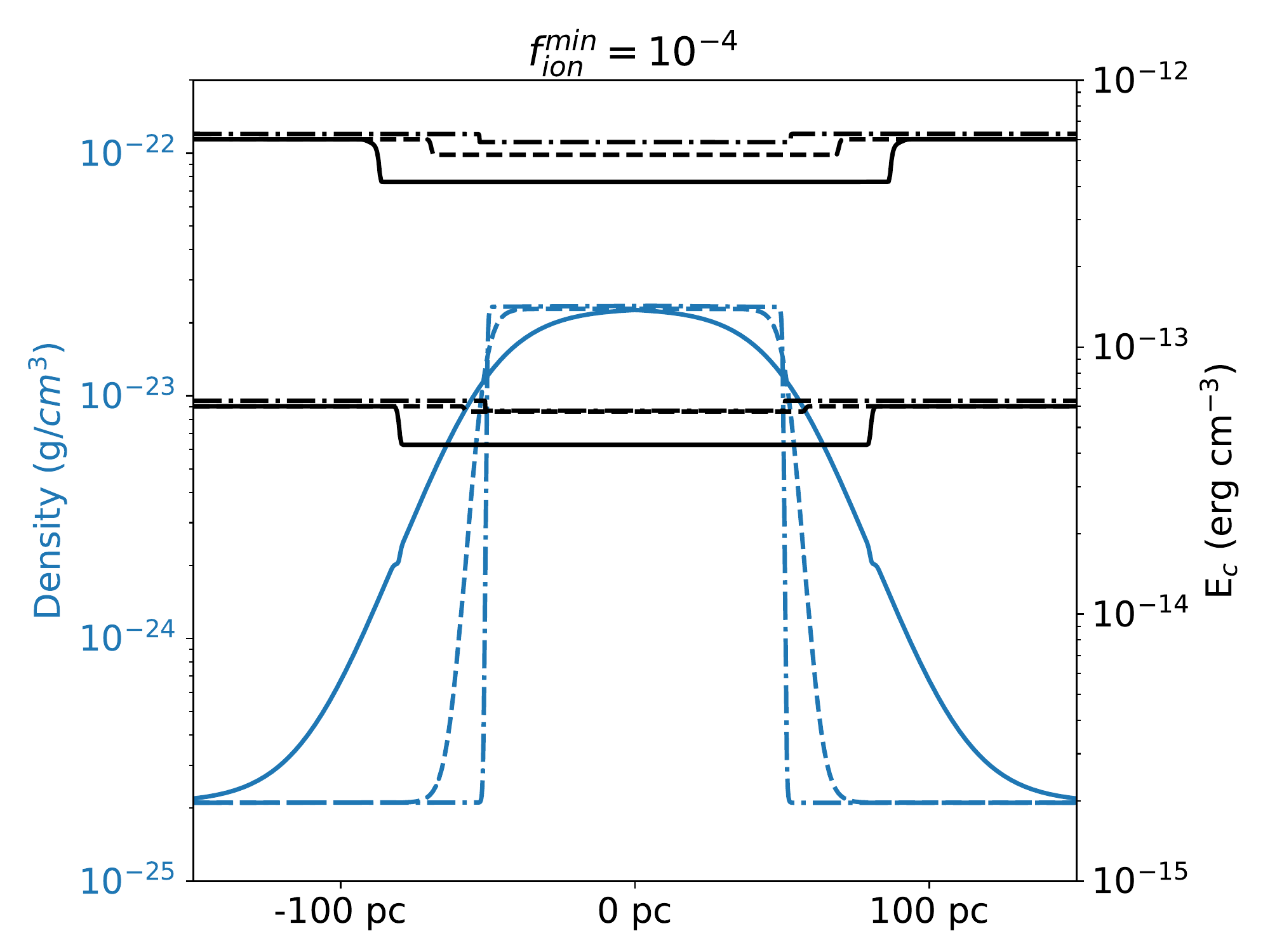}
\includegraphics[width = 0.46\textwidth]{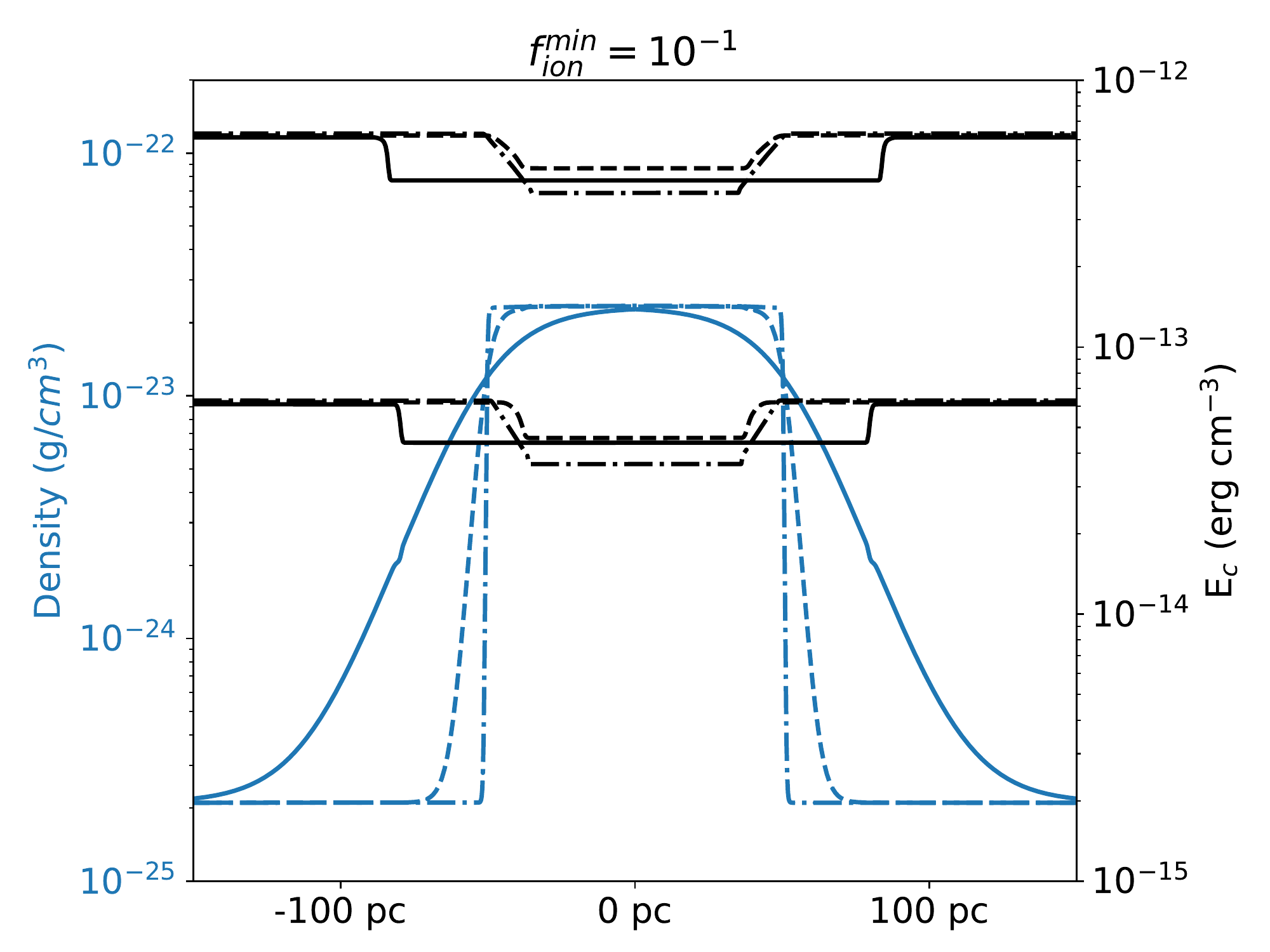}
\includegraphics[width = 0.46\textwidth]{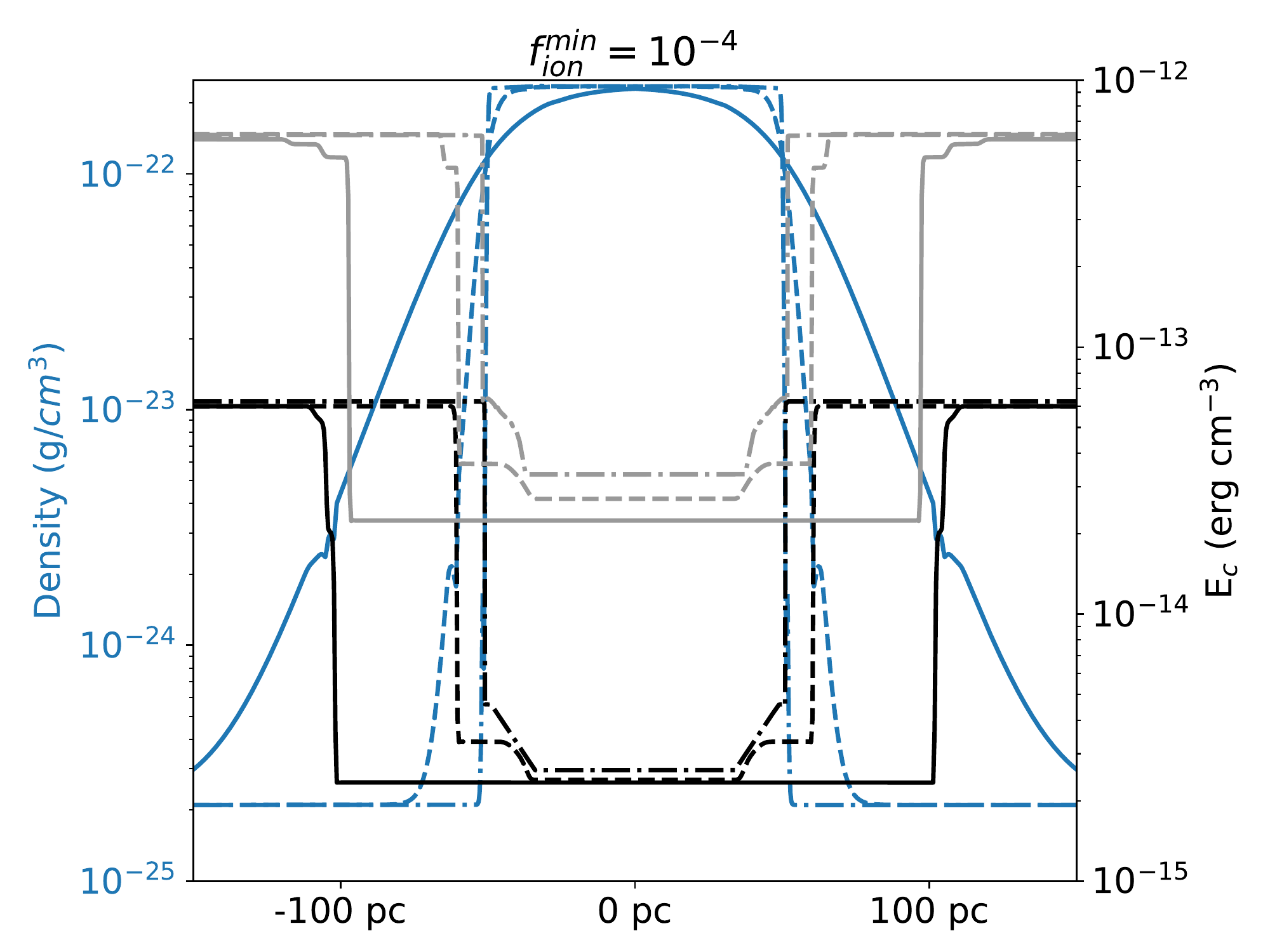}
\caption{Profiles of 1D clouds with peak densities of 10 cm$^{-3}$ (top row) and 100 cm$^{-3}$ (bottom row) initially bathed in a sea of cosmic rays with uniform pressure. Collisional losses decrease cosmic ray pressure within the cloud core, and the steady-state pressure in the cloud is governed by a balance between this loss rate and the influx of new cosmic rays through the interface, which we vary in width from 25 pc (solid lines), 5 pc (dashed lines), and 0.5 pc (dot-dashed lines). For the 10 cm$^{-3}$ cloud, the cosmic ray pressure within the cloud is almost the same as the outside pressure, especially for the $f_{ion}^{min} = 10^{-4}$ simulations with interfaces too thin to instigate significant bottlenecks. Wider and denser clouds show greater drops in cosmic ray pressure. }
\label{fig:cosmicraysea}
\end{figure*}

Figure \ref{fig:cosmicraysea} shows our results at snapshots when the cosmic ray energy density has ``bottomed out", i.e. when a steady-state has been reached between cosmic ray inflow through the interface and energy loss in the interior. The cloud morphology changes somewhat due to the pressure imbalance, but the cosmic ray pressure we start with is at least a few times smaller than the thermal pressure, so the effect is small. For the fiducial cloud density of 10 cm$^{-3}$ (top row) with minimum ion fractions of $10^{-4}$ and $10^{-1}$, the cosmic ray energy stays fairly level. As expected, the interface is the biggest bottleneck for inflowing cosmic rays. In the $f_{ion}^{min} = 10^{-4}$ simulations, a thicker interface leads to a larger drop in cosmic ray energy within the cloud. For $f_{ion}^{min} = 10^{-1}$, the diffusive flux term becomes more important, and the trend with interface width becomes more complicated, though the differences in cosmic ray pressure are quite small.

For a denser cloud (100 cm$^{-3}$), which is the fiducial value in \cite{Everett2011}, the energy drop is more substantial. There is a slight trend towards higher cosmic ray energy inside the cloud when the interface is thinner, but generally, the energy drops are very similar for different interface widths and different initial cosmic ray energies \emph{as long as the interface is well-resolved}. It's worth noting that, while we try to choose steady-state snapshots, the cloud interface is constantly changing due to collisionless heating, which in these simulations is not counteracted by radiative cooling. In the high initial cosmic ray energy cases, where the magnitude of this heating is greatest, the interface eventually widens and shows clear density ripples moving outwards from the cloud. This can throw the cosmic ray energy out of steady-state, so we've chosen early (within tens of Myrs) snapshots with seemingly steady-state profiles for these figures. Closer inspection of how the interface changes in the additional presence of radiative cooling and thermal conduction is left to future work.

%There are also more apparent differences when we start with a lower vs higher cosmic ray energy. For lower cosmic ray energy, the pressure drop is roughly the same regardless of interface width. The trend is a bit more clear for the higher cosmic ray energy runs: a wide bottleneck region induces a large drop in pressure, while the 0.5 pc interface doesn't suppress wave generation as much and allows slightly more cosmic ray penetration into the cloud. The 5 pc interface gives an anomalously large drop, however The lower cosmic ray energy leads to more decoupling and higher diffusive flux. 

%This is especially true when there is a thin interface. Without a bottleneck, the pressure gradient doesn't steepen and generate confining waves, so the diffusive flux remains large.

This study addresses the assertion in \cite{Everett2011} that cosmic ray energy will be roughly uniform inside and outside cold clouds and shows that this very much depends on interface properties. It also shows that the extent to which cosmic rays penetrate clouds and the relative importance of diffusive and advective fluxes depends somewhat on cosmic ray content already in the cloud. This is different from the single cosmic ray front simulations we have focused on in this paper, where the presence of a steep front very much negates the diffusive flux and locks cosmic rays to Alfv{\'e}n waves. In the now uniform pressure setup, the collisional loss time is not necessarily short enough to cause a steep pressure gradient, especially when the interface is narrow, and the diffusive flux can be significant. Future work will address more realistic scenarios with multiple cosmic ray bursts from different directions, as well as cosmic ray production within cold clouds.

\bibliographystyle{apj}
\bibliography{bibliography}

\end{document}